\documentclass{aa}  

\usepackage{graphicx}
\usepackage{xcolor}
\usepackage{txfonts}
\usepackage[normalem]{ulem}
\usepackage[pdftex,colorlinks=true,linkcolor=blue,citecolor=blue]{hyperref}
\newcommand{\aips}{\texttt{AIPS}}
\newcommand{\difmap}{\texttt{DIFMAP}}

\newcommand{\clean}{\texttt{CLEAN}}
\newcommand{\modelfit}{\texttt{MODELFIT}}

\newcommand{\vlba}{\mbox{VLBA}}
\newcommand{\fermi}{\textit{Fermi}/LAT}
\newcommand{\sma}{\mbox{SMA}}
\newcommand{\ovro}{\mbox{OVRO}}
\newcommand{\alma}{\mbox{ALMA}}

\newcommand{\source}{\mbox{4C\,$+$01.28}}
\newcommand{\gray}{\mbox{$\gamma$-ray}}

\makeatletter
\AtBeginDocument{%
  \@ifpackageloaded{lineno}{%
    \nolinenumbers
    \renewcommand\linenumbers{}%
    \renewcommand\runninglinenumbers{}%
  }{}
}
\makeatother

\begin{document}

   \title{Pinpointing the location of the \gray\, emitting region in the FSRQ \source}


   \author{F. R{\"o}sch\inst{1}
          \and
          M. Kadler\inst{1}
          \and
          E. Ros\inst{2} 
          \and
          L. Ricci\inst{1,2}
          \and
          M. A. Gurwell\inst{3}
          \and
          T. Hovatta\inst{4,5,6}
          \and
          N. R. MacDonald\inst{7}
          \and
          A. C. S. Readhead\inst{8}
          }

   \institute{Julius-Maximilians-Universit{\"a}t W{\"u}rzburg, Fakult{\"a}t f{\"u}r Physik und Astronomie, Institut f{\"u}r Theoretische Physik und Astrophysik, Lehrstuhl f{\"u}r Astronomie, Emil-Fischer-Str. 31, D-97074 W{\"u}rzburg, Germany \\ \email{florian.roesch@uni-wuerzburg.de}
   \and
   Max-Planck-Institut f{\"u}r Radioastronomie, Auf dem H{\"u}gel 69, 53121 Bonn, Germany
   \and
   Center for Astrophysics | Harvard \& Smithsonian, Cambridge, MA 02138, USA
   \and
   Finnish Centre for Astronomy with ESO (FINCA), University of Turku, FI-20014 Turku, Finland
   \and
   Aalto University Mets{\"a}hovi Radio Observatory, Mets{\"a}hovintie 114, FI-02540 Kylm{\"a}l{\"a}, Finland
   \and
   Aalto University Department of Electronics and Nanoengineering, PL~15500, FI-00076 Aalto, Finland
   \and
   The University of Mississippi, Department of Physics and Astronomy, Oxford, MS 38677, USA
   \and
   Owens Valley Radio Observatory, California Institute of Technology, Pasadena, CA 91125, USA
             }

   \date{Received ; accepted }

 
  \abstract
  {}
   {The flat-spectrum radio quasar (FSRQ) \source\,is a bright and highly variable radio and \gray\,emitter. We aim to pinpoint the location of the \gray\,emitting region within its jet in order to derive strong constraints on \gray\,emission models for blazar jets.}
   {We use radio and \gray\,monitoring data obtained with the Atacama Large Millimeter/submillimeter Array (\alma), the Owens Valley Radio Observatory (\ovro), the Submillimeter Array (\sma) and the Large Area Telescope on board the \textit{Fermi Gamma-ray Space Telescope} (\fermi) to study the cross-correlation between  \gray\,and multi-frequency radio light curves. Moreover, we employ Very Long Baseline Array (\vlba) observations at $43\,\mathrm{GHz}$ over a period of around nine years to study the parsec-scale jet kinematics of \source. To pinpoint the location of the \gray\,emitting region, we use a model in which outbursts shown in the \gray\,and radio light curves are produced when moving jet components pass through the \gray\,emitting and the radio core regions.}
   {We find two bright and compact newly ejected jet components that are likely associated with a high activity period visible
   in the \fermi\, \gray\,and different radio light curves. The kinematic analysis of the \vlba\, observations leads to a maximum apparent jet speed of $\beta_{\mathrm{app}}=19\pm10$ and an upper limit on the viewing angle of $\phi\lesssim4\degr$. Furthermore, we determine the power law indices that are characterizing the jet geometry, brightness temperature distribution and core shift to be $l=0.974\pm0.098$, $s=-3.31\pm0.31$ and $k_\mathrm{r}=1.09\pm0.17$, respectively, which are all in agreement with a conical jet in equipartition. A cross-correlation analysis shows that the radio light curves follow the \gray\ light curve. We pinpoint the location of the \gray\,emitting region with respect to the jet base to the range of $2.6\,\mathrm{pc}\leq d_{\mathrm{\gamma}}\leq20\,\mathrm{pc}$.
   } 
   {Our derived observational limits places the location of \gray\,production in \source\, beyond the expected extent of the broad-line region (BLR) and therefore challenges blazar-emission models that rely on inverse Compton up-scattering of seed photons from the BLR.}


   \maketitle

\section{Introduction}

Blazars, radio-loud active galactic nuclei (AGN) with jets pointing towards earth, represent the largest population of extra-galactic objects in the \gray\,band \citep{Abdo1}. They emit radiation throughout the entire electromagnetic spectrum from radio frequencies up to high \gray\,energies. However, the exact location where the \gray s are produced within the blazar jets is still under active discussion. While the radio emission is thought to be produced by synchrotron radiation from relativistic electrons, different leptonic \citep[e.g.,][]{Maraschi, Sikora94, Sikora09, Ghisellini09} and hadronic \citep[e.g.,][]{Mannheim} processes are discussed to explain the \gray\, emission. All these \gray\,emission models require seed photon fields to produce high-energy photons via inverse Compton (IC) scattering in the case of leptonic models or by photon-proton interactions in hadronic models. 
For IC models these seed photons can be provided by the jet's synchrotron photons that can be up-scattered to \gray\,energies by the same electron population that is responsible for the synchrotron radiation, which is called synchrotron self-Compton (SSC) emission \citep[e.g.,][]{Maraschi}. Alternatively, in external Compton (EC) and hadronic emission models, external photon fields are considered. These external photon fields can be provided by UV photons from the broad-line region \citep[BLR; e.g.,][]{Sikora94}, which is typically located $\lesssim1\,\mathrm{pc}$ away from the central super massive black hole \citep[SMBH; e.g.,][]{Zhang}, by IR photons between the BLR and the dust torus \citep[e.g.,][]{Sikora09} or by photons from the cosmic microwave background \citep[CMB; e.g.,][]{Ghisellini09}.

Due to the ubiquitous  intraday variability of the \gray\,emission of blazars which indicates a compact emission region, 
EC scattering of BLR photons is generally favored in the class of broad-line blazars. However, the same BLR photons used for EC scattering should become targets for the $\gamma-\gamma\rightarrow e^{\pm}$ process, leading to a strong cut-off in the \gray\,spectrum above $\sim20\,\mathrm{GeV}$, so that no emission at very high energies (VHE) $\geq100\,\mathrm{GeV}$ should be detectable \citep{Costamante}. Nevertheless, such VHE emission is observed in individual quasars \citep[e.g.,][]{Aleksic}. Moreover, \citet{Costamante} found no evidence for the expected spectral curvature due to BLR absorption in a sample of broad-line blazars suggesting that the \gray\,emitting region is located outside the BLR.

Studies of statistical samples with multi-wavelength and very long baseline interferometry (VLBI) information, support the idea of a \gray\,emitting region located well beyond the BLR. \citet{Jorstad01} found a connection between \gray\,outbursts and ejections of superluminal jet components from the radio core that places the \gray\,emitting region several parsecs downstream of the central engine. This scenario is supported by several cross-correlation studies between \gray\, and radio light curves \citep[e.g.,][]{Fuhrmann, Max-Moerbeck, Kramarenko2022} which found that the \gray\,light curves lead the radio light curves for many blazars, suggesting the \gray\,emitting region to be located upstream of the radio cores, and, in some cases, outside of the BLR's outer edges.

In this paper, we perform a detailed individual-source study to explore the described scenario using the blazar \source\, as a laboratory. \source\, also known as TXS\,1055+018, RGB\,J1058+015, and 4FGL\,J1058.4+0133) has a redshift of $z=0.89$ \citep{Jorstad} and is optically classified as a FSRQ \citep[e.g.,][]{Lister2005,Veron2010}. In the case of \source\, rich multi-frequency radio observational data are available with the light curves showing high variability alongside outbursts occurring every $1$-$2\,\mathrm{years}$ \citep{Lister2}. 
Interestingly, Very Long Baseline Array (\vlba) observations at $43\,\mathrm{GHz}$ of \source\, show the ejection of a superluminal jet component that seems to be associated with a bright \gray\,outburst \citep{MacDonald}. 
Overall, \source\, represents a perfect target to pinpoint the location of the \gray\,emitting region.

The paper is structured as follows. In Sect.~\ref{sec:obs} we present the data and the data-analysis methods. In Sect.~\ref{sec:results} we show the results of the analysis of the $43\,\mathrm{GHz}$ \vlba\, observations (Sect.~\ref{sec:vlba_results}) and of the cross-correlation of multi-frequency light curve data (Sect.~\ref{sec:cross}). In Sect.~\ref{sec:disc} we discuss how the combination of the results found by the analysis of the \vlba\, and the multi-frequency data can be used to pinpoint the location of the \gray\,emitting region in the jet of \source. Finally, we summarize the results and draw our conclusions in Sect.~\ref{sec:sum}.
Throughout the paper, we use a $\Lambda$CDM cosmological model with $H_{\mathrm{0}}=70\,\mathrm{km\,s^{-1}Mpc^{-1}}$, $\Omega_{\mathrm{m}}=0.30$ and $\Omega_{\mathrm{\Lambda}}=0.70$.

\section{Observations and data analysis}
\label{sec:obs}

\subsection{\vlba\, data}
\label{sec:vlba}

To study the parsec-scale jet structure of \source, we employed radio data from the Boston University (BU) Blazar Monitoring Program BEAM-ME\footnote{\url{http://www.bu.edu/blazars/BEAM-ME.html}}. 
This data, observed with the \vlba\, at $43\,\mathrm{GHz}$ since April 2009, has been calibrated and imaged by the BU-group using the Astronomical Image Processing System \citep[\aips;][]{Greisen} provided by the National Radio Astronomy Observatory (NRAO) and the \clean\, algorithm implemented in the program \difmap\,\citep{Shepherd}. More detailed information on the calibration and imaging process can be found in \citet{Jorstad05}, \citet{Jorstad} and \citet{Weaver2022}.

\begin{figure*}[hpt!]
   \centering
   \includegraphics[width=0.24\hsize]{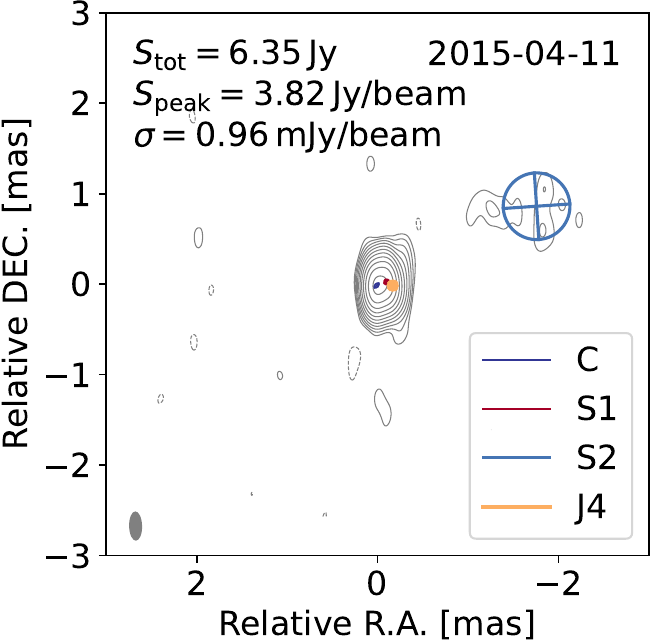}
   \includegraphics[width=0.24\hsize]{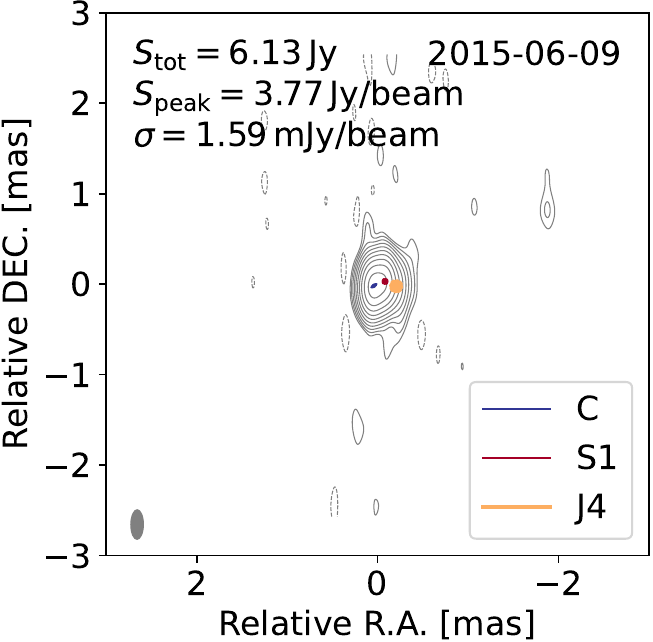}
   \includegraphics[width=0.24\hsize]{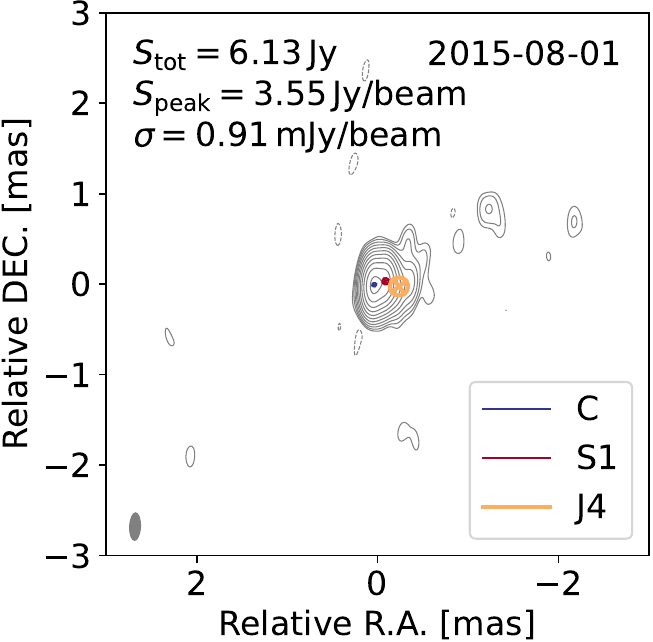}
   \includegraphics[width=0.24\hsize]{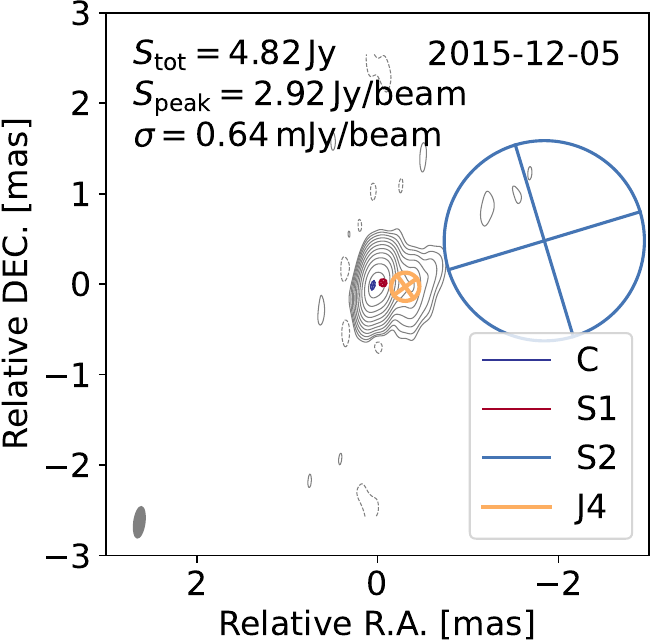}
   \includegraphics[width=0.24\hsize]{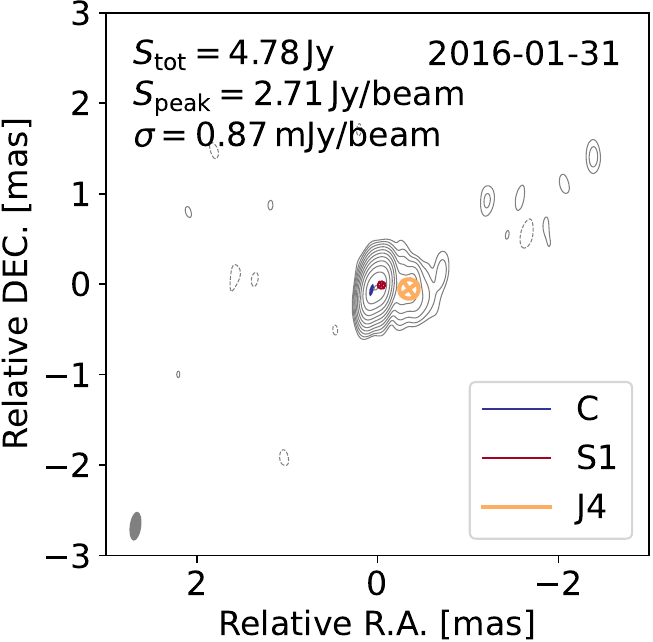}
   \includegraphics[width=0.24\hsize]{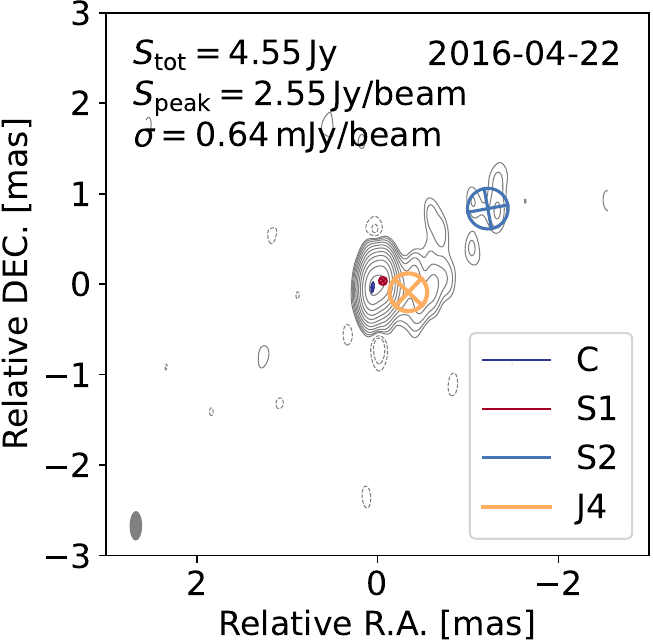}
   \includegraphics[width=0.24\hsize]{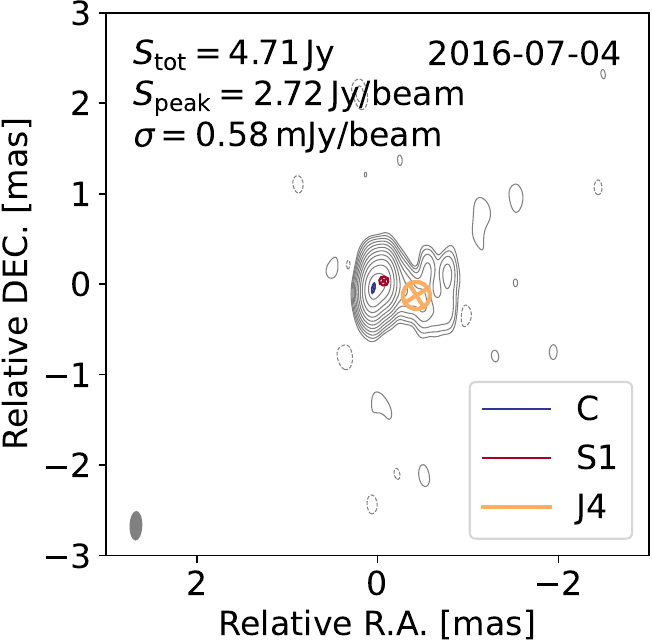}
   \includegraphics[width=0.24\hsize]{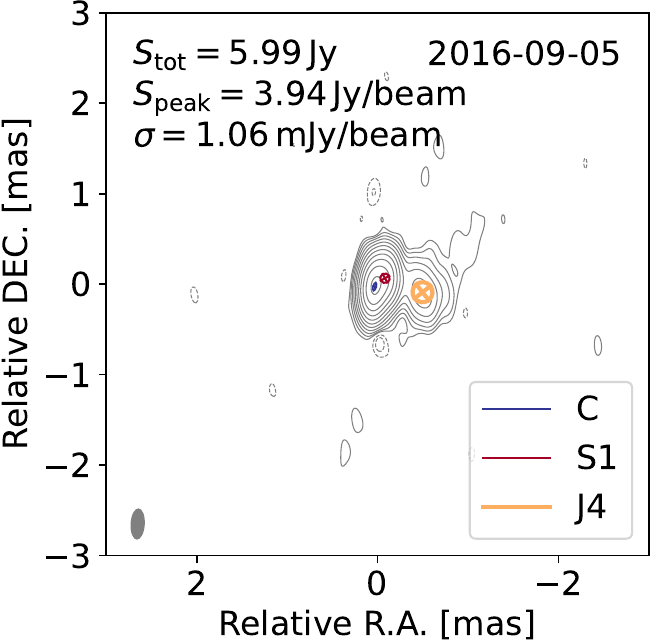}
   \includegraphics[width=0.24\hsize]{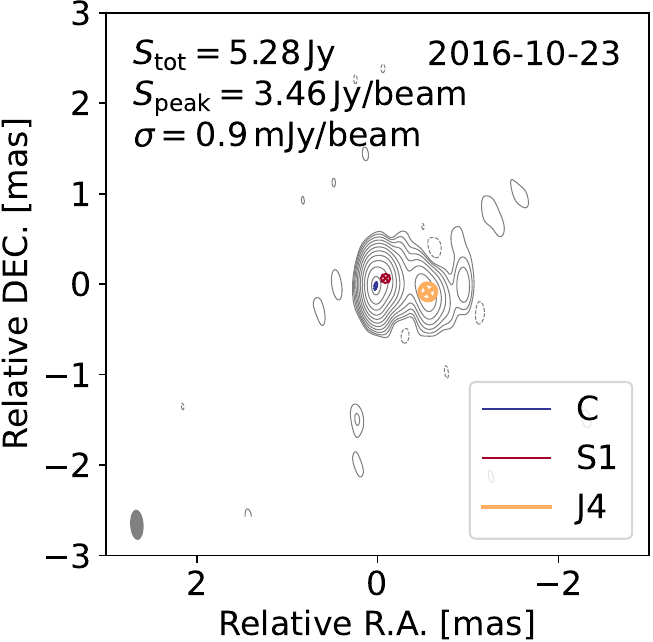}
   \includegraphics[width=0.24\hsize]{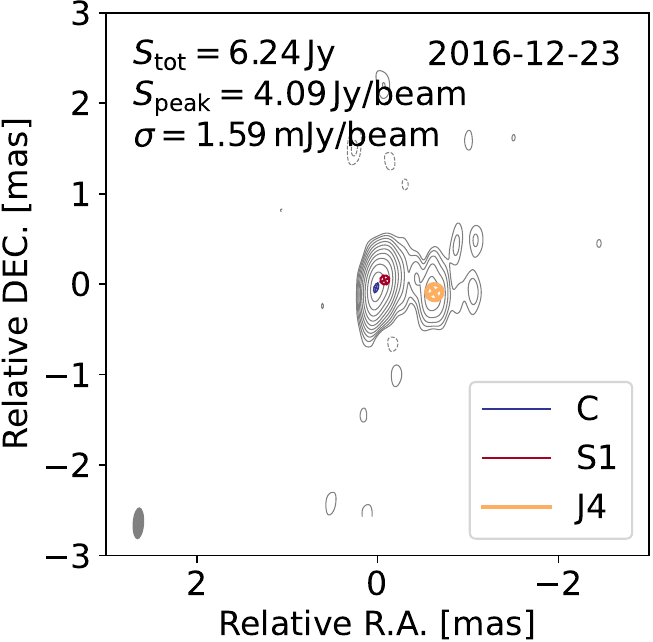}
   \includegraphics[width=0.24\hsize]{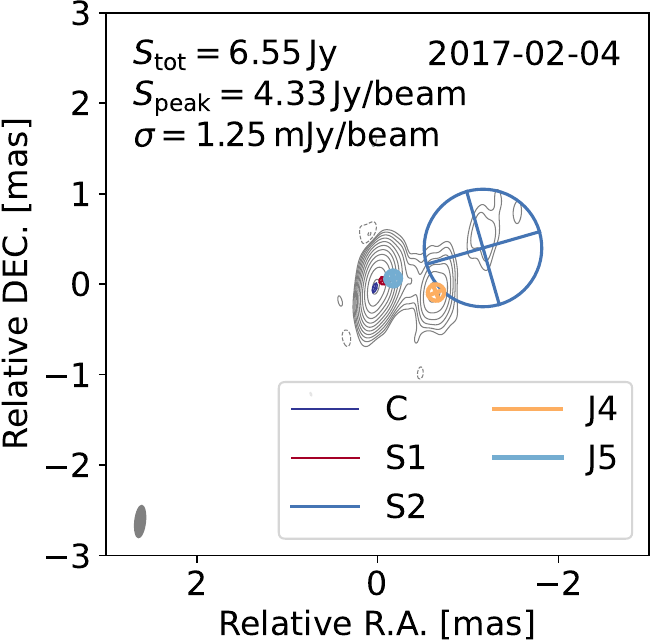}
   \includegraphics[width=0.24\hsize]{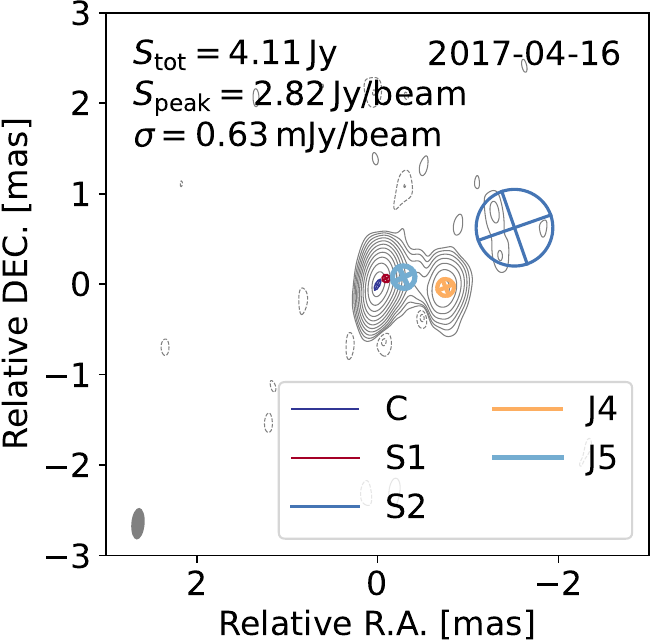}
   \includegraphics[width=0.24\hsize]{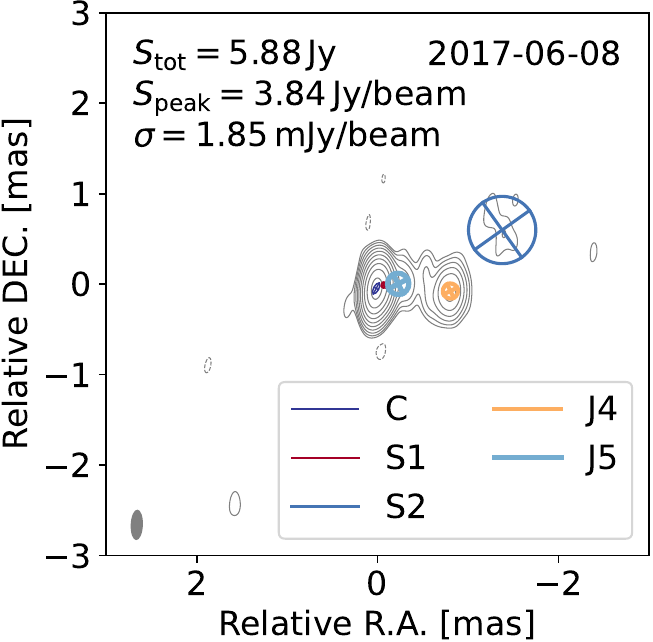}
   \includegraphics[width=0.24\hsize]{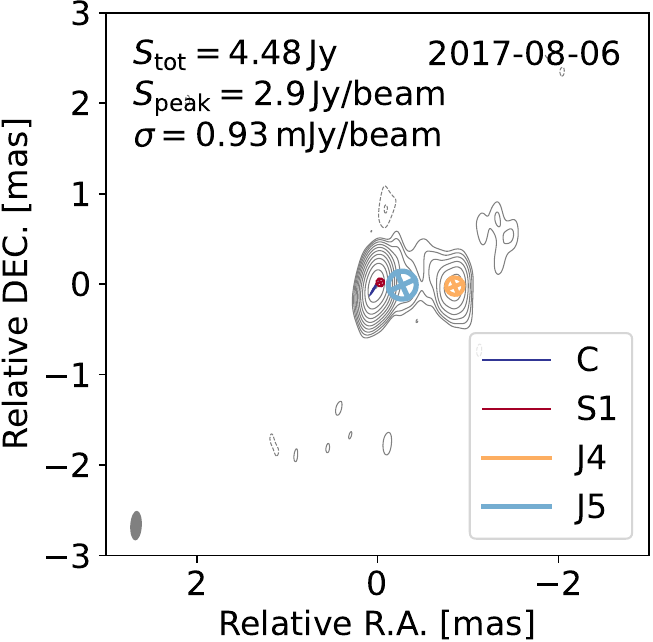}
   \includegraphics[width=0.24\hsize]{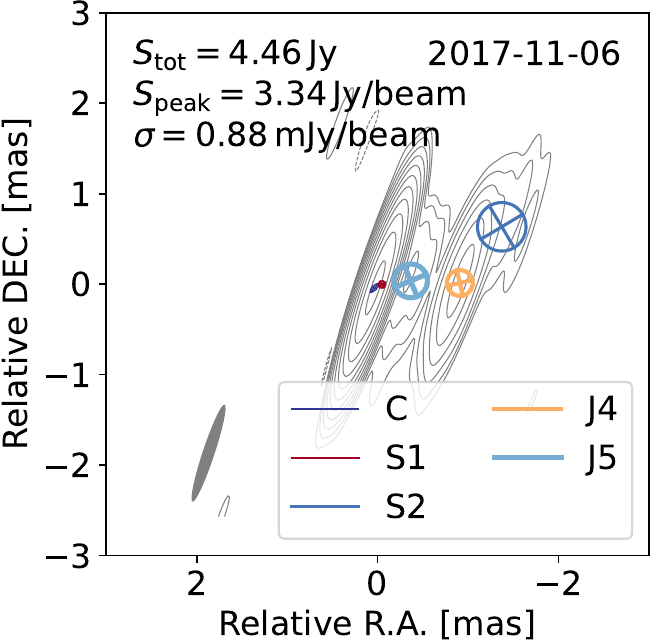}
   \includegraphics[width=0.24\hsize]{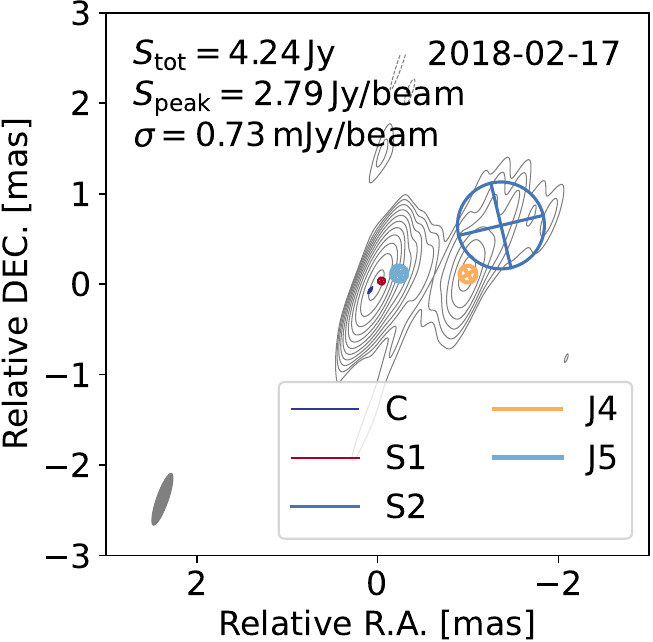}
   \includegraphics[width=0.24\hsize]{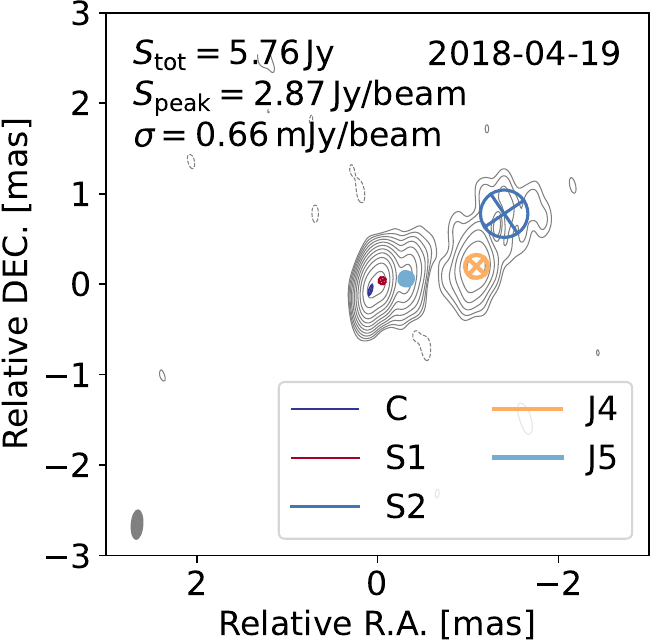}
   \includegraphics[width=0.24\hsize]{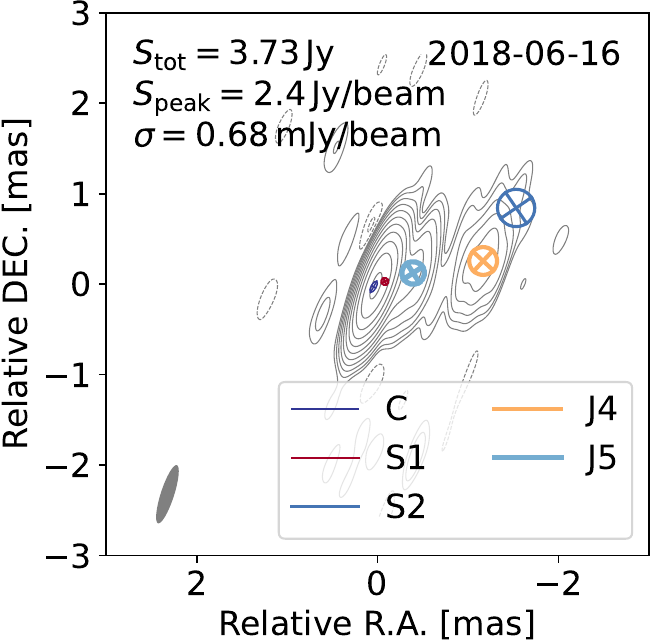}
   \includegraphics[width=0.24\hsize]{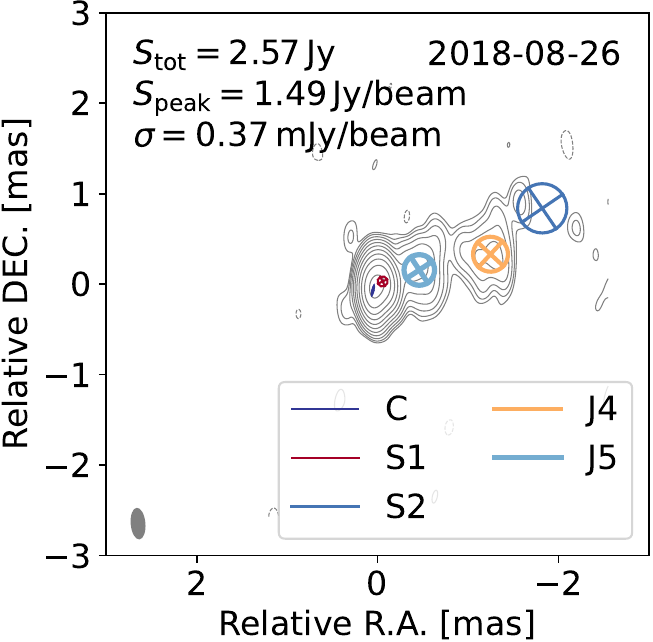}
   \includegraphics[width=0.24\hsize]{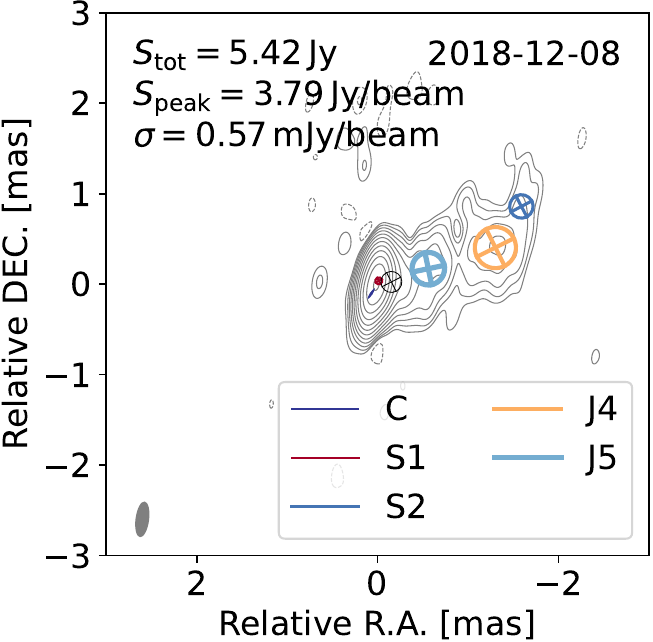}
      \caption{Selected uniformly weighted $43\,\mathrm{GHz}$ \vlba\, total intensity images of the FSRQ \source\, with the fitted Gaussian components overlaid. $S_{\mathrm{tot}}$ is the total integrated flux density, $S_{\mathrm{peak}}$ is the highest flux density per beam and $\sigma$ is the noise level. The gray ellipse in the bottom left corner corresponds to the beam. The contours begin at $3\sigma$ and increase logarithmically by a factor of 2. The image parameters are listed in Table~\ref{image}. The images show two newly ejected jet components appearing in August 2015 (J4) and June 2018 (J5) that can be tracked back until April 2015 and February 2017, respectively. In this work we focus on these two newly ejected components. Additional plots of the epochs observed before April 2015 are shown in Appendix~\ref{app:source} in Figs.~\ref{fig:app_images1} to~\ref{fig:app_images2}.}
         \label{fig:images1}
   \end{figure*}

Since in this work we focus on two newly ejected jet features that appeared in August 2015 and June 2018, and seem to be associated with a period of high \gray\, activity, 
we concentrate on 20 epochs observed on a similar time range, specifically between April 2015 and December 2018 (see Fig.~\ref{fig:images1}). 
However, since \citet{Weaver2022} presented an independent analysis of the $43\,\mathrm{GHz}$ BU data from April 2009 until December 2018 (consisting of 51 epochs), to guarantee comparability with their study, we consider the epochs starting from April 2009 as well.
These additional epochs are shown in Figs.~\ref{fig:app_images1} to~\ref{fig:app_images2}.

Although most of the 
images show one bright feature at the center and a jet in the northwest direction (see Fig.~\ref{fig:images1} and Figs.~\ref{fig:app_images1} to~\ref{fig:app_images2}), six epochs show unique and relatively bright features in the northern direction. Since such a structure is expected to be an artifact, in an attempt to improve them, we re-imaged them, obtaining images more in agreement with the structure seen at other epochs.

The parameters of all the analyzed epochs are reported in Table~\ref{image}.
In Fig.~\ref{fig:combkinvar}, we show the resultant light curve of the total flux density, focusing on the time range October 2012 - December 2018.
We assume an uncertainty of 5\,\% of the measured flux which corresponds to the typical amplitude calibration error \citep{Jorstad}. 

\subsubsection{Image modelling}
\label{sec:modelfitting}

To study the time evolution of the parsec-scale jet structure of \source, we fitted the fully calibrated visibility data of all 51 epochs with 2D Gaussian components using the \modelfit\, task within \difmap. While the core components are fitted with elliptical Gaussians, for the jet components we used circular Gaussians. The components, overlaid on the uniformly weighted images of \source, are plotted in Figs.~\ref{fig:app_images1} to~\ref{fig:app_images2} as well as Fig.~\ref{fig:images1} and their parameters 
are listed in Table~\ref{komp}. 
We filtered out the unresolved components, with the resolution limit calculated according to \citet{Lobanov}
 \begin{equation}
   \label{eq:limit}
   a_{\mathrm{lim}}=2^{2-\beta/2}b_{\mathrm{\psi}}\sqrt{\frac{\ln 2}{\pi}\ln\left(\frac{S/N}{S/N-1}\right)}\, ,
  \end{equation}    
where $b_{\mathrm{\psi}}$ is the Full Width at Half Maximum (FWHM) of the beam size along an arbitrary position angle $\psi$, $\mathrm{S/N}$ is the signal-to-noise ratio and $\beta=0$ for uniform weighting (for natural weighting $\beta=2$). The $\mathrm{S/N}$ is calculated as $S_{\mathrm{comp}}/\sigma_{\mathrm{comp}}$, where $S_{\mathrm{comp}}$ is the flux density of the fitted Gaussian component and $\sigma_{\mathrm{comp}}$ is the noise level of the area that is occupied by this component. 
In the case of circular Gaussian jet components, $b_{\mathrm{\psi}}$ represents the major axes of the corresponding beams. In contrast, for the elliptical Gaussian core components, $b_{\mathrm{\psi}}$ is determined by measuring the FWHM along the position angle of the major and minor axes of the fitted core component. This approach establishes resolution limits for both axes individually.
Whenever an axis is smaller than the corresponding resolution limit, the component is considered unresolved. 
By following this procedure, we found five unresolved circular jet components and two unresolved minor axes of elliptical core components. 

\subsubsection{Kinematic analysis}
\label{sec:kinanalysis}

To analyze the motion of the jet components, we cross-identified the components of the different epochs and fitted their distance to the core $d$ using the following equations:
\begin{eqnarray}
      d(t) & = & d_{\mathrm{mid}}+\mu(t-t_{\mathrm{mid}})\,, \label{eq:kin} \\
      \beta_{\mathrm{app}}        & = & \frac{\mu}{c}\cdot\frac{D_{\mathrm{L}}}{1+z}\,, \label{eq:betaapp}
   \end{eqnarray}
in which $t_{\mathrm{mid}}=(t_{\mathrm{max}}+t_{\mathrm{min}})/2$ is the midpoint of observation, $d_{\mathrm{mid}}$ is the distance of the jet component to the core at $t_{\mathrm{mid}}$, $\mu$ is the angular speed, $\beta_{\mathrm{app}}$ is the apparent speed in units of speed of light $c$ and $D_{\mathrm{L}}$ is the luminosity distance. For the uncertainties of the distances, we used the semi-major axes of the corresponding components, or their resolution limit for unresolved ones. 

Using the same classification scheme 
implemented in \citet{Jorstad} and \citet{Weaver2022}, we classified jet components with detections at $\geq10$ epochs and angular speed $\mu<2\sigma$ as stationary features, while all other jet components were classified as moving jet features. Stationary jet components are labeled with S, moving jet components are labeled with J and the core components are labeled with C. For moving jet components, we computed the ejection epoch $t_{\mathrm{0}}$ as the point in which the separation of the jet component to the core equals zero, namely
\begin{equation}
   \label{eq:t0}
   t_\mathrm{0} = -\frac{d_{\mathrm{mid}}}{\mu}+t_{\mathrm{mid}}\,.
  \end{equation}

\subsubsection{Brightness temperature}
\label{sec:tempanalysis}

Using the parameters of the Gaussian components given in Table~\ref{komp}, we calculated the brightness temperatures $T_{\mathrm{B}}$ of all components by 
 \begin{equation}
   \label{eq:tb}
   T_\mathrm{B} = \frac{2 \ln 2}{\pi k_{\mathrm{B}}}\frac{S_{\mathrm{comp}} \lambda^2 (1+z)}{a_{\mathrm{maj}} a_{\mathrm{min}}}\,,
  \end{equation}
in which $a_{\mathrm{maj}}$ and $a_{\mathrm{min}}$ are the FWHM of the major and minor axes of the Gaussian component, $\lambda$ is the wavelength of observation, $z$ is the redshift and $k_{\mathrm{B}}$ is the Boltzmann constant \citep{Kovalev}. Assuming relative uncertainties of 20\,\% for the major and minor axes and 5\,\% for the flux density, we calculated relative uncertainties of 29\,\% for the brightness temperatures. For unresolved components, the resolution limits are used, which leads to a lower limit on the brightness temperatures. The brightness temperatures obtained are also listed in Table~\ref{komp} and are plotted in the upper panel of Fig.~\ref{fig:kinmodbeschleunigt}.

\subsection{Other radio data}
\label{sec:radiodata}

\source\, was also observed at various other radio wavelengths. Data measured by the Atacama Large Millimeter/submilimeter Array (\alma) at band 3 (\alma\,3: $84-116\,\mathrm{GHz}$), band 6 (\alma\,6: $211-275\,\mathrm{GHz}$) and band 7 (\alma\,7: $275-373\,\mathrm{GHz}$) are available at the \alma\, Calibrator Source Catalogue\footnote{\url{https://almascience.eso.org/sc/}}. Furthermore, we used data observed by the Submillimeter Array (\sma) at a wavelength of $1.3\,\mathrm{mm}$ that are available at the Submillimeter Calibrator List\footnote{\url{http://sma1.sma.hawaii.edu/callist/callist.html}} \citep{Gurwell2007}. Because the flux densities of the individual \alma\, and \sma\, light curves were not measured at the same frequencies, we calculated the mean frequencies $\nu$, with uncertainties given by the standard deviation, for all four light curves respectively. We obtained $\nu=(97.0\pm6.2)\,\mathrm{GHz}$ (\alma\,3), $\nu=(225.6\pm4.8)\,\mathrm{GHz}$ (\sma), $\nu=(232.5\pm2.4)\,\mathrm{GHz}$ (\alma\,6) and $\nu=(341.9\pm8.5)\,\mathrm{GHz}$ (\alma\,7). These mean frequencies were used for further calculations.
In addition, we employ the $15\,\mathrm{GHz}$ light curve from the Owens Valley Radio Observatory (\ovro) blazar monitoring program \citep{Richards}. Note that we only used data observed from October 2012 to December 2018 in this study (see also Sect.~\ref{sec:corr} for more details). All these light curves are plotted in Fig.~\ref{fig:combkinvar} and show similar structures with flux density peaks in 2014, 2015 and 2017 and a prominent flux density minimum in 2013, in agreement with the \vlba\, light curve.

\subsection{\fermi\, data}
\label{sec:fermi}

\begin{figure}
            \includegraphics[width=\hsize]{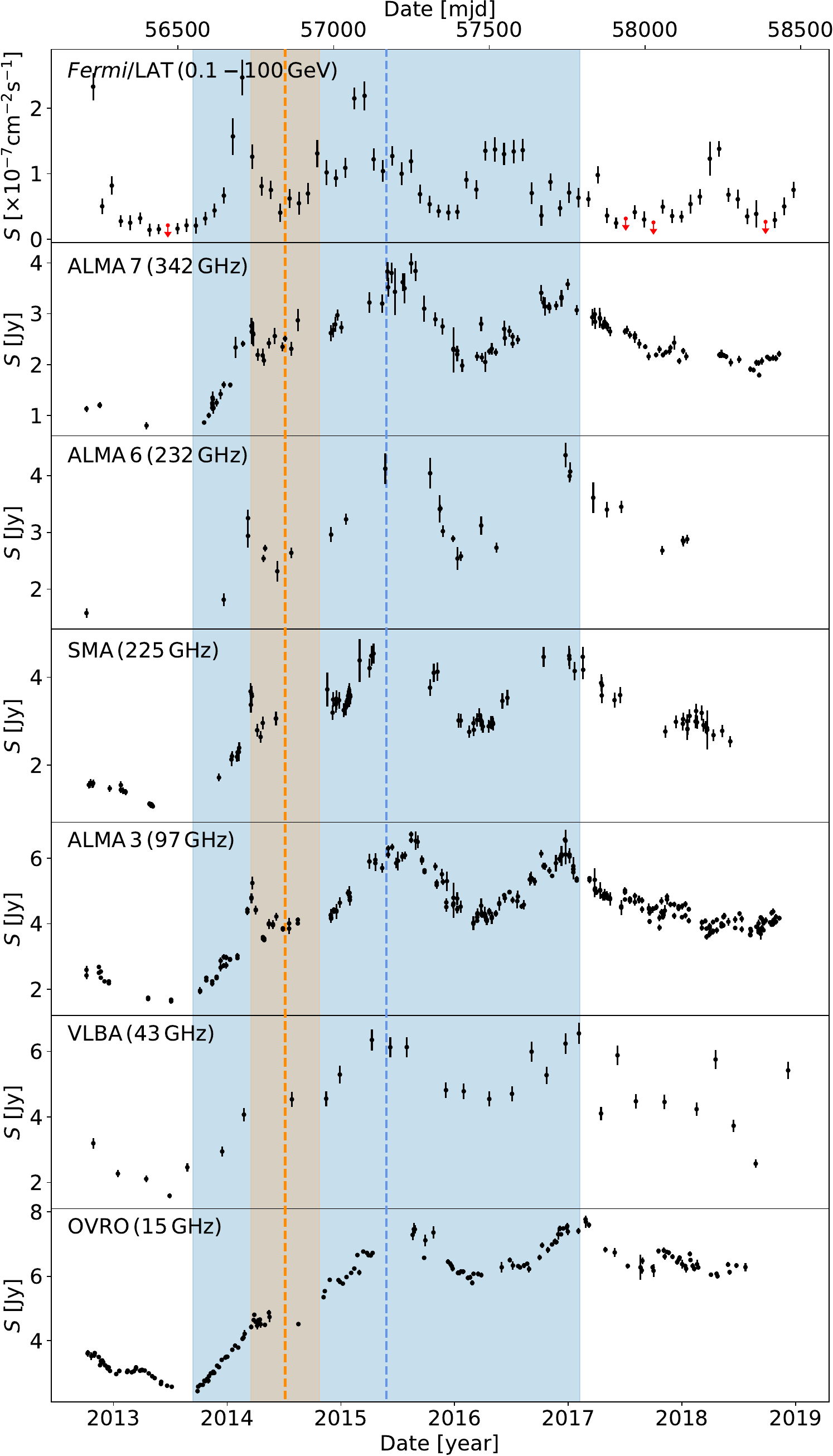}
      \caption{Monthly binned \gray\,and radio light curves observed by \fermi, \alma, \sma\, and \ovro\, and the \vlba\, total flux density light curve, showing similar variability behavior. In the upper panel red arrows indicate upper limits. The vertical dashed lines represent the ejection epochs of the jet components J4 (orange) and J5 (blue) with their $1\sigma$ uncertainties shown as the orange (J4) and blue (J5) bands. The ejection of the components J4 and J5 falls into the period of high activity starting in late 2013.}
         \label{fig:combkinvar}
   \end{figure}

In addition to radio light curves, we also used information from other wavelengths for our analysis. Specifically, \source\, was also observed by \fermi\,\citep{Atwood} at \gray\,energies between $0.1\,\mathrm{GeV}$ and $100\,\mathrm{GeV}$. Weekly and monthly binned \gray\, light curves are available at the \fermi\, Light Curve Repository\footnote{\url{https://fermi.gsfc.nasa.gov/ssc/data/access/lat/LightCurveRepository}.} \citep[LCR;][]{FermiLCR}. We used the monthly binned light curve of \source, which has only $\sim 4\%$ of upper-limit data points (defined by test statistic (TS) values $< 4$; for comparison, the weekly binned light curve contains $\sim 26\%$ of upper limits), which is ideal for the further analysis. Moreover, several radio light curves have relatively large gaps leading to mean sampling times more comparable to monthly binning. The \gray\, light curve measured from October 2012 to December 2018 is shown in Fig.~\ref{fig:combkinvar}. 
It shows high variability with prominent bright outbursts in 2014 and 2015 and a prominent flux minimum in 2013, as can also be seen in the radio light curves. 


\subsection{Cross-correlation analysis}
\label{sec:corr}

Because of their similar behaviour, we assume that the light curves measured by \alma, \ovro, \sma, \vlba\, and \fermi\, are correlated (see Fig.~\ref{fig:combkinvar}).
To test this assumption, we performed a cross-correlation analysis between the \fermi\, \gray\,light curve and the radio ones. We neglected the \alma\,6 and \vlba\, light curves in this study because of their poor sampling rate compared to the other radio observations. To ensure comparability of the cross-correlation results obtained using the different radio light curves, we only used flux densities measured during the time range form October 2012 to December 2018, during which every involved light curve has measurements (Fig.~\ref{fig:combkinvar}).

Several \gray\,data points are only upper limits (indicated as red arrows in the upper panel of Fig.~\ref{fig:combkinvar}) and were neglected for the analysis \citep[similar to the approach shown in][]{Max-Moerbeck}. 
Nonetheless, to assess the impact of their exclusion we performed the analysis treating them as actual data points finding no significant discrepancies. 


\subsubsection{Cross-correlation functions}
\label{sec:corr_function}

The cross-correlation function (CCF) of two evenly sampled light curves $x(t_{\mathrm{i}})$ and $y(t_{\mathrm{i}})$ as function of the time lag $\tau$ is given by
\begin{equation}
   \label{eq:ccf}
   \mathrm{CCF}(\tau) = \frac{1}{N}\sum^\mathrm{N}_\mathrm{i}\frac{(x(t_\mathrm{i})-\bar{x})(y(t_\mathrm{i}-\tau)-\bar{y})}{\sigma_\mathrm{x}\sigma_\mathrm{y}}\,,
\end{equation}
in which $\bar{x}$, $\bar{y}$ and $\sigma_\mathrm{x}$, $\sigma_\mathrm{y}$ are the mean values and standard deviations of $x(t_{\mathrm{i}})$ and $y(t_{\mathrm{i}})$, respectively \citep[e.g.,][]{Fuhrmann}.

Due to the unevenly sampled radio light curves, we used two different methods for the cross-correlation analysis, namely the Discrete Cross-Correlation Function (DCF) according to \citet{Edelson} as well as the Interpolated Cross-Correlation Function \citep[ICF;][]{White}.

According to \citet{Edelson}, at first, the set of unbinned discrete correlations $(\mathrm{UCCF}_\mathrm{ij})$ of all measured pairs of data points from the two investigated light curves $(x(t_{\mathrm{i}}), y(t_{\mathrm{j}}))$, associated with the pairwise lag $\Delta t_{\mathrm{ij}}=t_{\mathrm{j}}-t_{\mathrm{i}}$, is computed as 
\begin{equation}
   \label{eq:uccf}
   \mathrm{UCCF}_\mathrm{ij} = \frac{(x(t_\mathrm{i})-\bar{x})(y(t_\mathrm{j})-\bar{y})}{\sigma_\mathrm{x}\sigma_\mathrm{y}}\,.
\end{equation}
Binning these unbinned correlations in time and averaging over the pairs for which $\tau-\frac{\Delta\tau}{2}\leq\Delta t_{\mathrm{ij}}<\tau+\frac{\Delta\tau}{2}$ then leads to the DCF correlation coefficients at time lag $\tau$. For this we used the average sampling time of both investigated light curves as the bin size $\Delta\tau$ \citep[see, e.g.,][]{Markowitz2003}. We first calculated the median sampling times of the two individual light curves separately and then used the mean value of those as $\Delta\tau$.

For the ICF method introduced by \citet{White}, each measured data point $x(t_{\mathrm{i}})$ was paired with the interpolated values $y(t_{\mathrm{i}}+\tau)$. In a second pass, the measured data points $y(t_{\mathrm{j}})$ were paired with the interpolated values $x(t_{\mathrm{j}}-\tau)$. For this purpose, we used a piece-wise linear interpolation with an interpolation unit of 0.5 times the average sampling time of the two light curves used to compute the ICF \citep[e.g.,][]{Markowitz2003}. Finally, averaging over both results leads to the ICF correlation coefficients at time lag $\tau$. 

In order to obtain correlation coefficients that are identical to the standard linear correlation coefficient known as Pearson's r, we calculated the DCF and ICF with local normalization. For this, we only used data points that contribute to the DCF or ICF at any particular time lag to calculate the means and standard deviations used in Eqs.~(\ref{eq:ccf}) and~(\ref{eq:uccf}) \citep[see, e.g.,][]{White,Welsh1999,Fuhrmann,Max-Moerbeck}. Note that the CCF of two unrelated red-noise light curves can result in spurious correlation coefficients higher than expected \citep{Welsh1999,Markowitz2003}, especially for time lags greater than 1/3 of the duration of the investigated light curves \citep{Press1978}. Because of the red-noise nature of our light curves, we only calculated correlation coefficients for time lags between $\pm$1/3 of the duration where both light curves overlap in sampling to avoid such spurious correlation coefficients.

\subsubsection{Correlation significance}
\label{sec:corr_significance}

To determine the significance of the cross-correlation between the \fermi\,\gray\,and the different radio light curves, we utilized the method of mixed source correlations \citep[e.g.,][]{Fuhrmann}. For this purpose we calculated the DCF and ICF between each radio light curve of \source\, and 226 unrelated \gray\,light curves of other blazars available at the \fermi\, LCR in the same way as described in Sect.~\ref{sec:corr_function}. Comparing these mixed correlation coefficients with those derived using only the light curves of \source\, for each time lag, we are able to estimate the probability that the latter are produced by chance correlations.

According to \citet{FermiLCR}, the \fermi\, LCR provides light curves of 1525 variable 4FGL-DR2 catalog sources with variability indices $>21.67$, which corresponds to the chance of $<1\%$ of being steady. Most of the sources are of the blazar type. However, there are also a few light curves of other source classes available at the LCR. To guarantee comparability, we only selected blazar light curves for the significance analysis. Furthermore, since we neglected upper limits in the correlation analysis (see Sect.~\ref{sec:corr}), we set a selection cut on the average significance given by the average TS values of the \gray\,detections of the LCR blazars to ensure a high enough sampling of the \gray\,light curves. Therefore, we only chose light curves with an average significance $>47.4$, which corresponds to 1/2 of the average significance of \source\, of $94.8$. 
With these selection criteria, we found a sample of 226 blazars (without \source) consisting of 108 FSRQs, 110 BL Lac objects and 8 blazar candidates of uncertain type. 

For all these \gray\,light curves we 
calculated the mixed cross-correlations neglecting upper limits. 
Here, we used the same bin size and interpolation unit as used for calculating the DCF and ICF using the \source\,\gray\,light curve. We then calculated the probability density function (PDF) of the distribution of the derived 226 different mixed correlation coefficients for every time lag. From these PDFs we then estimated the two sided $68.27\%$, $95.45\%$ and $99.73\%$ confidence intervals which we refer to in the following as Gaussian equivalent $1\sigma$, $2\sigma$ and $3\sigma$ confidence levels, respectively. 

\subsubsection{Time lag uncertainties}
\label{sec:corr_uncertainties}

To estimate the uncertainties of the time lags of the DCF and ICF peak coefficients, we used the Monte Carlo simulation method introduced by \citet{Peterson} for both methods. According to these authors, the uncertainties of the time lags depend mainly on the measurement uncertainties of the flux densities and the sampling rate of the light curves. To take the uncertainties in flux density into account, we modified each flux density measurement by random Gaussian deviates based on its individual uncertainty, which is referred to as flux randomization (FR). Moreover, to account for the sampling rate of the light curves, we used the method of random subset selection (RSS), in which we randomly discarded up to $37\%$ of the data points of the light curve. Combining both methods, FR and RSS, into one Monte Carlo run, we simulated 1000 randomly modified pairs of light curves and calculated the DCF and ICF as explained in Sect.~\ref{sec:corr_function}. For this purpose, we used the same bin size and interpolation unit as for the cross-correlation analysis between the original light curves. From these 1000 correlation functions we calculated the distribution of the time lags of the corresponding peak coefficients using only peak coefficients that are higher than the corresponding  $1\sigma$, $2\sigma$ or $3\sigma$ confidence level above which the peak coefficient derived from the original light curves lies. For example, if the peak coefficient from the original light curves lies between the corresponding $2\sigma$ and $3\sigma$ confidence levels, we only used time lags at peak coefficients higher than the $2\sigma$ confidence level to calculate the distribution. We then estimated the uncertainty of the time lag at the peak coefficient from the original light curves to be the standard deviation of the derived time lag distribution.

\section{Results}
\label{sec:results}

\subsection{$43\,\mathrm{GHz}$ VLBA data}
\label{sec:vlba_results}

To analyze the motion of the jet components, we studied their position relative to the core component. Their time evolution between April 2015 and December 2018 is shown in Fig.~\ref{fig:images1}, in which the fitted Gaussian components are plotted overlaid on the uniformly weighted images of \source. 
We highlight how in August 2015, a new bright and compact jet feature appeared moving outwards in the jet that is identified as component J4 and can be tracked back until April 2015. In June 2018, a second moving jet component emerged.
We were able to track it down to February 2017 and we identified it as component J5. 
In this work, we focus on these two newly ejected jet components. 
However, (see Sect.~\ref{sec:kin}), we also studied the kinematics of the epochs observed before April 2015 to ensure the comparability with an independent model-fit analysis of all 51 epochs spanning around 9 years of observations, from April 2009 until December 2018 \citep{Weaver2022}. 
The time evolution of the jet components of these additional epochs is shown in Figs.~\ref{fig:app_images1} to~\ref{fig:app_images2}. 
With the results of the kinematic analysis, we were able to calculate an upper limit for the viewing angle for \source\, (see Sect.~\ref{sec:orientation}). 
Furthermore, we also used the size and brightness temperatures of the jet components to study the jet geometry (see Sect.~\ref{sec:geometry}) and brightness-temperature gradient (see Sect.~\ref{sec:Tb_gradient}), respectively.

\subsubsection{Kinematic analysis}
\label{sec:kin}

\begin{figure}
   \centering
   \includegraphics[width=\hsize]{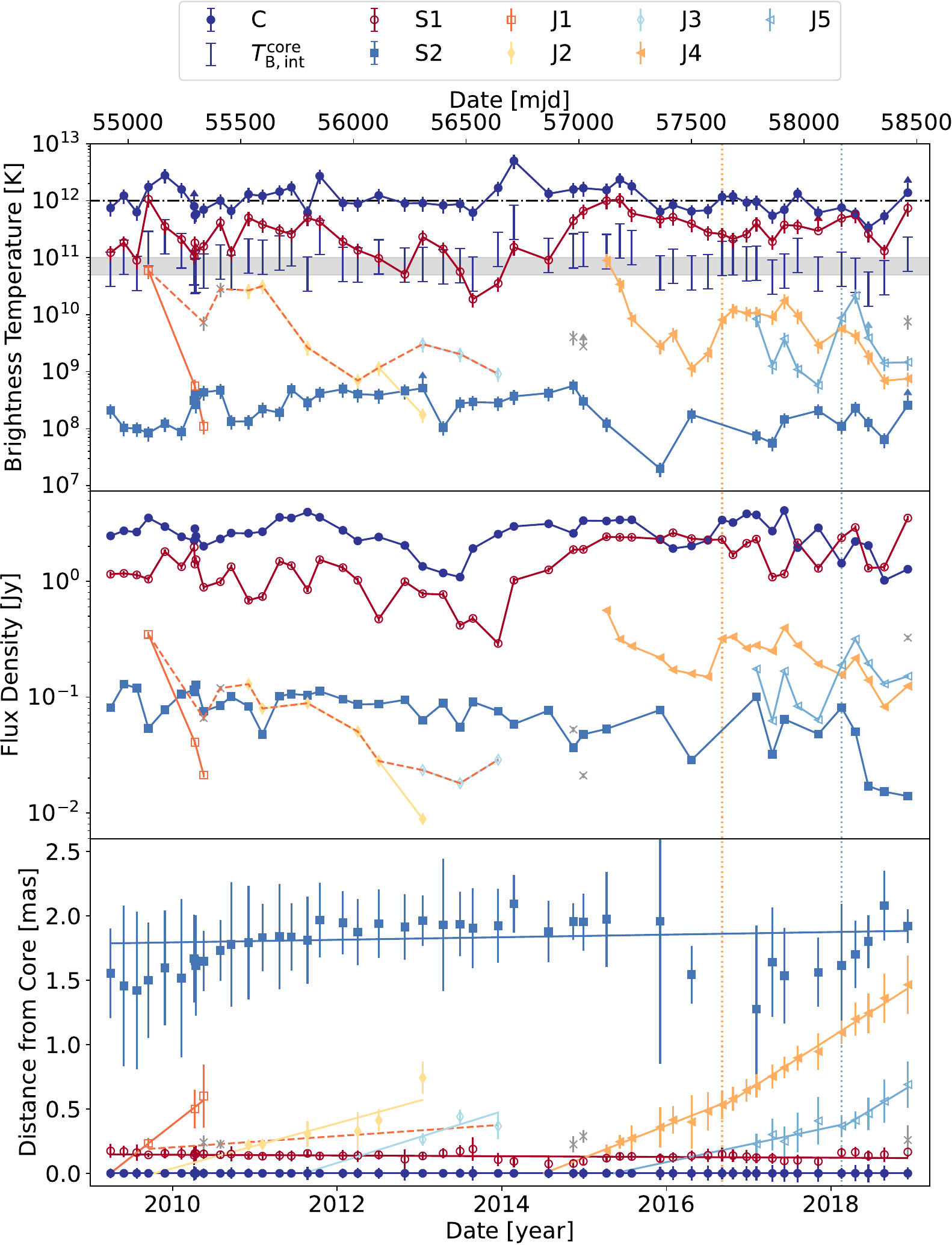}
      \caption{Upper panel: Observed brightness temperatures of the jet components with relative uncertainties of $29\%$ plotted over time. The arrows denote lower limits for unresolved components. The horizontal dashed-dotted line represents the inverse Compton limit of $10^{12}\,\mathrm{K}$. The gray-shaded area denotes the range of brightness-temperature values for a jet in equipartition. The intrinsic brightness temperature of the core (blue bars) is mostly consistent with equipartition and lies only above equipartition close to the ejection epochs of moving jet components. Middle panel: Flux densities of the jet components with relative uncertainties of $5\%$ plotted over time. Lower panel: Distance of the jet components relative to the core component plotted over time. The solid lines are fitted via linear regression and their gradients represent the angular speed of the corresponding component. The dashed fitted line represents an alternative kinematics model for the components J1, J2 and J3 (see Sect.~\ref{sec:kin}). Components J4 and J5 seem to accelerate and are therefore fitted by two separated linear regressions each with their transition points indicated by the two vertical dotted lines. At these transition points their brightness temperatures and flux densities show a steep increase. 
              }
         \label{fig:kinmodbeschleunigt}
   \end{figure}

To calculate the apparent speeds and ejection epochs of the jet components, we fitted their distances from the core with respect to the observation time as described in Sect.~\ref{sec:kinanalysis} \citep[also see][for a brief discussion of this kinematics analysis]{Roesch2022}{}{}. In the bottom panel of Fig.~\ref{fig:kinmodbeschleunigt} the distance of the jet components to the core is plotted using the fitted Gaussians' semi-major axes or the resolution limits for unresolved components as uncertainties. 
In all 51 epochs, we found a stationary component (S1) close to the core alongside a second stationary one in almost all epochs at distances of around $1.5\,\mathrm{mas}$ to $2\,\mathrm{mas}$ from the core. 
The position of these two components is consistent to $15\,\mathrm{GHz}$ MOJAVE observations \citep{Lister, Lister2021} which also show two stationary features at similar distances.

In the epochs observed before April 2015, we also detected some additional components within the bright feature at the core region or at distances out to about $0.6\,\mathrm{mas}$ from the core. Because it is difficult to identify these jet components unambiguously, we considered two different models to describe their kinematics. On the one hand, we considered them as one single component moving with an apparent speed of $\beta_\mathrm{app}=2.00\pm0.61$, represented by the dashed line in Fig.~\ref{fig:kinmodbeschleunigt}. This speed would be much lower then the speeds derived for J4 and J5 (see Table~\ref{speedbeschleunigt} and the discussion below). On the other hand, we identified them with three different jet components J1, J2 and J3 moving at apparent speeds comparable to those of 
J4 and J5. This second model is represented by the solid lines in Fig.~\ref{fig:kinmodbeschleunigt}. However, J1 and J2 could only be tracked across three epochs. Therefore, adopting the convention of \citet{Lister1} in which at least five epochs are required to build a robust kinematics model and considering the uncertain identification of these three components, we did not use their speeds for further analysis. Previous independent kinematic analyses \citep{Jorstad,Weaver2022} identified these components as one moving jet component and a trailing feature that was formed behind it. 
While the speed of the moving component is comparable to the speeds of J1, J2 and J3 of our second kinematics model, the speed of the trailing feature is similar to that of the component from our first kinematics model. Therefore, this identification seems to be a combination of both models presented in this work.
   
\begin{table}
\caption{\label{speedbeschleunigt}Apparent speeds and ejection epochs of the jet components.}
\centering
\small
\setlength{\tabcolsep}{4.5pt}
\begin{tabular}{ccccc}
\hline\hline
ID & $d$ & $\mu$ & $\beta_{\mathrm{app}}$ & $t_{\mathrm{0}}$   \\
   &    $[\mathrm{mas}]$ & $[\mathrm{mas\,yr^{-1}}]$ & $[c]$ & $[\mathrm{yr}]$          \\
       (1) & (2) & (3) & (4) & (5) \\
\hline
S1 & ----- & $-0.0029\pm0.0018$ & $-0.139\pm0.084$ & -----  	 \\
S2 & ----- & $0.010\pm0.014$ & $0.48\pm0.67$ & ----- 	 \\
J2 & ----- & $0.175\pm0.042$ & $8.4\pm2.0$ & $2009.78\pm0.37$    \\
J4 & $\lesssim0.53$ & $0.250\pm0.066$ & $12.0\pm3.2$ & $2014.51\pm0.30$  	 \\
J4 & $\gtrsim0.53$ & $0.407\pm0.052$ & $19.5\pm2.5$ & -----  	 \\
J5 & $\lesssim0.37$ & $0.14\pm0.11$ & $6.5\pm5.1$ & $2015.4\pm1.7$  	 \\
J5 & $\gtrsim0.37$ & $0.39\pm0.21$ & $19\pm10$ & -----  	 \\
\hline
\end{tabular}
\tablefoot{Col.(1): ID of the jet components; Col.(2): Distance range of the jet components with respect to the core corresponding to the transition point of the kinematics fits; Col.(3): Angular speed of the jet components; Col.(4): Apparent speed of the jet components in units of speed of light; Col.(5): Ejection epoch of the components. }
\end{table}

The two newly ejected jet features J4 and J5, whose kinematics were used for further analysis, seem to accelerate to higher speeds at distances downstream of $\sim0.53\,\mathrm{mas}$ and $\sim0.37\,\mathrm{mas}$ from the core, respectively. At smaller distances upstream of these transition points, both components seem to travel at lower constant speeds\footnote{For the discussion (Sect.~\ref{sec:disc}), we assume that these constant speeds even persisted further upstream.} (see lower panel of Fig.~\ref{fig:kinmodbeschleunigt}). Therefore, we fitted both components with two separate linear regressions each using Eq.~(\ref{eq:kin}). Their transition points are indicated by the two vertical dotted lines in Fig.~\ref{fig:kinmodbeschleunigt}. 
The choice of the transition point was based on the fact that when crossing this region, the brightness temperature and flux density of J4 and J5, shown in Fig.~\ref{fig:kinmodbeschleunigt}, underwent a steep increase. \citet{Weaver2022} found similar results for J4 leading to comparable angular speeds, also using a broken linear fit with two transition points, located at distances consistent with our findings. \citet{Weaver2022} only identified Gaussian components with J5 at epochs between June 2018 and December 2018. In contrast to that, we also identified Gaussian components fitted to the remaining flux density shown in epochs between February 2017 and February 2018 with this component. However, the angular speeds of J5 between June 2018 and December 2018 are similar in both studies. 
Finally, we calculated the ejection epochs of these two components as described in Sect.~\ref{sec:kinanalysis} and for J4 we found similar results to \citet{Weaver2022}.

The speeds and ejection epochs derived for robust components are listed in Table~\ref{speedbeschleunigt}. 
Due to the large $1\sigma$ uncertainty of the ejection epoch of J5, this component could be ejected before J4, although J5 appears clearly after J4 (see Fig.~\ref{fig:images1}). However, because J5 travels at a lower speed through the inner part of the jet, this scenario would be plausible. In that case, J4 would cross J5 upstream of S1. Since both components were only detected downstream of S1, this would lead to similar images.

\subsubsection{Viewing angle}
\label{sec:orientation}
   
The apparent speed in units of speed of light $\beta_{\mathrm{app}}$ depends on the intrinsic jet speed in units of speed of light $\beta$ and the viewing angle $\phi$ and is given by
 \begin{equation}
   \label{eq:betaapp2}
   \beta_{\mathrm{app}}=\frac{\beta\sin\phi}{1-\beta\cos\phi}.
  \end{equation} 
Therefore, an upper limit on $\phi$ can be estimated by setting $\beta=1$ and solving Eq.~(\ref{eq:betaapp2}) for $\phi$, which leads to
 \begin{equation}
   \label{eq:winkel}
   \phi\leq\arccos\left(\frac{\beta_{\mathrm{app,\,max}}^2-1}{\beta_{\mathrm{app,\,max}}^2+1}\right) ,
  \end{equation} 
in which $\beta_{\mathrm{app,\,max}}$ is the largest possible apparent jet speed.

To calculate this upper limit on the viewing angle, we neglected J1, J2, and J3 because of their unclear identification and kinematic analysis. 
Therefore, the largest possible apparent speed within the $1\sigma$ uncertainties is $\beta_{\mathrm{app,\,max}}=29$ derived 
for J5. Using this value for $\beta_{\mathrm{app,\,max}}$ together with Eq.~(\ref{eq:winkel}), we obtained the upper limit on the viewing angle to be $\phi\lesssim4\degr$.
This upper limit on the viewing angle is consistent to the typical viewing angle for blazar jets of $\phi<5\degr$ derived by \citet{Jorstad}. Furthermore, it is comparable to other estimates of the viewing angle of \source. While \citet{Weaver2022} determined the viewing angle to be $\phi=(3.2\pm1.0)\degr$, using $43\,\mathrm{GHz}$ \vlba\, observations and taking the variability Doppler factor into account, \citet{Pushkarev09} used $15\,\mathrm{GHz}$ \vlba\, observations to determine the viewing angle to be $\phi=4.4\degr$. 

\subsubsection{Jet geometry}
\label{sec:geometry}

\begin{figure}
   \centering
   \includegraphics[width=\hsize]{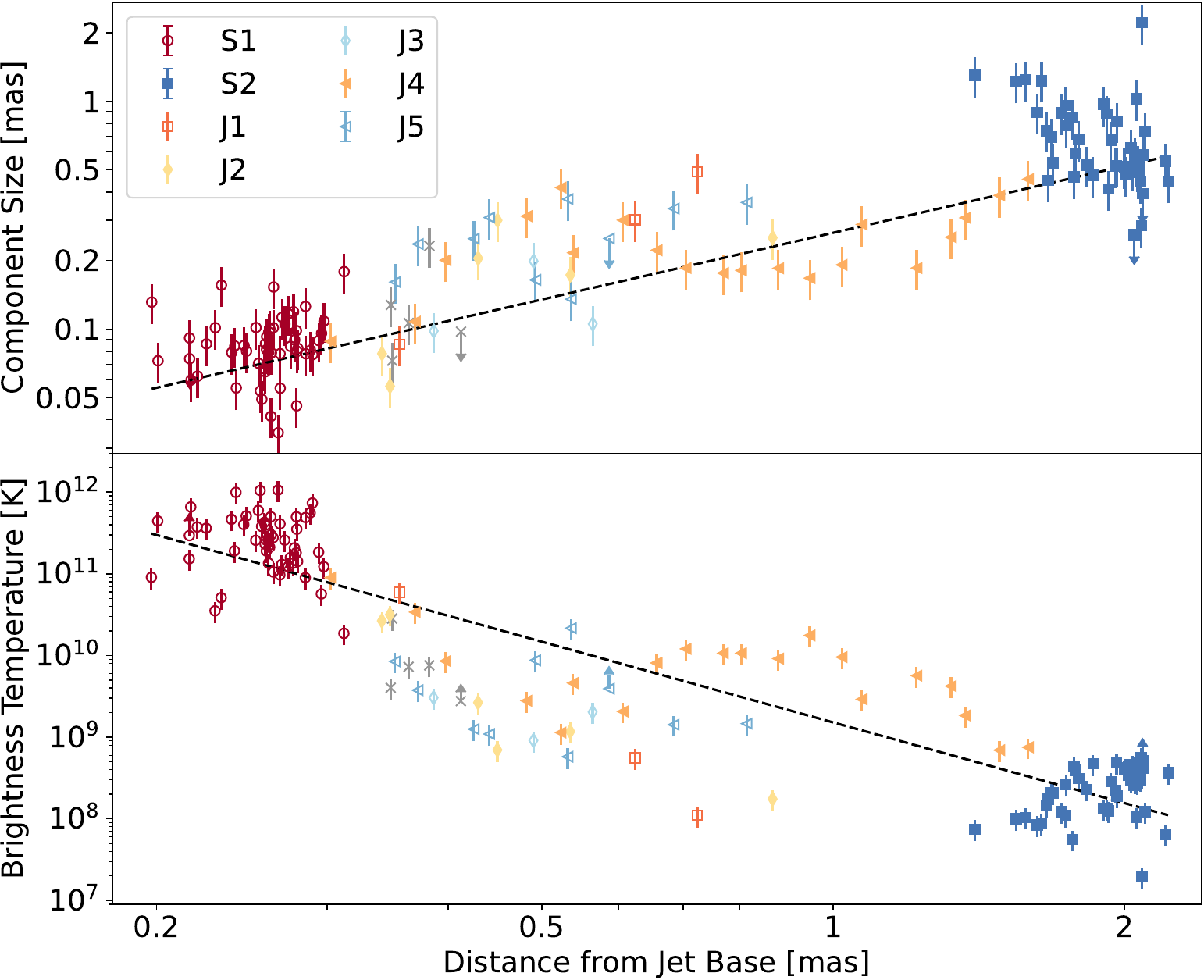}
      \caption{Upper panel: Jet width $D$, given by the FWHM size of the jet components, plotted as a function of their distance from the jet base. The dashed line is fitted via $D=C(d+d_{\mathrm{c,\,43,\,app}})^l$, where $d_{\mathrm{c,\,43,\,app}}$ is the apparent distance of the $43\,\mathrm{GHz}$ core to the jet base, $d$ is the distance from the core to the jet component, $C$ is a constant and $l$ is the power law index representing the jet geometry. The best fit results in $l=0.974\pm0.098$ which is consistent with a conical jet. Lower panel: Observed brightness temperature $T_\mathrm{B}$ of the jet components plotted as a function of the components' distances to the jet base. The dashed line is fitted via $\log(T_\mathrm{B})=s\cdot\log(d+d_{\mathrm{c,\,43,\,app}})+\log(C)$, in which $d_{\mathrm{c,\,43,\,app}}$ is the apparent distance of the $43\,\mathrm{GHz}$ core to the jet base and $s$ is the brightness-temperature gradient. For $d_{\mathrm{c,\,43,\,app}}$ we used the value derived in Sect.~\ref{sec:geometry}. The best fit results in $s=-3.31\pm 0.31$ which corresponds to a conical jet in equilibrium between magnetic field strength density and electron energy density \citep{Burd2022}.} 
         \label{fig:geometry}
   \end{figure}

Following the jet model introduced by \citet{Blandford} and \citet{Konigl} we assumed that the jet diameter $D$ along the jet axis $d$ can be described by a power law: $D\propto d^l$.
To investigate the jet width profile, we used the FWHM of the aforementioned circular Gaussian components \citep[see, e.g.,][]{Burd2022}, with the uncertainties defined as 20\% of the major axis of the respective component.
The results of these measurements are presented in Fig.~\ref{fig:geometry}. 

Subsequently, we employed a single power law fit expressed as $D=C(d+d_{\mathrm{c,\,43,\,app}})^l$, to model the data \citep[see also][]{Kravchenko2025,Kovalev2020}. Here, $d$ is the distance between the $43\,\mathrm{GHz}$ core and the jet components and $d_{\mathrm{c,\,43,\,app}}$ denotes the apparent position of the $43\,\mathrm{GHz}$ core from the jet base. The radio core is the most compact part of a jet and should be located at the transition region between optically thick and thin emission, where the optical depth is $\approx 1$ \citep{Blandford,Konigl,Marscher1}. Note that we only used resolved components for the fit. The corresponding best-fit parameters are as follows:
$C=(0.265\pm0.022)\,\mathrm{mas}^{1-l}$,
$d_{\mathrm{c,\,43,\,app}}=(0.124\pm0.059)\,\mathrm{mas}$ and
$l=0.974\pm0.098$.
The derived power law index $l$ is consistent with $l=1$ within its $1\sigma$ uncertainty, indicating a conical jet geometry \citep[e.g.,][]{Kadler}. However, it differs from $l=1.55\pm0.07$ found by \citet{Kravchenko2025} using $15\,\mathrm{GHz}$ MOJAVE data. They probably found a steeper gradient because they are putting more weight on larger jet scales rather than the scales probed by the $43\,\mathrm{GHz}$ BU data.

\subsubsection{Brightness temperature}
\label{sec:Tb_gradient}

The brightness temperatures of the core and jet components are plotted in the upper panel of Fig.~\ref{fig:kinmodbeschleunigt}. All jet components have brightness temperatures below the inverse Compton limit of $10^{12}\,\mathrm{K}$ \citep{Kellermann1} within their uncertainties while the brightness temperatures of the core are sometimes significantly above this limit for short time intervals. Brightness temperatures above the inverse Compton limit can probably be explained by Doppler boosting. The observed brightness temperatures, $T_\mathrm{B,obs}$, are Doppler boosted with $T_\mathrm{B,obs}=\delta T_\mathrm{B,int}$, where $T_\mathrm{B,int}$ is the intrinsic brightness temperature and $\delta=\sqrt{1-\beta^2}/(1-\beta\cos{\phi})$ is the Doppler factor \citep[e.g.,][]{Kovalev}. Using Eq.~(\ref{eq:betaapp2}) the Doppler factor can be written as $\delta=\sqrt{1-\beta_\mathrm{app}^2+2\beta_\mathrm{app}/\tan\phi}$. With this formula, we calculated the range of possible Doppler factors. We used the derived upper limit for the viewing angle of $\phi\lesssim4$ and the lowest possible speed at which components travel through the core region of $\beta_\mathrm{app}=1.4$ measured for J5 within its $1\sigma$ uncertainty to estimate the lower limit. For the upper limit, we used $\beta_\mathrm{app}=15.2$ derived for J4 within its $1\sigma$ uncertainty and the lower limit of the viewing angle of $\phi=2.2$ ($\phi=3.2\pm1.0$) presented by \citet{Weaver2022}. We used the obtained range of $3.6\leq\delta\leq24$ to calculate the possible range of the intrinsic brightness temperatures of the core which are plotted in the upper panel of Fig.~\ref{fig:kinmodbeschleunigt} (blue bars). They are clearly below the inverse Compton limit and mostly lie at values that are expected for a jet in equipartion (indicated in gray). This equipartition brightness temperature lies around $5\cdot10^{10}\,\mathrm{K}$ \citep{Readhead} with an upper limit of $10^{11}\,\mathrm{K}$ \citep{Singal2009}. Note that the intrinsic core brightness temperatures show values that are significantly above equipartition only at four epochs that can be associated with the ejection of J2, J3, J4 and J5, respectively.

Following the jet model introduced by \citet{Blandford} and \citet{Konigl} and assuming that the magnetic field $B$, the electron density $N$ and the jet diameter $D$ are given by power laws ($B\propto d^b$, $N\propto d^n$ and $D\propto d^l$), it can be shown that the brightness temperature can also be described by a power law: $T_\mathrm{B}\propto d^s$. The brightness-temperature gradient $s$ is given by \citep[see e.g.,][]{Kravchenko2025,Burd2022,Kadler}
 \begin{equation}
   \label{eq:Tb_gradient}
   s=l+n+b(1-\alpha) ,
  \end{equation} 
where $\alpha$ is the spectral index ($S\propto\nu^\alpha$) of the optically thin jet emission. To investigate this gradient for \source, we fitted the brightness temperatures of all resolved jet components by $\log(T_\mathrm{B})=s\cdot\log(d+d_{\mathrm{c,\,43,\,app}})+\log(C)$\footnote{Since the brightness temperatures of the jet components range over $\sim 7$ orders of magnitude and we assume relative uncertainties of $29\%$, higher brightness temperatures have larger uncertainties. A simple weighted ($\frac{1}{\sigma^2}$) power-law fit of the form $T_\mathrm{B}=C(d+d_\mathrm{c,43,app})^s$ would therefore provide an $s$-value that is too small, because smaller brightness temperatures would have larger weights which would pull the fit towards smaller brightness temperatures.} using $d_{\mathrm{c,\,43,\,app}}=(0.124\pm0.059)\,\mathrm{mas}$ derived in Sect.~\ref{sec:geometry} as illustrated in the lower panel of Fig.~\ref{fig:geometry}. To take the uncertainty of $d_{\mathrm{c,\,43,\,app}}$ into account, we used a Monte Carlo simulation method in which we altered the value of $d_{\mathrm{c,\,43,\,app}}$ randomly within its uncertainty. With this method, we calculated 1000 different $d_{\mathrm{c,\,43,\,app}}$ values and performed 1000 different fits. We then calculated the best-fit parameters and their uncertainties to be the mean and standard deviation of the parameters derived from the different fits, which results in
$C=(15.4\pm 4.4)\cdot 10^8\,\mathrm{K\,mas}^{-s}$ and
$s=-3.31\pm 0.31$. The derived brightness-temperature gradient $s$ is not consistent with $s=-3.75\pm0.11$ found by \citet{Kravchenko2025} using $15\,\mathrm{GHz}$ MOJAVE data. Similar to $l$ (see Sect.~\ref{sec:geometry}), this steeper gradient is likely due to the fact that \citet{Kravchenko2025} are putting more weight to larger jet scales that are not probed by the $43\,\mathrm{GHz}$ BU data.
Assuming equipartition between the magnetic field strength density and the electron energy density, the derived brightness-temperature gradient corresponds to a conical jet \citep[$-6\leq s\leq -2.5$;][]{Burd2022} and differs from a parabolic jet \citep[$-3\leq s\leq -1.25$;][]{Burd2022}, confirming the results presented in Sec.\,\ref{sec:geometry}.




\subsection{Cross-correlation analysis}
\label{sec:cross}

\begin{figure}[htpb]
   \centering
   \includegraphics[width=\hsize]{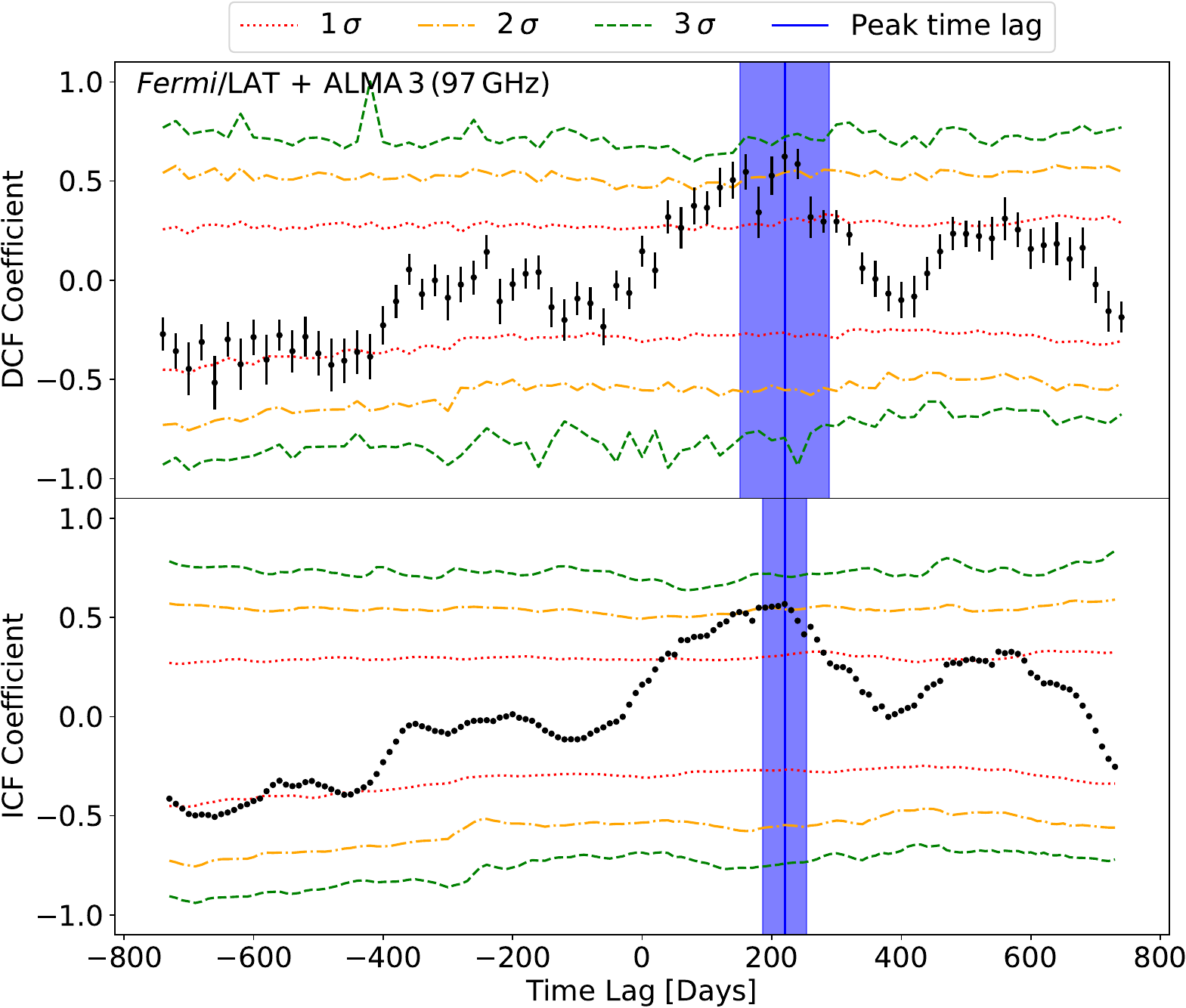}
      \caption{DCF (upper panel) and ICF (lower panel) cross-correlation coefficients between the \fermi\, \gray\,and \alma\,3 light curves plotted over time lag. For the DCF the bin size is chosen to be $20\,\mathrm{days}$, while the interpolation unit for the ICF is chosen to be $10\,\mathrm{days}$ (both calculated as explained in Sect.~\ref{sec:corr_function}). Positive time lags mean that the radio light curve follows the \fermi\, \gray\,light curve. The time lags for the peak cross-correlation coefficients are marked by solid blue lines, with their $1\sigma$ uncertainties given by the shaded blue area. The dotted red, dashed-dotted orange and dashed green lines correspond to the two sided Gaussian equivalent $1\sigma$, $2\sigma$ and $3\sigma$ confidence intervals. The DCF and ICF cross-correlation coefficients between the \fermi\, and \alma\,7, \sma\, and \ovro\, light curves are plotted in Fig.~\ref{fig:corr_fermi-alma7}, Fig.~\ref{fig:corr_fermi-sma} and Fig.~\ref{fig:corr_fermi-ovro}, respectively.}
         \label{fig:corr_fermi-alma3}
   \end{figure}

Because of the similar behavior of the different light curves shown in Fig.~\ref{fig:combkinvar}, we computed DCF and ICF cross-correlation coefficients between the \fermi\,\gray\,and the \alma, \sma\, and \ovro\, radio light curves as described in Sect.~\ref{sec:corr}.
The results of the DCF and ICF between the \fermi\, and the \alma\,3 light curves are plotted in Fig.~\ref{fig:corr_fermi-alma3}. Additional plots of the cross-correlation between the \fermi\, and the \alma\,7, \sma\, and \ovro\, light curves are presented in Figs.~\ref{fig:corr_fermi-alma7} to~\ref{fig:corr_fermi-ovro}. One can see that the cross-correlation functions between all four pairs of light curves have their peak cross-correlation coefficients at positive time lags indicated by the solid blue lines with their uncertainties, calculated as discussed in Sect.~\ref{sec:corr_uncertainties}, shown by the blue shaded areas. These positive time lags indicate that the radio light curves follow the \gray\,light curve. Furthermore, the DCF and ICF between each individual pair of light curves peaks at comparable time lags that are consistent with each other within their $1\sigma$ uncertainties. 
Note that the global peak coefficient of the DCF between the \fermi\, and \ovro\, light curves of $0.54\pm0.14$ is located at a larger time lag of $(630\pm219)\,\mathrm{days}$ that is not consistent with the corresponding ICF time lag and with those found for the other three radio light curve. However, there is another clear peak at a time lag of $(252\pm219)\,\mathrm{days}$ that is comparable with the other ones (see Fig.~\ref{fig:corr_fermi-ovro}). While the global peak time lag corresponds to the correlation of the radio flares in 2015 and 2017 with the \gray\, flares in 2014 and 2015, the second peak time lag corresponds to the correlation of these two radio flares to the \gray\, flares in 2015 and 2016 (see Fig.~\ref{fig:combkinvar}). Therefore, we used the second time lag for further calculations.

The peak cross-correlation coefficients derived for both methods are listed in Table~\ref{lags}, together with their corresponding time lags and significances. Here, one can see that the peak coefficients show a slight shift towards longer time lags with decreasing frequency of the radio light curves which could probably be explained by opacity effects (see discussion in Sect.~\ref{sec:coreshift} for further details).

\begin{table}
\caption{\label{lags}Cross-correlation results derived by the DCF (top) and ICF (bottom) between the \fermi\, \gray\,and different radio light curves.}
\centering
\small
\setlength{\tabcolsep}{4pt}
\begin{tabular}{ccccc}
\hline\hline
Light curve & $\nu$ & $r_{\mathrm{corr}}$ & Significance & $\tau_{\mathrm{\gamma,\nu}}$ \\
     & $[\mathrm{GHz}]$ &  & $[\sigma]$ & $[\mathrm{days}]$ \\
    (1) & (2) & (3) & (4) & (5) \\
\hline
\multicolumn{5}{l}{DCF:} \\  
    \alma\, 7 & $341.9\pm8.5$ & $0.63\pm0.13$ & $2.73$ & $84\pm102$ \\
    \sma & $225.6\pm4.8$ & $0.57\pm0.15$ & $1.91$ & $144\pm224$ \\
    \alma\, 3 & $97.0\pm6.2$ & $0.623\pm0.079$ & $2.39$ & $220\pm69$ \\
    \ovro & $15$ & $0.45\pm0.14$ & $1.16$ & $252\pm219$ \\
    \hline
    \multicolumn{5}{l}{ICF:} \\
    \alma\, 7 & $341.9\pm8.5$ & $0.54$ & $2.07$ & $120\pm46$ \\
    \sma & $225.6\pm4.8$ & $0.48$ & $1.88$ & $198\pm82$ \\
    \alma\, 3 & $97.0\pm6.2$ & $0.57$ & $2.15$ & $220\pm34$ \\
    \ovro & $15$ & $0.48$ & $1.34$ & $225\pm153$ \\
\hline
\end{tabular}
\tablefoot{Col.(1): Radio light curve; Col.(2): Mean frequency of the radio light curve; Col.(3): Peak cross-correlation coefficient; Col.(4): Significance of the peak cross-correlation coefficient in units of the Gaussian equivalent standard deviation; Col.(5): Time lag corresponding to the peak cross-correlation coefficient.}
\end{table}

\section{Discussion}
\label{sec:disc}

From our study of 51-epochs of \vlba\, observations of \source\, at $43\,\mathrm{GHz}$, we found two newly ejected jet features likely associated with the high activity period starting in late 2013 shown 
in the \fermi\, \gray\, and in several radio light curves observed at different frequencies by \alma, \ovro\, and \sma. 
To illustrate that, the ejection epochs derived in Sect.~\ref{sec:kin} are plotted as vertical dashed lines in Fig.~\ref{fig:combkinvar}, together with the different \gray\,and radio light curves. The orange and blue bands represent the $1\sigma$ uncertainties of the ejection epochs of J4 (orange) and J5 (blue). 

In the following, we will use a model in which the flares shown by the \gray\,and radio light curves occur due to opacity effects when a moving jet component passes through the \gray\,emitting region and the regions where the radio cores are located \citep[also see][]{Fuhrmann, Max-Moerbeck}. With this model, we will analyze the core shift (see Sect.~\ref{sec:coreshift}) and determine the location of the \gray\,emitting region (see Sect.~\ref{sec:gammaposition}) in the jet of \source.

\subsection{Core shift}
\label{sec:coreshift}
   
 The absolute position of the radio core $d_{\mathrm{c,\,\nu}}$ depends on the frequency $\nu$ and is given by 
 \begin{equation}
   \label{eq:coreshift}
    d_{\mathrm{c,\,\nu}}\propto\nu^{-\frac{1}{k_{\mathrm{r}}}},
  \end{equation}
in which the power law index $k_{\mathrm{r}}$ depends on the spectral index as well as on the magnetic field and particle density distributions \citep{Konigl}. Hence, the location of the radio core shifts upstream towards the central engine with increasing frequency \citep{Marscher1}.

Assuming a jet in which the electron energy density $N$, the magnetic field strength density $B$, the jet diameter $D$ and the brightness temperature $T_\mathrm{B}$ are given by power laws ($N\propto d^n$, $B\propto d^b$, $D\propto d^l$, $T_\mathrm{B}\propto d^s$) the power law index representing the core shift is given by
 \begin{equation}
   \label{eq:coreshift_index}
    k_{\mathrm{r}}=\frac{(2\alpha-3)b-2n-2}{5-2\alpha},
  \end{equation}
where $\alpha$ is the optically thin spectral index \citep{Lobanov1}. Further assuming a jet in equipartition between electron energy density and magnetic field strength density, which is consistent to our findings on the jet geometry (Sect.~\ref{sec:geometry}) and brightness temperature gradient (Sect.~\ref{sec:Tb_gradient}), $N\propto B^2\propto d^{2b}$ leading to $n=2b$ \citep[e.g.][]{Burd2022}. 
Inserting $n=2b$ into Eq.~\ref{eq:Tb_gradient} and Eq.~\ref{eq:coreshift_index} and solving these equations for $b$ and $k_\mathrm{r}$ results in
 \begin{equation}
   \label{eq:B_index}
    b=\frac{s-l}{3-\alpha},
  \end{equation}
  
 \begin{equation}
   \label{eq:coreshift_index2}
    k_{\mathrm{r}}=\frac{(2\alpha-7)(s-l+1)+1}{(5-2\alpha)(3-\alpha)}.
  \end{equation}

To calculate these two power law indices, the optically thin spectral index $\alpha$ of the jet emission is needed. Since this $\alpha$ can not be determined from the data we used in this work, we used the jet spectral index presented by \citet{Hovatta2014}. They calculated the spectral index at parsec scales between $8\,\mathrm{GHz}$ and $15\,\mathrm{GHz}$ \vlba\, observations along the ridge line of the jet of a sample of AGN. For \source, they found a jet spectral index that varies around $-1$ with its maximum and minimum at around $-0.3$ and $-1.6$ within $1\sigma$ uncertainties. Therefore, to be conservative, we used $\alpha=-1.00\pm0.70$ to calculate $k_\mathrm{r}$ and $b$. Note that this $\alpha$ is also consistent with the jet spectral index of \source\, at smaller scales derived between quasi-simultaneous VLBI observations at $43\,\mathrm{GHz}$ and $86\,\mathrm{GHz}$ (Ricci et al. in prep.).   
Together with our results on the power law indices of the jet geometry, $l=0.974\pm0.098$ (see Sect.~\ref{sec:geometry}), and the brightness temperature gradient, $s=-3.31\pm0.31$ (see Sect.~\ref{sec:Tb_gradient}), we obtain
$b=-1.07\pm0.20$ and
$k_\mathrm{r}=1.09\pm0.17.$
These values are both comparable with a conical jet in equipartition having a toroidal magnetic field. For such a jet $b=-1$, since a toroidal field scales with the jet diameter via $B\propto D^{-1}\propto d^{-l}$ ($l=1$ for a conical jet). Note that a poloidal magnetic field scales as $B\propto D^{-2}\propto d^{-2l}$, leading to $b=-2$ \citep{Burd2022}. Furthermore, for a freely expanding jet in equipartition between magnetic-field energy and jet particle density $k_\mathrm{r}=1$ \citep{Blandford}.

Using the aforementioned model, 
we are able to explain the behavior of the time lags derived by the cross-correlation analysis between \gray\,and radio light curves measured at different frequencies presented in Sect.~\ref{sec:cross} using the core shift.

The location of the radio core at frequency $\nu$ with respect to the jet base can be calculated by
 \begin{equation}
   \label{eq:core_location}
    d_{\mathrm{c,\,\nu}}=d_\mathrm{\gamma}+d_\mathrm{\gamma,\,\nu}=d_\mathrm{\gamma}+\mu\tau_\mathrm{\gamma,\,\nu}\,,
  \end{equation}
where $d_\mathrm{\gamma}$ is the location of the \gray\,emitting region with respect to the jet base and $d_\mathrm{\gamma,\,\nu}$ is the distance between $d_\mathrm{\gamma}$ and $d_{\mathrm{c,\,\nu}}$.  
Inserting Eq.~(\ref{eq:core_location}) into Eq.~(\ref{eq:coreshift}), we obtain $d_{\mathrm{c,\,\nu}}=d_{\mathrm{\gamma}}+\mu\tau_{\mathrm{\gamma,\,\nu}}\propto\nu^{-\frac{1}{k_{\mathrm{r}}}}$. 
Using $d_{\mathrm{\gamma}}$ as reference position and using a model in which the jet component producing the \gray\,and radio outbursts travels at a constant speed through the region of the jet where the radio cores and the \gray\,emitting region are located\footnote{In Sect.~\ref{sec:kin} we found that the two newly ejected components J4 and J5 moved at constant speeds in the inner jet and accelerated to higher speeds downstream at distances $\gtrsim0.37\,\mathrm{mas}$ from the $43\,\mathrm{GHz}$ core. Here, we assume that these constant speeds even persisted further upstream, from the jet base up to the $43\,\mathrm{GHz}$ core.}, we obtain
 \begin{equation}
   \label{eq:coreshifttimelag}
    \tau_{\mathrm{\gamma,\,\nu}}\propto\nu^{-\frac{1}{k_{\mathrm{r}}}},
  \end{equation}
which explains the fact that the time lags derived with the cross-correlation analysis presented in Sect.~\ref{sec:cross} increase with decreasing frequency across the observed radio light curves.




\subsection{Location of the \gray\, emitting region}
\label{sec:gammaposition}

\begin{figure}
   \centering
   \includegraphics[width=\hsize]{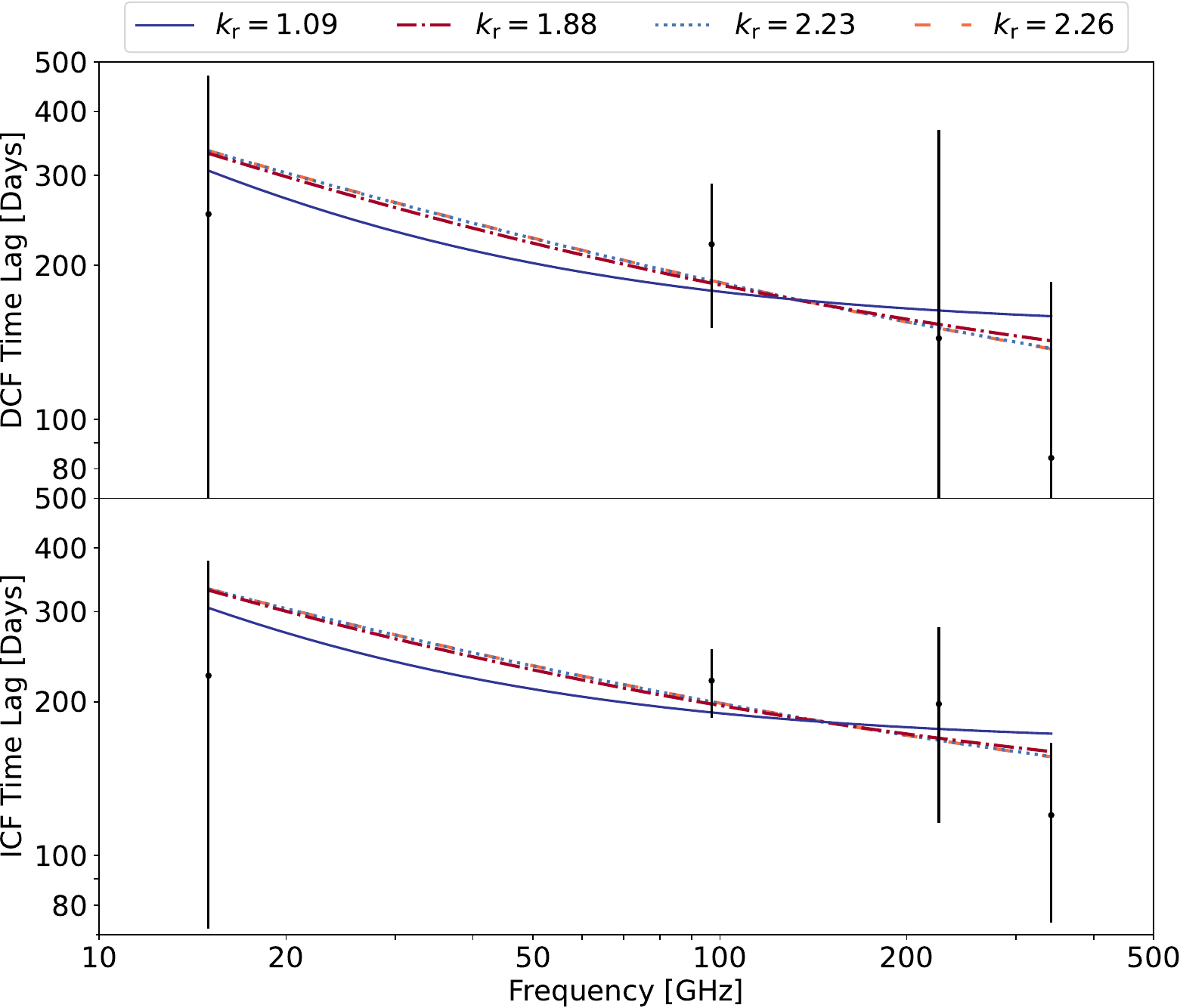}
      \caption{Time lags $\tau_\mathrm{\gamma,\nu}$ derived from the DCF (upper panel) and the ICF (lower panel) plotted as a function of frequency $\nu$. Positive time lags mean that the radio light curve follows the \fermi\, \gray\,light curve. To investigate the core shift, the time lags are fitted by $\tau_\mathrm{\gamma,\nu}=\tau_\mathrm{\gamma}+C\nu^{-\frac{1}{k_\mathrm{r}}}$, where $\tau_\mathrm{\gamma}$ is the time a newly ejected component needs to travel from the jet base to the \gray\,emitting region at speed $\mu$ and $C$ is a constant. Using the calculated power law index $k_\mathrm{r}=1.09$, the best fits (solid dark blue lines) results in the de-projected location of the \gray\,emitting region of $2.6\,\mathrm{pc}\leq d_\mathrm{\gamma} = \frac{\tau_\mathrm{\gamma}\mu}{\sin{\phi}}\leq20\,\mathrm{pc}$. For greater $k_\mathrm{r}$ the lower limit of the \gray\,region would shift upstream in the jet to $d_\mathrm{\gamma, min}\leq1\,\mathrm{pc}$ for $k_\mathrm{r}\geq1.88$ (dashed-dotted red lines), $d_\mathrm{\gamma, min}\leq0.1\,\mathrm{pc}$ for $k_\mathrm{r}\geq2.23$ (dotted light blue lines) and $d_\mathrm{\gamma, min}\leq0.01\,\mathrm{pc}$ for $k_\mathrm{r}\geq2.26$ (dashed orange lines). } 
         \label{fig:coreshift}
   \end{figure}

To investigate the core shift in more detail we plotted the time lags listed in Table~\ref{lags} as a function of the frequency in Fig.~\ref{fig:coreshift}. Using $d_\mathrm{\gamma}$ as reference position, we fitted the time lags derived by the DCF (upper panel of Fig.~\ref{fig:coreshift}) as well as the time lags obtained by the ICF (bottom panel of Fig.~\ref{fig:coreshift}) by 
\begin{equation}
   \label{eq:coreshift_fit}
    \tau_\mathrm{\gamma,\,\nu}=\tau_\mathrm{\gamma}+C\nu^{-\frac{1}{k_\mathrm{r}}},
  \end{equation}
where $\tau_\mathrm{\gamma}$ is the time a newly ejected jet component needs to travel from the jet base to the \gray\,emitting region and $k_\mathrm{r}=1.09$ derived from Eq. (\ref{eq:coreshift_index2}). The best fits of the DCF and ICF time lags are both plotted as solid dark blue lines in Fig.~\ref{fig:coreshift} and the parameters of both fits are listed in Table~\ref{coreshift_parameter}.


\begin{table}
\caption{\label{coreshift_parameter}Best fit parameter and apparent locations of the \gray\,emitting region.}
\centering
\small
\begin{tabular}{cccc}
\hline\hline
Method & $C$ & $\tau_\mathrm{\gamma}$ & $d_\mathrm{\gamma, app}$ \\
        & $[\mathrm{days\,GHz}^{\frac{1}{k_\mathrm{r}}}]$ & $[\mathrm{days}]$ & $[\mathrm{mas}]$ \\
    (1) & (2) & (3) & (4) \\
\hline
    DCF & $1869\pm3026$ & $150\pm73$ & $0.058\pm0.053$ \\
    ICF & $1675\pm1983$ & $165\pm36$ & $0.063\pm0.052$ \\
\hline
\end{tabular}
\tablefoot{Col.(1): Cross-correlation method; Col.(2): Constant of proportionality for the best fit; Col.(3): Time a newly ejected jet component needs to travel from the jet base to the \gray\,emitting region; Col.(4): Apparent location of the \gray\,emitting region.}
\end{table}

Assuming that a newly ejected jet component travels with constant speed from the jet base to the \gray\,emitting region and produces a \gray\, flare when crossing this region, we calculated the apparent location of the \gray\,emitting region $d_\mathrm{\gamma,app}$ with respect to the jet base by 
\begin{equation}
   \label{eq:gammapositionapp}
    d_\mathrm{\gamma,app}=\mu\tau_\mathrm{\gamma},
  \end{equation}
using the results from the fits on the DCF and ICF time lags, respectively, and the speed derived by the kinematic analysis of the $43\,\mathrm{GHz}$ VLBA data presented in Sect.~\ref{sec:kin} and listed in Table~\ref{speedbeschleunigt}.

Since the cross-correlation is driven by the flare in 2015, which can most likely be associated with the ejection of J5 (see Fig.~\ref{fig:combkinvar}), we used the angular speed of $\mu=(0.14\pm0.11)\,\mathrm{mas\,yr^{-1}}$ of J5 to calculate the apparent location of the \gray\,emitting region. Furthermore, this angular speed also corresponds to the maximum angular speed of \source\, of $\mu=(0.150\pm0.014)\,\mathrm{mas\,yr^{-1}}$ derived by using MOJAVE data observed with the \vlba\, at $15\,\mathrm{GHz}$ \citep{Lister}. The determined values of $d_\mathrm{\gamma,app}$ are listed in  Table~\ref{coreshift_parameter} for both methods, respectively. Both values are consistent with each other within their uncertainties. Therefore, we calculated the weighted mean of the apparent location of the \gray\,emitting region to be 
$d_{\mathrm{\gamma,\,app}}=0.061\pm0.037\,\mathrm{mas}$.
At a redshift of $z=0.89$ \citep{Jorstad}, this equals to 
$d_{\mathrm{\gamma,\,app}}=0.47\pm0.29\,\mathrm{pc}$.


To determine the de-projected location of the \gray\,emitting region $d_{\mathrm{\gamma}}$, the viewing angle $\phi$ of the jet can be used, as in
 \begin{equation}
   \label{eq:gammapositiondeprojected}
    d_{\mathrm{\gamma}}=\frac{d_{\mathrm{\gamma,\,app}}}{\sin\phi}.
  \end{equation}
Using a similar method as for the Doppler factor (see Sect.~\ref{sec:Tb_gradient}), we calculated the range of $d_{\mathrm{\gamma}}$. For this, we used the upper limit of the viewing angle of $\phi\lesssim4\degr$ derived in Sect.~\ref{sec:orientation} and the smallest possible location of the \gray\,emitting region of 
$d_{\mathrm{\gamma,\,app,\,min}}=0.18\,\mathrm{pc}$ within its $1\sigma$ uncertainty to calculate the lower limit. Furthermore, using the largest possible value for $d_{\mathrm{\gamma,\,app}}$ within its $1\sigma$ uncertainty of 
$d_{\mathrm{\gamma,\,app,\,max}}=0.76\,\mathrm{pc}$ and the smallest possible viewing angle of $\phi=2.2$ ($\phi=3.2\pm1.0$) derived by \citet{Weaver2022}, we calculated the upper limit of $d_{\mathrm{\gamma}}$. With this method we can pinpoint the location of the \gray\, emitting region in the jet of \source\, to the range of    
$$2.6\,\mathrm{pc}\leq d_{\mathrm{\gamma}}\leq20\,\mathrm{pc}.$$
We highlight how such a distance is far beyond the BLR, that typically extends up to $\lesssim1\,\mathrm{pc}$ \citep[see, e.g.,][]{Zhang}.

This result is consistent with the findings from previous studies on several other blazars by, for example, \citet{Fuhrmann, Max-Moerbeck, Kramarenko2022} using similar methods. Moreover, recent studies on different other blazars also found \gray\, emitting regions beyond the BLR using the \gray\,/\,optical ratio of flare energy dissipated  \citep{Kundu2025} and via modeling the spectral energy distribution \citep[][]{Naseef2025, Thekkoth2024}. However, there are also studies that found \gray\, absorption signatures in the light curves of several other FSRQs indicating \gray\, emitting regions inside the BLR \citep{Agarwal2024, Das2023, Dmytriiev2025},
supported also by the variability time scales observed \citep{Das2023}.


\subsection{Implications for \gray\, emission models}
\label{sec:emission_models}

The \gray\, emission in FSRQs is generally explained by EC scattering of seed photons originated from the BLR \citep{Costamante}. However, given the derived range of the de-projected location of the \gray\,emitting region of 
$2.6\,\mathrm{pc}\leq d_{\mathrm{\gamma}}\leq20\,\mathrm{pc}$ and assuming a typical extend of the BLR of $\lesssim1\,\mathrm{pc}$ \citep[e.g.,][]{Zhang}, we can rule out this emission model for the case of \source. 

This result is also supported by the derived value of the power law index representing the core shift of $k_\mathrm{r}=1.09\pm0.17$ (see Sect.~\ref{sec:coreshift}). For a freely expanding jet in equipartition between magnetic-field energy and jet particle density $k_\mathrm{r}=1$ \citep{Blandford}. However, $k_{\mathrm{r}}$ can reach $2.5$ in regions with steep pressure gradients and can become even larger in the presence of external density gradients which are present in the BLR \citep{Lobanov1}. Therefore, we investigated how $k_\mathrm{r}$ would change if the \gray\,emitting region was located closer to the jet base inside the BLR. The BLR typically extends from a few to a few tens of light-days, translating into $0.01\,\mathrm{pc}$ to $0.1\,\mathrm{pc}$, for low-luminosity AGN and from several tens to hundreds of light-days ($0.1\,\mathrm{pc}$ to $1\,\mathrm{pc}$) for high-luminosity quasars \citep[e.g.,][]{Raimundo,Kaspi2000,Zhang}. Hence, we calculated the hypothetical values that $k_\mathrm{r}$ would have if the \gray\,emitting region was located at distances of $1\,\mathrm{pc}$, $0.1\,\mathrm{pc}$ and $0.01\,\mathrm{pc}$ from the jet base, respectively. For this purpose, we increased $k_\mathrm{r}=1.09$ incrementally by steps of $0.001$, fitted the DCF and ICF time lags, respectively, by Eq.~(\ref{eq:coreshift_fit}) using the increased $k_\mathrm{r}$-values, and calculated the lower limit of the de-projected location of the \gray\,emitting region $d_\mathrm{\gamma}$ as explained in Sect.~\ref{sec:gammaposition} until $d_\mathrm{\gamma, min}$ reached the above mentioned distances. With this method we found that the \gray\,emitting region would shift upstream in the jet to $d_\mathrm{\gamma, min}\leq1\,\mathrm{pc}$ for 
$k_\mathrm{r}\geq1.88$ (dashed-dotted red lines in Fig.~\ref{fig:coreshift}), $d_\mathrm{\gamma, min}\leq0.1\,\mathrm{pc}$ for 
$k_\mathrm{r}\geq2.23$ (dotted light blue lines in Fig.~\ref{fig:coreshift}) and $d_\mathrm{\gamma, min}\leq0.01\,\mathrm{pc}$ for 
$k_\mathrm{r}\geq2.26$ (dashed orange lines in Fig.~\ref{fig:coreshift}), meaning that $k_\mathrm{r}$ would have been greater than the derived value of $k_\mathrm{r}=1.09\pm0.17$, if the \gray\,emitting region was located inside the BLR, reaching values of $k_\mathrm{r}\gtrsim2$, in accordance with \citet{Lobanov1}. 

However, alternatively to EC scattering on BLR photons, the \gray\,emission could be produced via SSC scattering \citep[e.g.,][]{Maraschi}, EC scattering on IR photons originated between the BLR and the dusty torus \citep[e.g.,][]{Sikora09}, EC scattering on CMB photons \citep[e.g.,][]{Ghisellini09} or via hadronic emission models \citep[e.g.,][]{Mannheim}. Furthermore, given the spine-sheath structure of the jet of \source\, detected by \citet{Attridge} and \citet{Pushkarev}, the seed photon field could also be provided by the jet sheath. \citet{MacDonald2015} developed a model in which synchrotron electrons from an emitting region in the jet sheath are inverse-Compton scattered by electrons of a jet component that propagates relativistically along the jet spine. Within the scope of this model, \citet{MacDonald} were able to explain an orphan \gray\,flare by \source\, in February 2014, as well as different other orphan \gray\,flares by several other luminous blazars. 

\section{Summary}
\label{sec:sum}

In this study, we combined a kinematic analysis of $43\,\mathrm{GHz}$ \vlba\, observations of \source\, with a cross-correlation analysis between the \fermi\,\gray\, and several radio light curves observed with \alma, \sma, and \ovro\, at different frequencies, to calculate the location of the \gray\, emitting region in the jet of \source. Our findings are as follows: 
\begin{itemize}
    \item Investigating 51 epochs of \source\, observed with the \vlba\, at $43\,\mathrm{GHz}$ over a period of around nine years from April 2009 until December 2018, we found two new prominent jet features, J4 and J5, that were ejected in $2014.51\pm0.30$ and $2015.4\pm1.7$, respectively. These newly ejected components can be associated with two bright flares shown in the \fermi\,\gray\, and several radio light curves observed by \alma, \sma\, and \ovro\, at different frequencies.
    \item Using the maximum speed of $\beta_\mathrm{app}=19\pm10$ derived for J5, we calculated the upper limit of the viewing angle of the jet of \source\, to be $\phi\lesssim4^\circ$, in agreement with findings from \citet{Pushkarev}, \citet{Jorstad} and \citet{Weaver2022}.
    \item To analyze the jet geometry and brightness-temperature gradient, we fitted the FWHM of the Gaussian components, parameterizing the jet width, and the brightness temperatures with respect to their distance to the jet base in terms of power laws, leading to power law indices of $l=0.974\pm0.098$ and $s=-3.31\pm0.31$, respectively. Both values are consistent with a conical jet ($l=1$) in equipartition between the magnetic field strength density and the electron energy density \citep[$-6\leq s \leq -2.5$;][]{Burd2022}.
    \item The cross-correlation analysis between the \fermi\, \gray\,light curve and several radio light curves observed at different frequencies, resulted in positive time lags, which means that the \gray\,light curve leads the radio light curves. 
    \item Using the derived power law indices representing the jet geometry and the brightness-temperature gradient, we calculated the power law index for the core shift as $k_\mathrm{r}=1.09\pm0.17$. This power law index is also consistent with a conical jet in equipartition \citep[$k_\mathrm{r}=1$;][]{Lobanov1}. 
    \item Combining the results of the kinematic analysis of the $43\,\mathrm{GHz}$ \vlba\, observations with those of the cross-correlation analysis we were able to pinpoint the location of the \gray\, emitting region within the jet of \source. For this we used a model similar to \citet{Fuhrmann} and \citet{Max-Moerbeck} in which the outbursts shown by the different light curves were produced when J4 and J5 passed through the \gray\,emitting region and the frequency-dependent radio cores. We obtained the possible range of de-projected locations of the \gray\,emitting region with respect to the jet base of 
    $2.6\,\mathrm{pc}\leq d_{\mathrm{\gamma}}\leq20\,\mathrm{pc}$, which is far beyond the BLR which typically extends up to $\lesssim1\,\mathrm{pc}$ \citep[see, e.g.,][]{Zhang}. This result is in contradiction with blazar-emission models that rely on inverse Compton up-scattering of seed photons from the BLR. 
\end{itemize}

\begin{acknowledgements}
We thank the anonymous referee for their helpful comments and suggestions which improved the manuscript.
We thank S. G. Jorstad, A. P. Marscher and Z. R. Weaver for helpful discussions and comments on the kinematic analysis. 
F. R., M. K. and L. R. acknowledge support from the Deutsche Forschungsgemeinschaft (DFG, grants 434448349 and 443220636 [FOR5195: Relativistic Jets in Active Galaxies]).
T. H. was supported by Academy of Finland projects 317383, 320085, 345899, and 362571.
This study makes use of the following ALMA data: ADS/JAO.ALMA\#2011.0.00001.CAL. ALMA is a partnership of ESO (representing its member states), NSF (USA) and NINS (Japan), together with NRC (Canada), MOST and ASIAA (Taiwan), and KASI (Republic of Korea), in cooperation with the Republic of Chile. The Joint ALMA Observatory is operated by ESO, AUI/NRAO and NAOJ.
This study makes use of VLBA data from the VLBA-BU Blazar Monitoring Program (BEAM-ME and VLBA-BU-BLAZAR;
\url{http://www.bu.edu/blazars/BEAM-ME.html}), funded by NASA through the Fermi Guest Investigator Program. The VLBA is an instrument of the National Radio Astronomy Observatory, which is a facility of the National Science Foundation operated by Associated Universities, Inc.
The Submillimeter Array is a joint project between the Smithsonian Astrophysical Observatory and the Academia Sinica Institute of Astronomy and Astrophysics and is funded by the Smithsonian Institution and the Academia Sinica. We recognize that Maunakea is a culturally important site for the indigenous Hawaiian people; we are privileged to study the cosmos from its summit.
\end{acknowledgements}

\bibliographystyle{aa} 
\bibliography{bibtex} 

\begin{appendix}

\section{Additional plots of the cross-correlation analysis}
\label{app:cross-corr}


\begin{figure}[h!]
   \centering
   \includegraphics[width=\hsize]{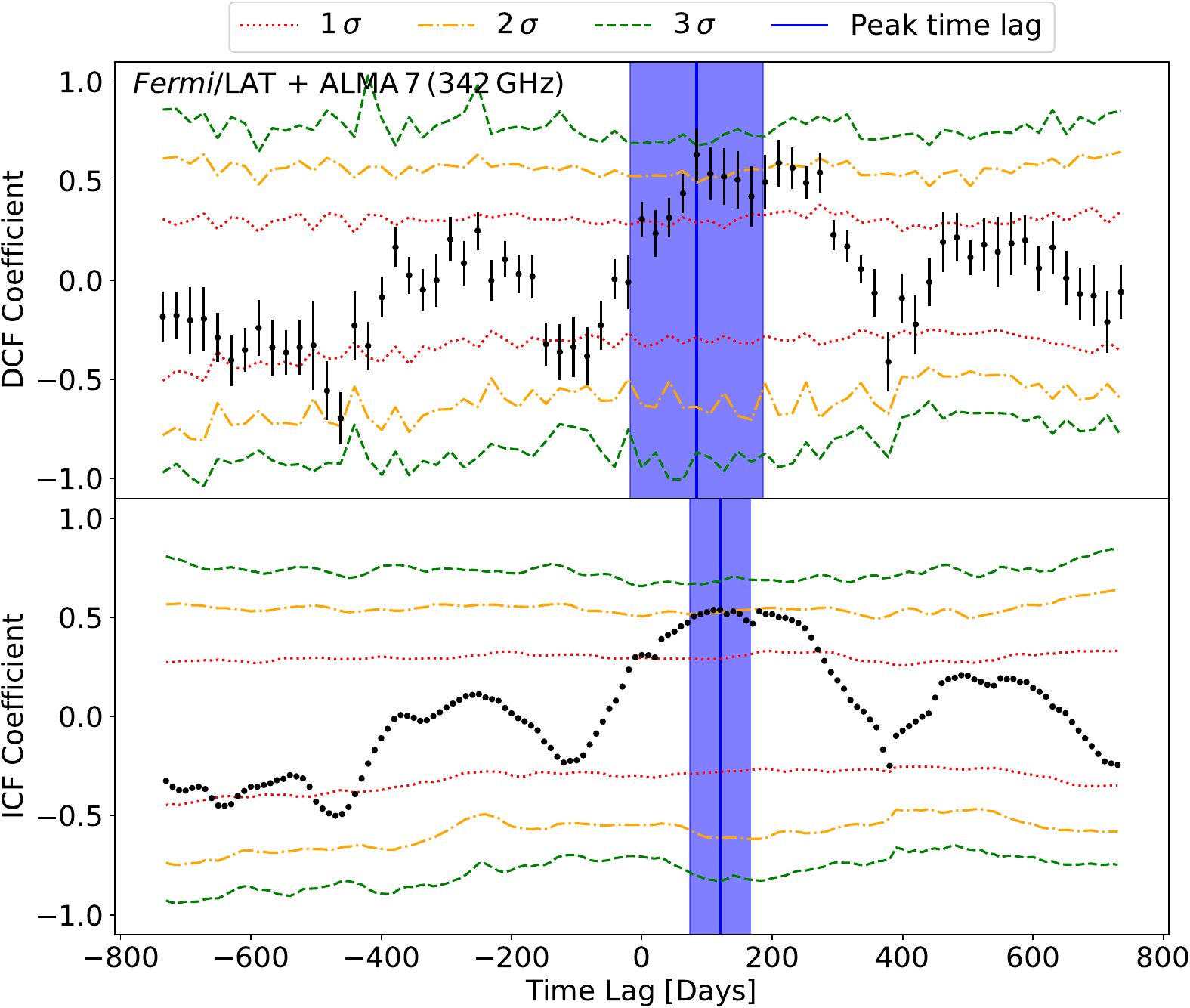}
      \caption{DCF (upper panel) and ICF (lower panel) cross-correlation coefficients between the \fermi\, \gray\, and \alma\,7 light curves plotted over time lag. For the DCF the bin size is chosen to be $21\,\mathrm{days}$, while the interpolation unit for the ICF is chosen to be $10\,\mathrm{days}$ (both calculated as explained in Sect.~\ref{sec:corr_function}). Positive time lags mean that the radio light curve follows the \fermi\, \gray\,light curve. The time lags for the peak cross-correlation coefficients are marked by solid blue lines, with their $1\sigma$ uncertainties given by the shaded blue area. The dotted red, dashed-dotted orange and dashed green lines correspond to the two sided Gaussian equivalent $1\sigma$, $2\sigma$ and $3\sigma$ confidence intervals.}
         \label{fig:corr_fermi-alma7}
   \end{figure}


\begin{figure}[h!]
   \centering
   \includegraphics[width=\hsize]{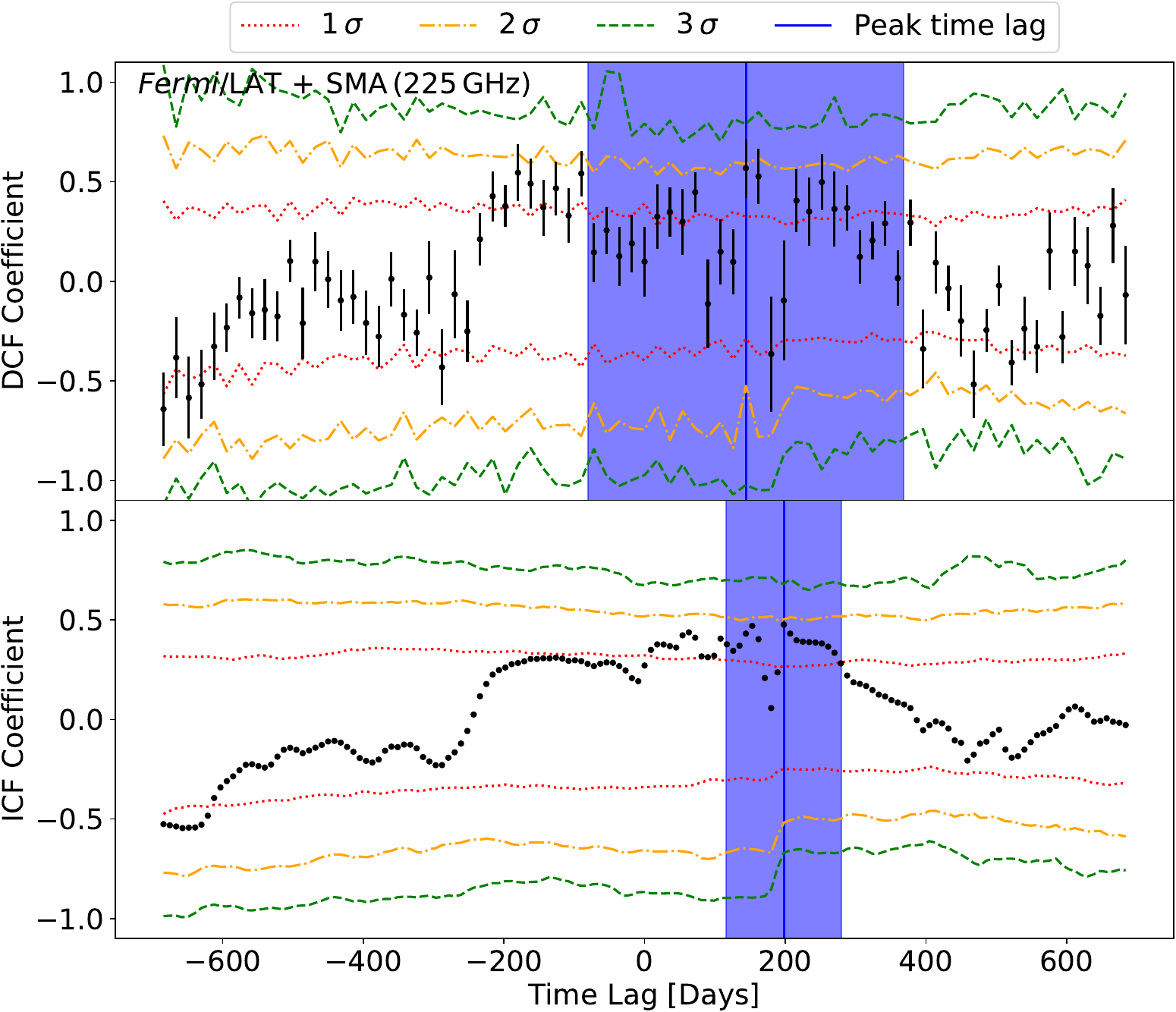}
      \caption{DCF (upper panel) and ICF (lower panel) cross-correlation coefficients between the \fermi\, \gray\, and \sma\, light curves plotted over time lag. For the DCF the bin size is chosen to be $18\,\mathrm{days}$, while the interpolation unit for the ICF is chosen to be $9\,\mathrm{days}$ (both calculated as explained in Sect.~\ref{sec:corr_function}). Positive time lags mean that the radio light curve follows the \fermi\, \gray\,light curve. The time lags for the peak cross-correlation coefficients are marked by solid blue lines, with their $1\sigma$ uncertainties given by the shaded blue area. The dotted red, dashed-dotted orange and dashed green lines correspond to the two sided Gaussian equivalent $1\sigma$, $2\sigma$ and $3\sigma$ confidence intervals.}
         \label{fig:corr_fermi-sma}
   \end{figure}


\begin{figure}[h!]
   \centering
   \includegraphics[width=\hsize]{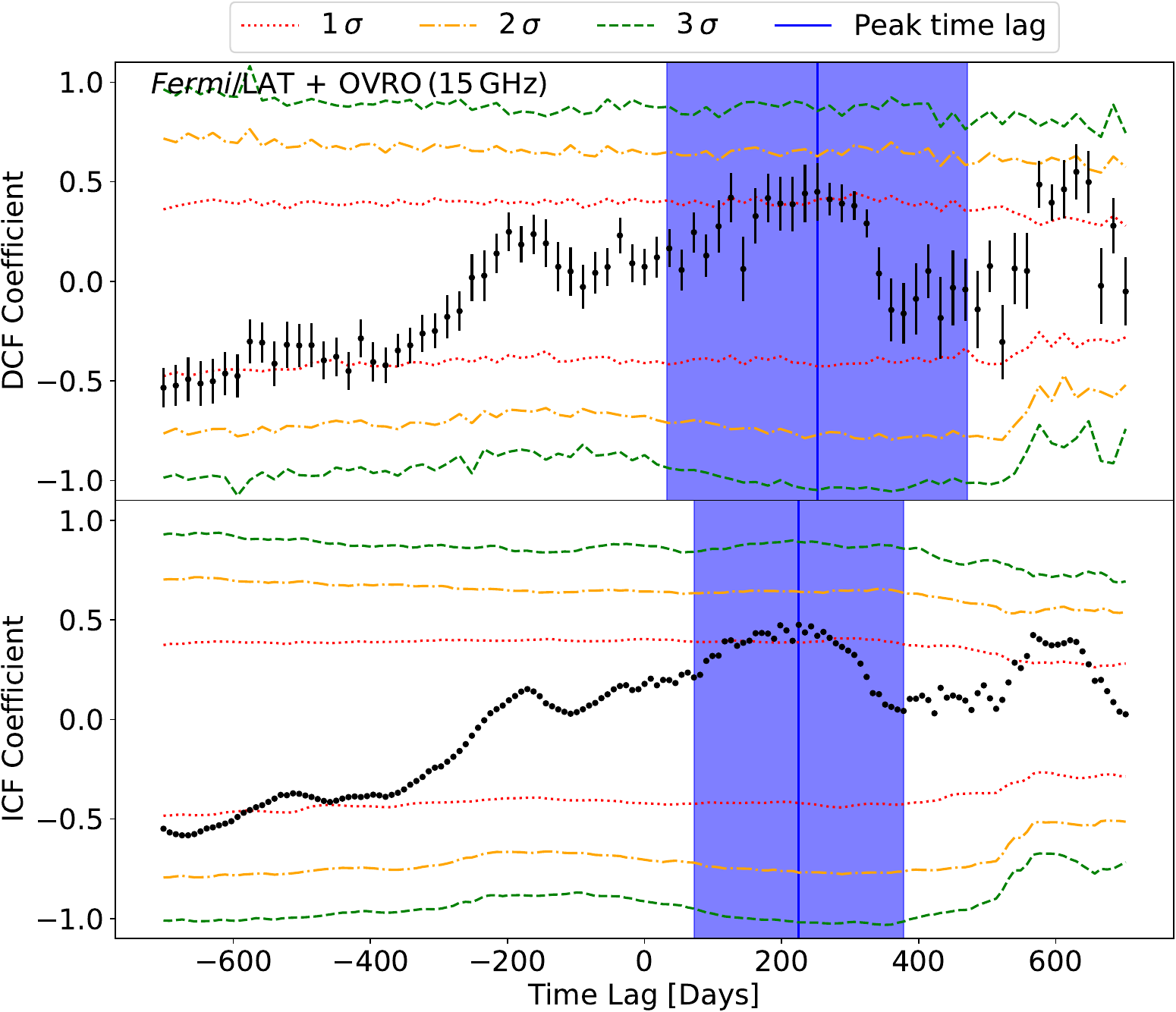}
      \caption{DCF (upper panel) and ICF (lower panel) cross-correlation coefficients between the \fermi\, \gray\, and \ovro\, light curves plotted over time lag. For the DCF the bin size is chosen to be $18\,\mathrm{days}$, while the interpolation unit for the ICF is chosen to be $9\,\mathrm{days}$ (both calculated as explained in Sect.~\ref{sec:corr_function}). Positive time lags mean that the radio light curve follows the \fermi\, \gray\,light curve. The time lags for the peak cross-correlation coefficients are marked by solid blue lines, with their $1\sigma$ uncertainties given by the shaded blue area. The dotted red, dashed-dotted orange and dashed green lines correspond to the two sided Gaussian equivalent $1\sigma$, $2\sigma$ and $3\sigma$ confidence intervals. Note that we did not choose the DCF time lag of the global peak, since this time lag would not be consistent with the one of the ICF as well as those derived for the other light curves (see Sect.~\ref{sec:cross} for details).}
         \label{fig:corr_fermi-ovro}
   \end{figure}

\onecolumn

\section{Additional $43\,\mathrm{GHz}$ VLBA images}
\label{app:source}


\begin{figure*}[h!]
   \centering
   \includegraphics[width=0.24\hsize]{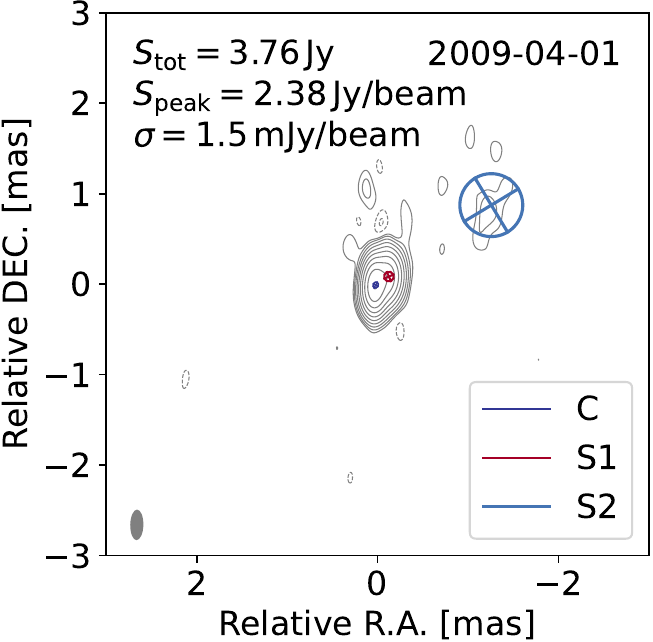}
   \includegraphics[width=0.24\hsize]{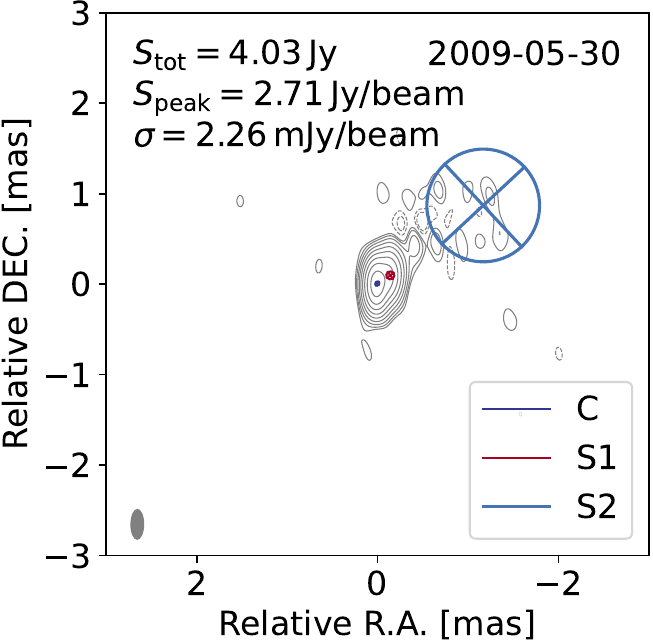}
   \includegraphics[width=0.24\hsize]{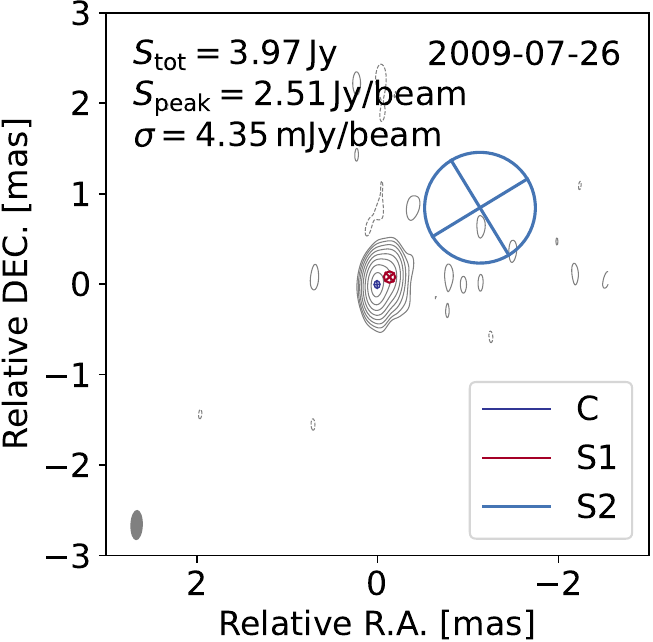}
   \includegraphics[width=0.24\hsize]{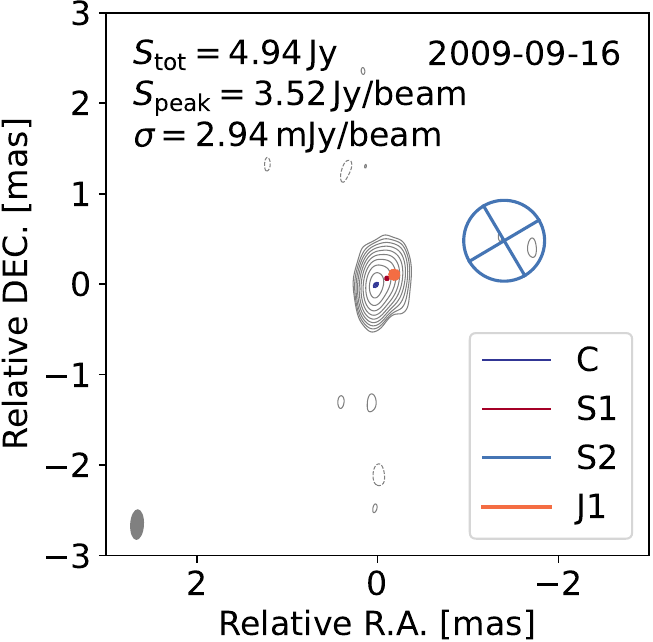}
   \includegraphics[width=0.24\hsize]{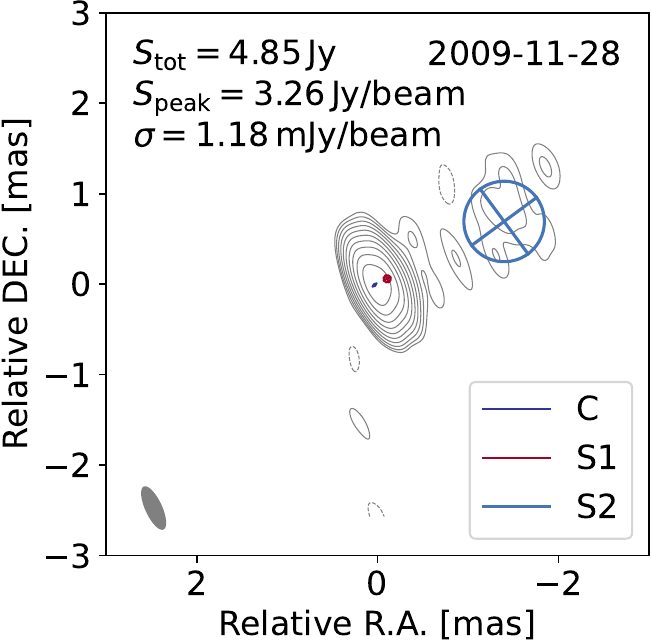}
   \includegraphics[width=0.24\hsize]{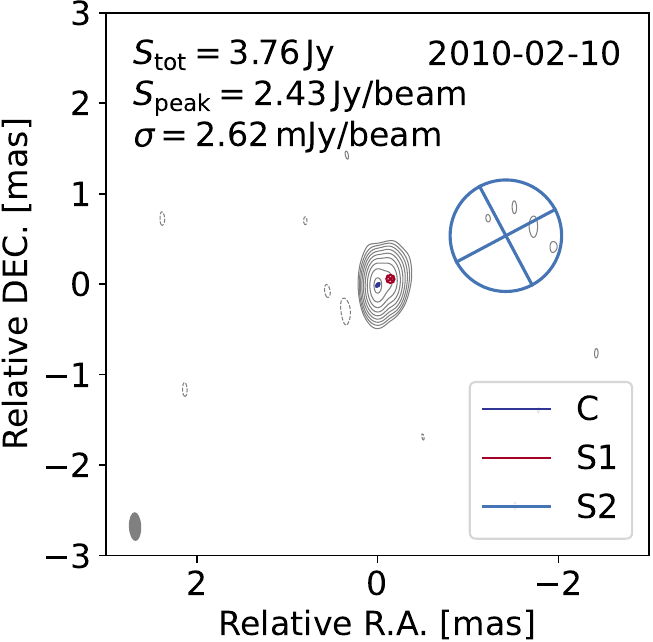}
   \includegraphics[width=0.24\hsize]{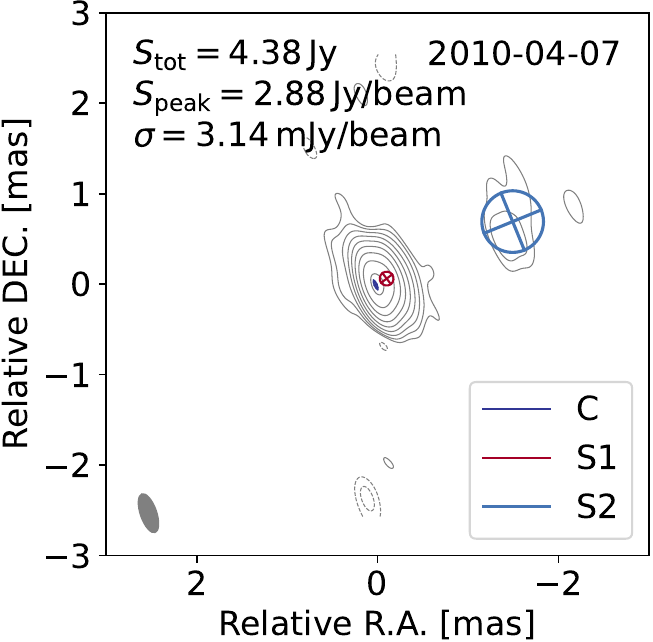}
   \includegraphics[width=0.24\hsize]{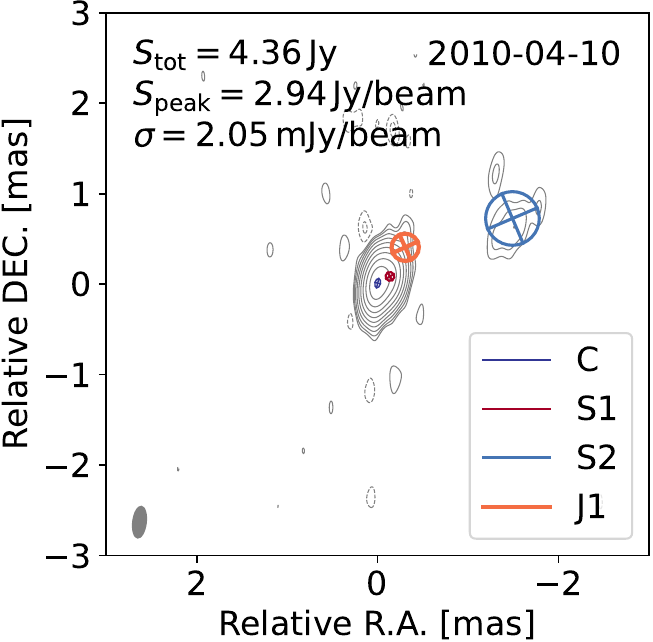}
   \includegraphics[width=0.24\hsize]{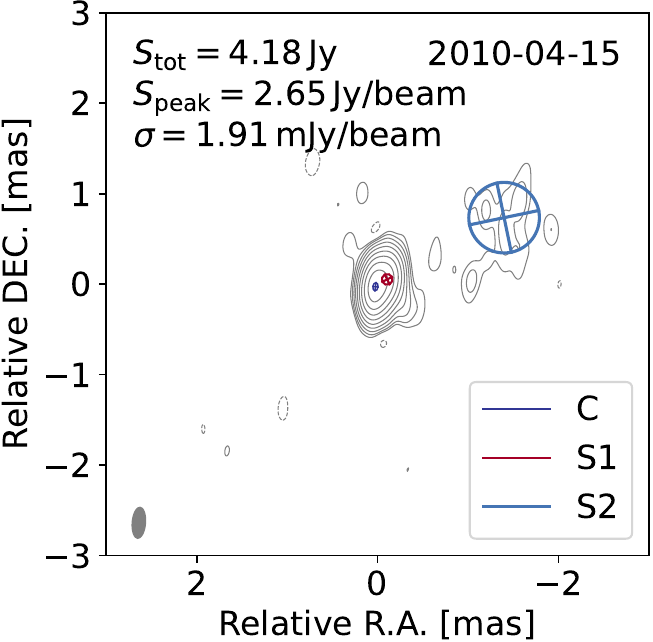}
   \includegraphics[width=0.24\hsize]{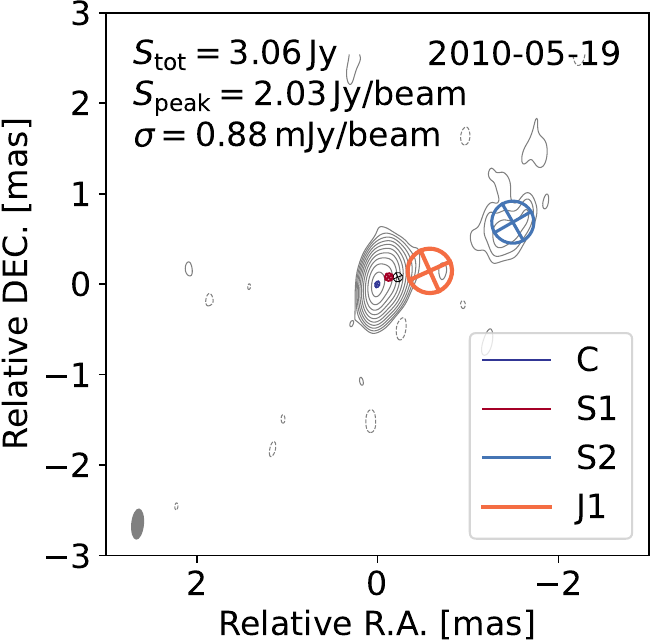}
   \includegraphics[width=0.24\hsize]{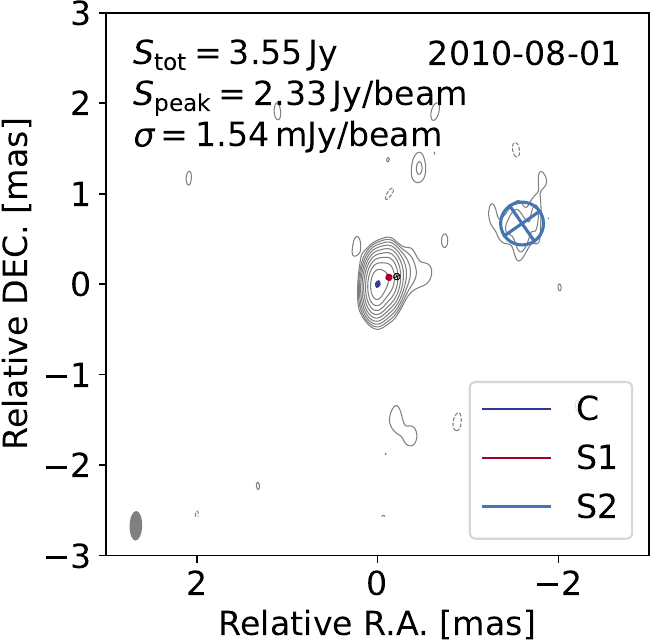}
   \includegraphics[width=0.24\hsize]{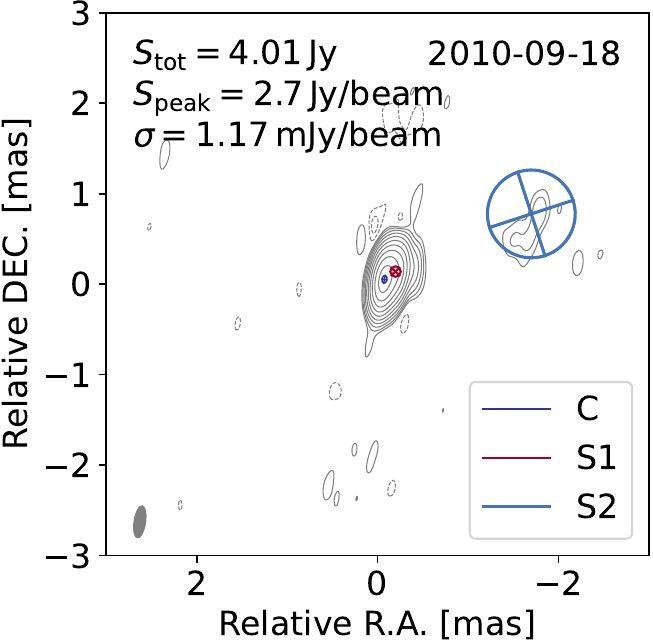}
   \includegraphics[width=0.24\hsize]{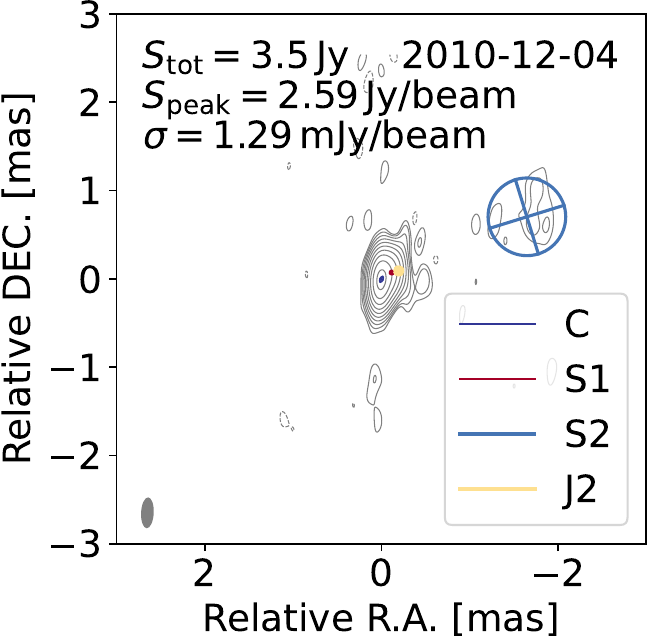}
   \includegraphics[width=0.24\hsize]{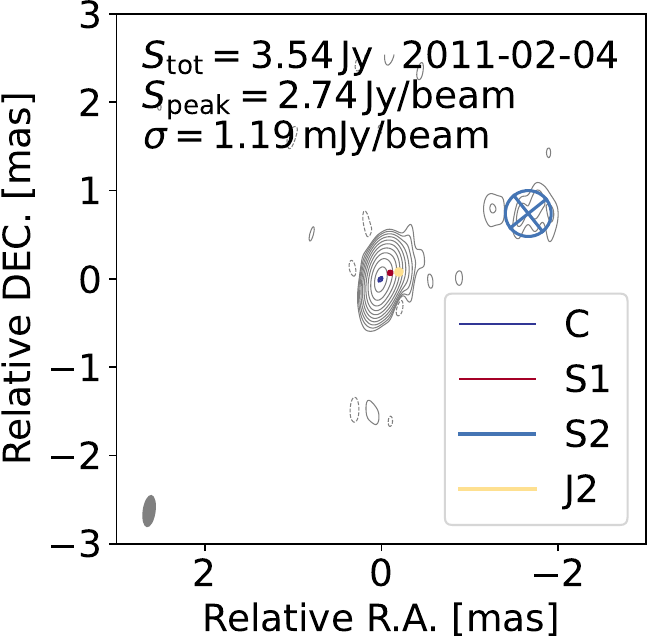}
   \includegraphics[width=0.24\hsize]{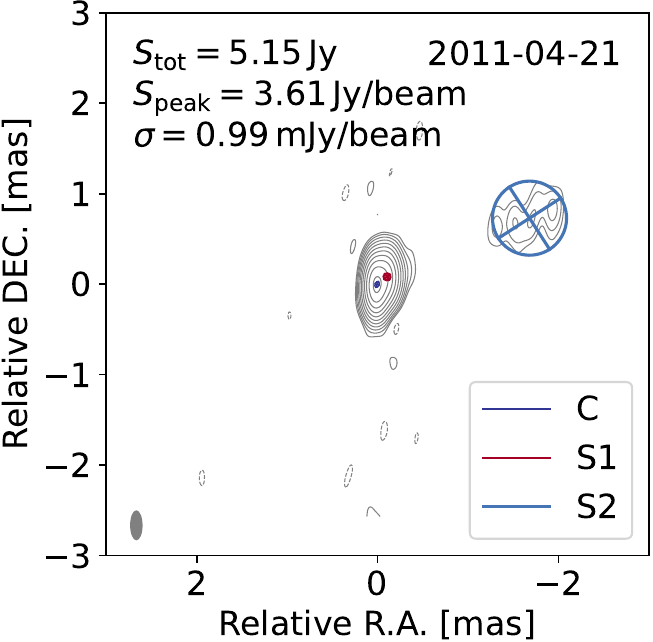}
   \includegraphics[width=0.24\hsize]{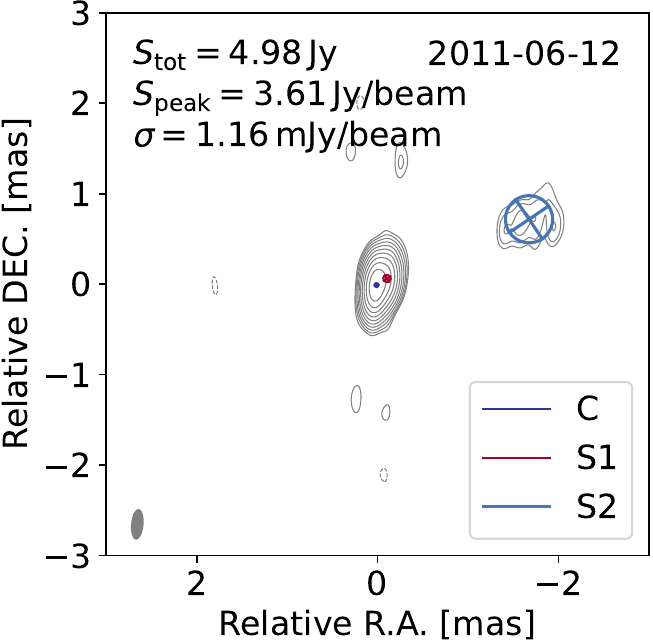}
   \includegraphics[width=0.24\hsize]{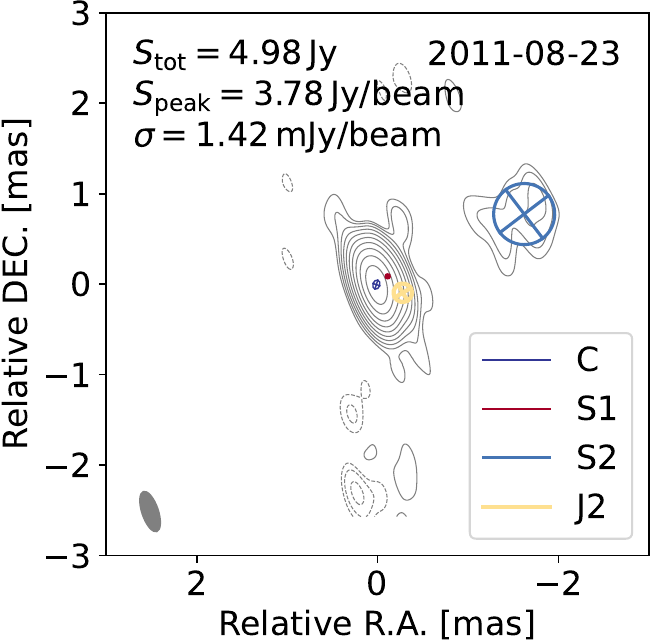}
   \includegraphics[width=0.24\hsize]{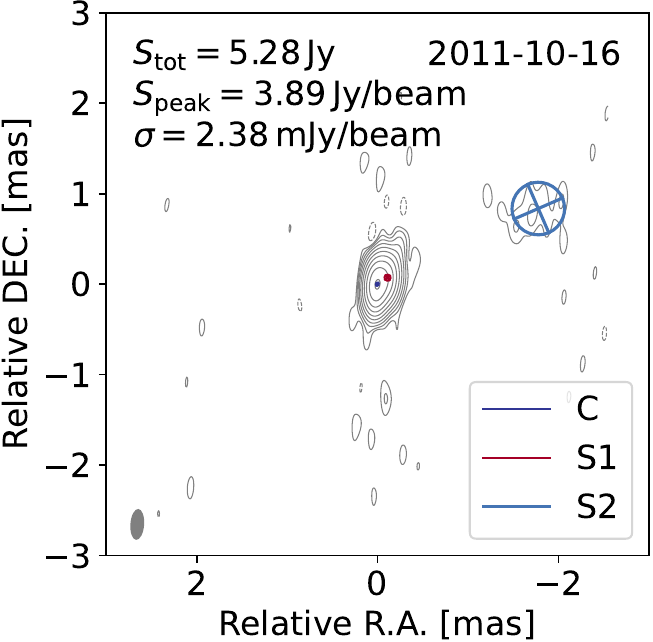}
   \includegraphics[width=0.24\hsize]{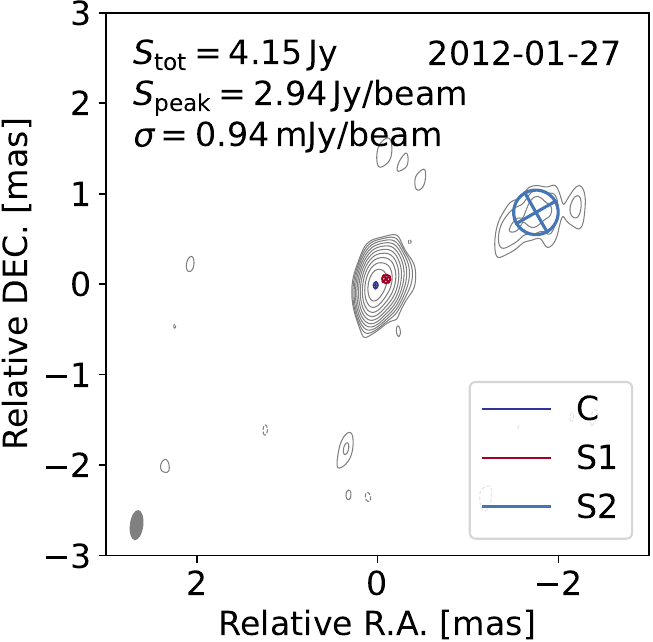}
   \includegraphics[width=0.24\hsize]{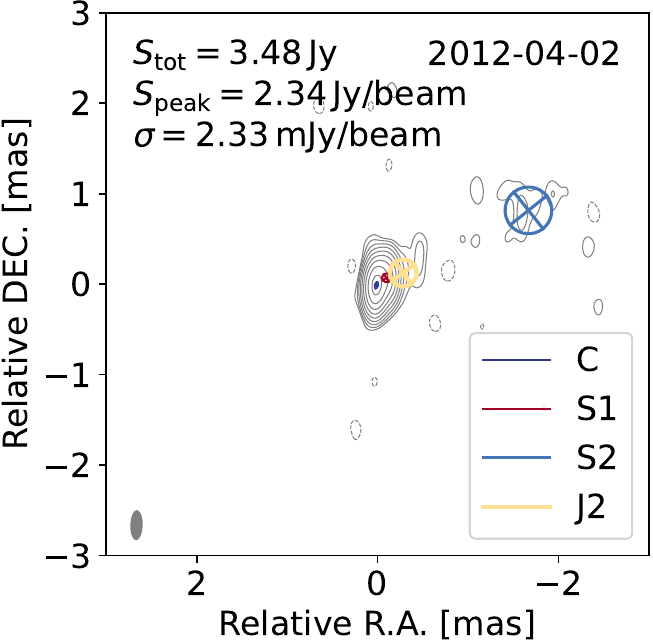}
      \caption{Uniformly weighted $43\,\mathrm{GHz}$ \vlba\, total intensity images of the FSRQ \source\, with the fitted Gaussian components overlaid. $S_{\mathrm{tot}}$ is the total integrated flux density, $S_{\mathrm{peak}}$ is the highest flux density per beam and $\sigma$ is the noise level. The gray ellipse in the bottom left corner corresponds to the beam. The contours begin at $3\sigma$ and increase logarithmically by a factor of 2. The image parameters are listed in Table~\ref{image}.}
         \label{fig:app_images1}
   \end{figure*}

\begin{figure*}[h!]
   \centering
   \includegraphics[width=0.24\hsize]{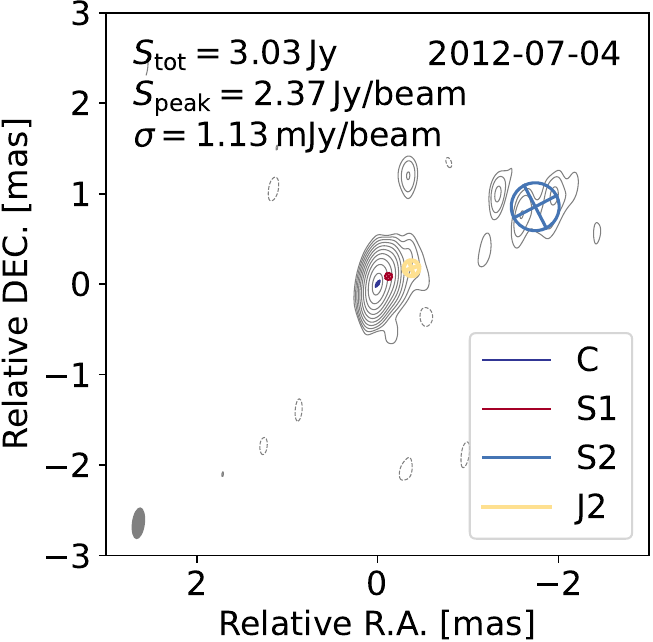}
   \includegraphics[width=0.24\hsize]{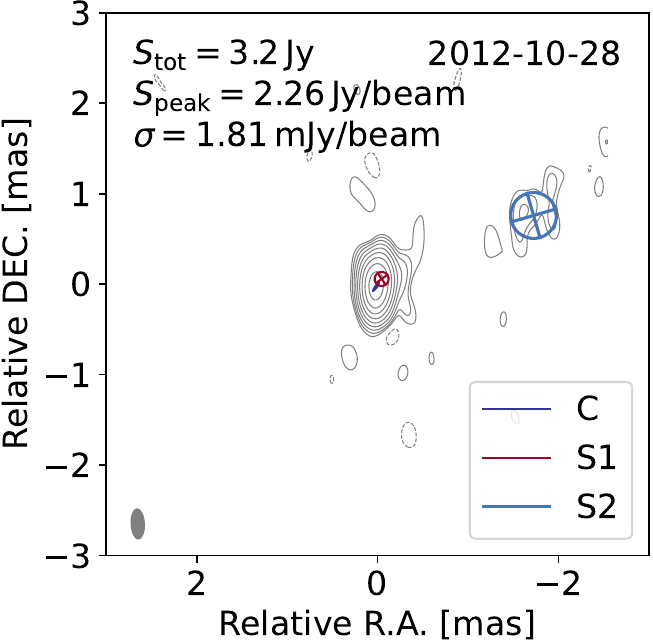}
   \includegraphics[width=0.24\hsize]{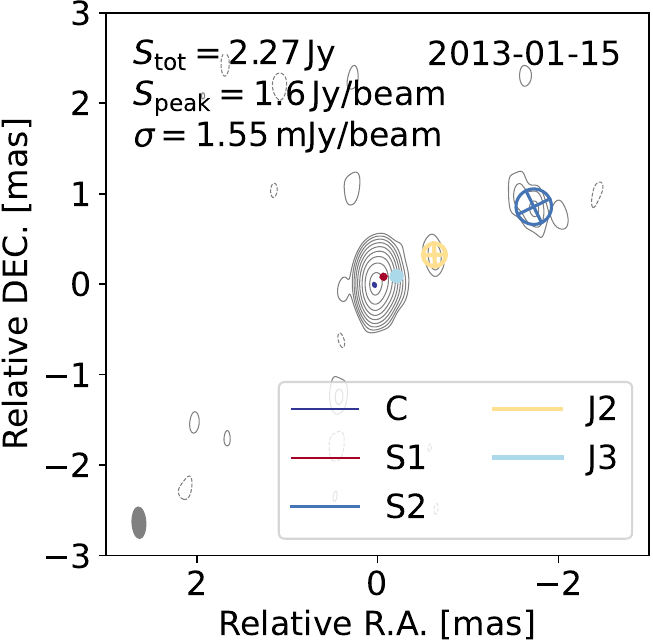}
   \includegraphics[width=0.24\hsize]{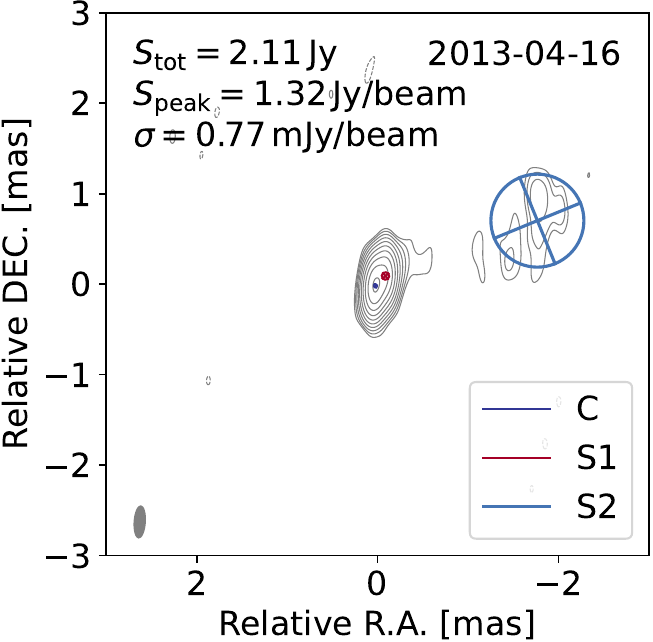}
   \includegraphics[width=0.24\hsize]{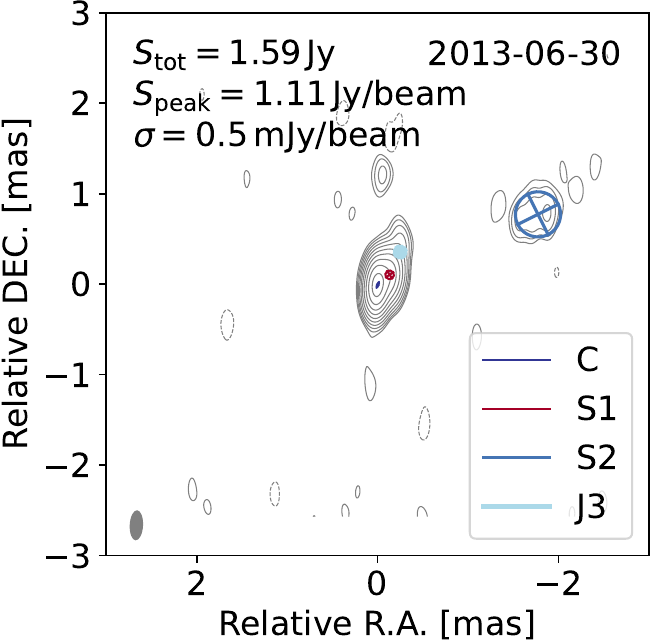}
   \includegraphics[width=0.24\hsize]{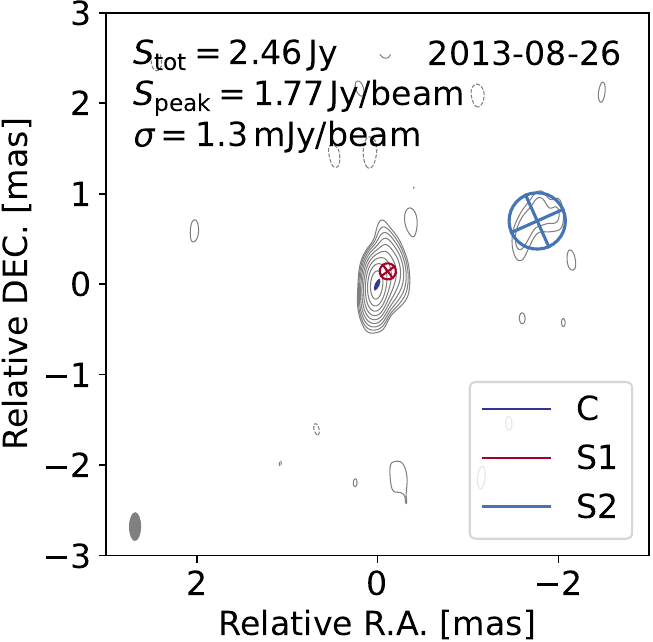}
   \includegraphics[width=0.24\hsize]{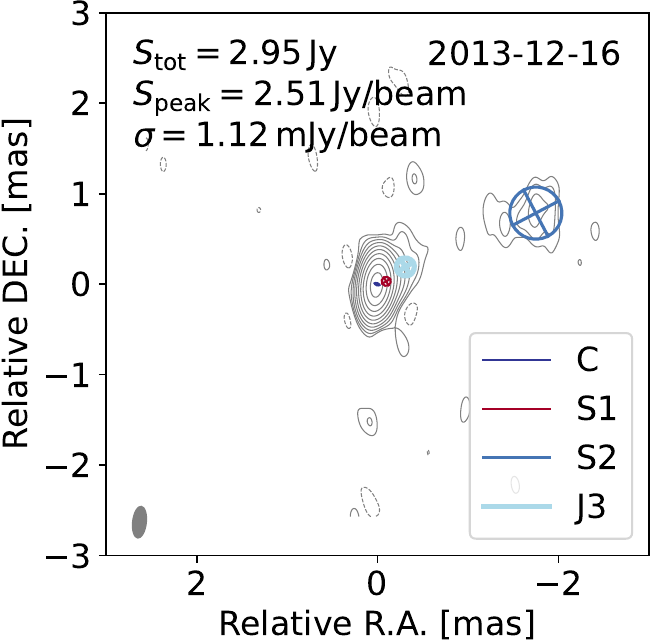}
   \includegraphics[width=0.24\hsize]{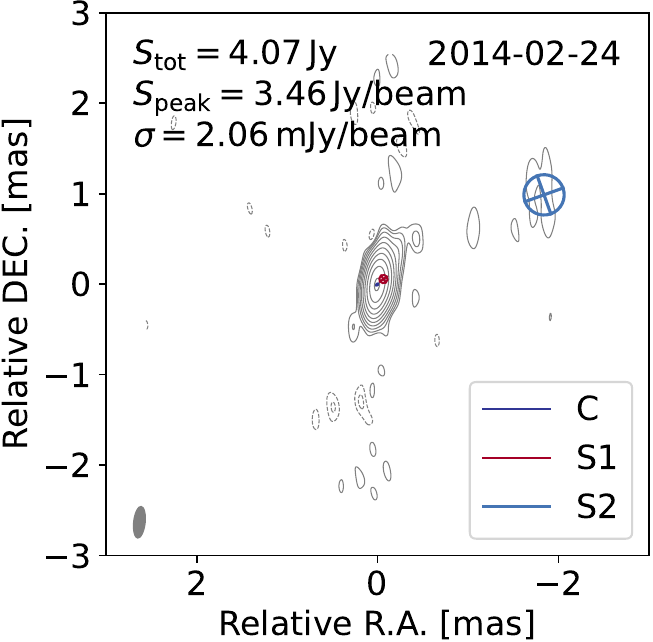}
   \includegraphics[width=0.24\hsize]{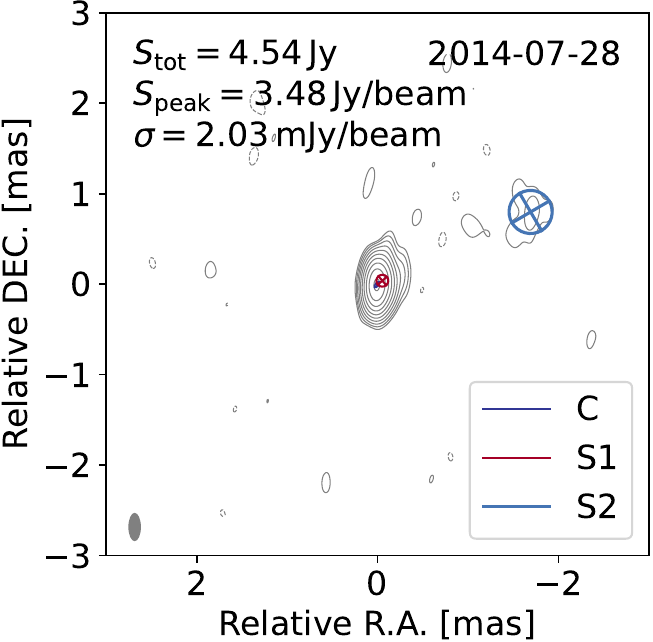}
   \includegraphics[width=0.24\hsize]{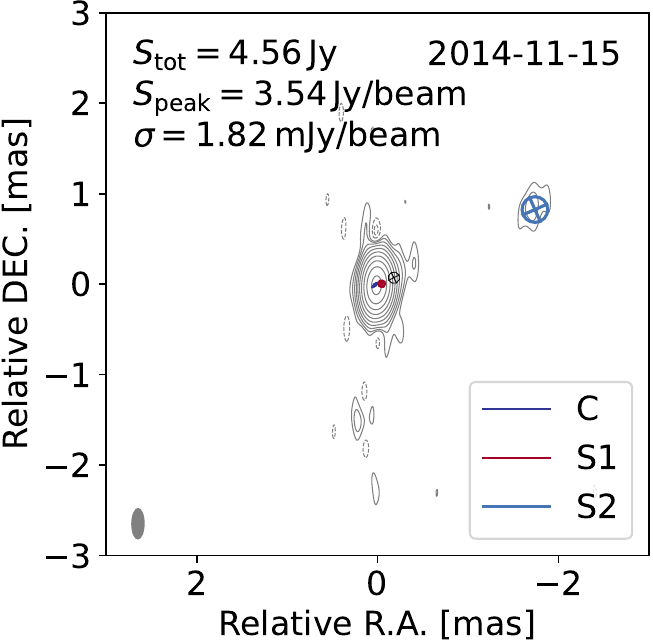}
   \includegraphics[width=0.24\hsize]{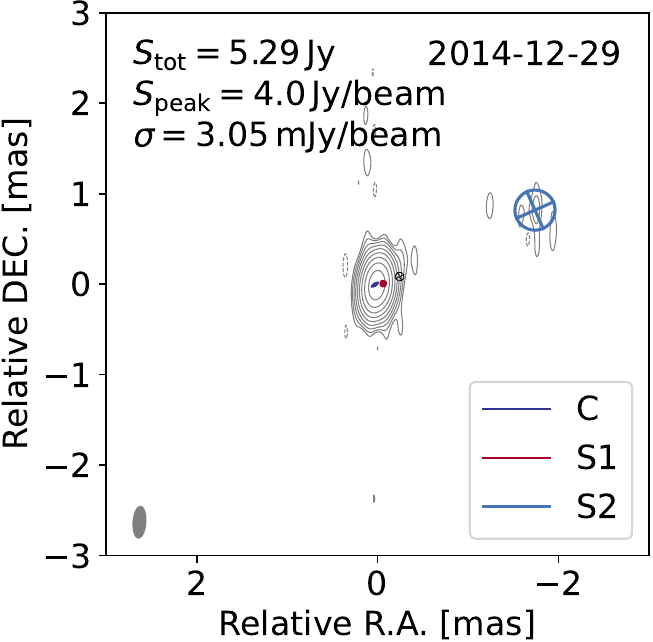}
      \caption{Fig.~\ref{fig:app_images1} continued.}
         \label{fig:app_images2}
   \end{figure*}


\section{Additional tables}
\label{app:tables}

\begin{table*}[h!]
\caption{\label{image}Image parameters of the uniformly weighted $43\,\mathrm{GHz}$ \vlba\, observations.}
\centering
\begin{tabular}{cccccccc}
\hline \hline
Date & Array & $S_\mathrm{tot}$ & $S_\mathrm{peak}$ & $\sigma_\mathrm{rms}$ & $b_\mathrm{maj}$ & $b_\mathrm{min}$ & P.A. \\
YYYY-MM-DD &  & [Jy] & [Jy/beam] & [mJy/beam] & [mas] & [mas] & [deg] \\
(1) & (2) & (3) & (4) & (5) & (6) & (7) & (8) \\
\hline
2009-04-01 & I & 3.76 & 2.38 & 1.5 & 0.338 & 0.145 & -1.071 \\
2009-05-30 & I & 4.03 & 2.71 & 2.26 & 0.343 & 0.153 & -0.412 \\
2009-07-26 & II & 3.97 & 2.51 & 4.35 & 0.336 & 0.138 & -1.849 \\
2009-09-16 & I & 4.94 & 3.52 & 2.94 & 0.341 & 0.153 & -3.034 \\
2009-11-28 & III & 4.85 & 3.26 & 1.18 & 0.524 & 0.188 & 24.735 \\
2010-02-10 & IV & 3.76 & 2.43 & 2.62 & 0.319 & 0.136 & 2.036 \\
2010-04-07 & III & 4.38 & 2.88 & 3.14 & 0.467 & 0.195 & 18.959 \\
2010-04-10 & I & 4.36 & 2.94 & 2.05 & 0.368 & 0.164 & -6.834 \\
2010-04-15 & I & 4.18 & 2.65 & 1.91 & 0.361 & 0.162 & -4.525 \\
2010-05-19 & I & 3.06 & 2.03 & 0.88 & 0.347 & 0.142 & -5.373 \\
2010-08-01 & I & 3.55 & 2.33 & 1.54 & 0.327 & 0.139 & -1.637 \\
2010-09-18 & I & 4.01 & 2.7 & 1.17 & 0.371 & 0.14 & -8.405 \\
2010-12-04 & I & 3.5 & 2.59 & 1.29 & 0.35 & 0.143 & -2.78 \\
2011-02-04 & II & 3.54 & 2.74 & 1.19 & 0.37 & 0.147 & -7.963 \\
2011-04-21 & V & 5.15 & 3.61 & 0.99 & 0.333 & 0.138 & 0.202 \\
2011-06-12 & VI & 4.98 & 3.61 & 1.16 & 0.343 & 0.138 & -5.094 \\
2011-08-23 & III & 4.98 & 3.78 & 1.42 & 0.485 & 0.198 & 18.559 \\
2011-10-16 & I & 5.28 & 3.89 & 2.38 & 0.343 & 0.152 & -4.113 \\
2012-01-27 & VII & 4.15 & 2.94 & 0.94 & 0.335 & 0.149 & -7.086 \\
2012-04-02 & I & 3.48 & 2.34 & 2.33 & 0.334 & 0.137 & -1.749 \\
2012-07-04 & I & 3.03 & 2.37 & 1.13 & 0.355 & 0.144 & -7.291 \\
2012-10-28 & I & 3.2 & 2.26 & 1.81 & 0.348 & 0.159 & 2.956 \\
2013-01-15 & I & 2.27 & 1.6 & 1.55 & 0.361 & 0.165 & 2.634 \\
2013-04-16 & I & 2.11 & 1.32 & 0.77 & 0.37 & 0.137 & -3.809 \\
2013-06-30 & VII & 1.59 & 1.11 & 0.5 & 0.333 & 0.147 & -2.776 \\
2013-08-26 & I & 2.46 & 1.77 & 1.3 & 0.319 & 0.134 & -0.502 \\
2013-12-16 & IV & 2.95 & 2.51 & 1.12 & 0.367 & 0.166 & -6.317 \\
2014-02-24 & VII & 4.07 & 3.46 & 2.06 & 0.367 & 0.141 & -6.262 \\
2014-07-28 & I & 4.54 & 3.48 & 2.03 & 0.314 & 0.139 & 0.071 \\
2014-11-15 & I & 4.56 & 3.54 & 1.82 & 0.351 & 0.149 & -0.601 \\
2014-12-29 & I & 5.29 & 4.0 & 3.05 & 0.367 & 0.154 & -4.098 \\
2015-04-11 & I & 6.35 & 3.82 & 0.96 & 0.325 & 0.143 & 1.312 \\
2015-06-09 & I & 6.13 & 3.77 & 1.59 & 0.341 & 0.152 & -0.303 \\
2015-08-01 & I & 6.13 & 3.55 & 0.91 & 0.319 & 0.132 & -2.485 \\
2015-12-05 & I & 4.82 & 2.92 & 0.64 & 0.366 & 0.137 & -7.627 \\
2016-01-31 & I & 4.78 & 2.71 & 0.87 & 0.323 & 0.127 & -7.567 \\
2016-04-22 & I & 4.55 & 2.55 & 0.64 & 0.327 & 0.138 & -0.611 \\
2016-07-04 & I & 4.71 & 2.72 & 0.58 & 0.328 & 0.141 & -1.759 \\
2016-09-05 & I & 5.99 & 3.94 & 1.06 & 0.347 & 0.158 & -3.308 \\
2016-10-23 & I & 5.28 & 3.46 & 0.9 & 0.338 & 0.148 & 3.487 \\
2016-12-23 & I & 6.24 & 4.09 & 1.59 & 0.355 & 0.127 & -3.698 \\
2017-02-04 & I & 6.55 & 4.33 & 1.25 & 0.375 & 0.132 & -6.121 \\
2017-04-16 & I & 4.11 & 2.82 & 0.63 & 0.351 & 0.147 & -4.838 \\
2017-06-08 & I & 5.88 & 3.84 & 1.85 & 0.338 & 0.136 & -2.272 \\
2017-08-06 & II & 4.48 & 2.9 & 0.93 & 0.329 & 0.128 & -3.145 \\
2017-11-06 & VIII & 4.46 & 3.34 & 0.88 & 1.131 & 0.131 & -18.329 \\
2018-02-17 & IX & 4.24 & 2.79 & 0.73 & 0.621 & 0.141 & -19.259 \\
2018-04-19 & I & 5.76 & 2.87 & 0.66 & 0.341 & 0.141 & -4.124 \\
2018-06-16 & IX  & 3.73 & 2.4 & 0.68 & 0.677 & 0.147 & -17.942 \\
2018-08-26 & I & 2.57 & 1.49 & 0.37 & 0.352 & 0.163 & 4.936 \\
2018-12-08 & I & 5.42 & 3.79 & 0.57 & 0.399 & 0.151 & -7.454 \\
\hline
\end{tabular}
\tablefoot{Col.(1): UTC observation date; Col.(2): Array configuration: I: full VLBA array, II: without Pie Town, III: without Mauna Kea, IV: without Hancock, V: without North Liberty, VI: without Owens Valley, VII: without Fort Davis, VIII: without Brewster and St.~Croix, IX: without St.~Croix; Col.(3): Total flux density with relative uncertainties of 5\,\%; Col.(4): Peak flux density with relative uncertainties of 5\,\%; Col.(5): Noise level; Col.(6): FWHM of the major axis of the beam; Col.(7): FWHM of the minor axis of the beam; Col.(8): Position angle of the beam measured north through east.}
\end{table*}

\longtab[2]{
\begin{longtable}{ccccccccc}
\caption{\label{komp} \modelfit\, parameters of the fitted Gaussian jet components.}\\
\hline\hline
Date & ID & Relative R.A. & Relative Dec. & $a_\mathrm{maj}$ & $a_\mathrm{min}$ & P.A. & $S$ & $T_\mathrm{B}$ \\
YYYY-MM-DD &  & [mas] & [mas] & [mas] & [mas] & [deg] & [Jy] & [K] \\
(1) & (2) & (3) & (4) & (5) & (6) & (7) & (8) & (9) \\
\hline
\endfirsthead
\caption{Continued.}\\
\hline\hline
Date & ID & Relative R.A. & Relative Dec. & $a_\mathrm{maj}$ & $a_\mathrm{min}$ & P.A. & $S$ & $T_\mathrm{B}$ \\
YYYY-MM-DD &  & [mas] & [mas] & [mas] & [mas] & [deg] & [Jy] & [K] \\
(1) & (2) & (3) & (4) & (5) & (6) & (7) & (8) & (9) \\
\hline
\endhead
\hline
\endfoot
\hline
\multicolumn{9}{l}{Col.(1): UTC observation date; Col.(2): Identification of the component, C denotes the core component, S denotes}\\
\multicolumn{9}{l}{stationary jet components and J denotes moving jet components. If two IDs are given, the first ID corresponds to}\\
\multicolumn{9}{l}{kinematic model 1 and the second one to kinematic model 2 (see Sect.~\ref{sec:kin}); Col.(3): R.A. of the components relative}\\
\multicolumn{9}{l}{to the designated phase center; Col.(4): Declination of the components relative to the designated phase center;}\\
\multicolumn{9}{l}{Col.(5): FWHM of the major axis of the components (relative uncertainties of $20\,\%$ are assumed). For unresolved}\\
\multicolumn{9}{l}{axes, the corresponding resolution limit is given as an upper limit; Col.(6): FWHM of the minor axis of the}\\
\multicolumn{9}{l}{components (relative uncertainties of $20\,\%$ are assumed). For unresolved axes, the corresponding resolution limit is}\\
\multicolumn{9}{l}{given as an upper limit; Col.(7): Position angle of the components measured north through east; Col.(8): Flux density}\\
\multicolumn{9}{l}{of the components (relative uncertainties of $5\,\%$ are assumed); Col.(9): Observed brightness temperature of the}\\
\multicolumn{9}{l}{components (relative uncertainties of $29\,\%$ are calculated). For unresolved components, their brightness temperature}\\
\multicolumn{9}{l}{was calculated using the resolution limits of the corresponding unresolved axes and is therefore given as an lower}\\
\multicolumn{9}{l}{limit.}\\
\endlastfoot
2009-04-01 & C & 0.021 & -0.01 & 0.072 & 0.058 & -27.9 & 2.467 & $0.74\cdot 10^{12}$ \\
 & S1 & -0.126 & 0.083 & 0.108 & 0.108 & -156.0 & 1.152 & $1.22\cdot 10^{11}$ \\
 & S2 & -1.258 & 0.874 & 0.696 & 0.696 & -148.9 & 0.081 & $2.09\cdot 10^{8}$ \\
\hline
2009-05-30 & C & 0.003 & 0.006 & 0.058 & 0.048 & -37.9 & 2.729 & $1.22\cdot 10^{12}$ \\
 & S1 & -0.141 & 0.097 & 0.089 & 0.089 & -143.6 & 1.171 & $1.85\cdot 10^{11}$ \\
 & S2 & -1.169 & 0.87 & 1.247 & 1.247 & -137.0 & 0.129 & $1.04\cdot 10^{8}$ \\
\hline
2009-07-26 & C & 0.005 & -0.004 & 0.084 & 0.063 & 3.1 & 2.662 & $0.63\cdot 10^{12}$ \\
 & S1 & -0.133 & 0.078 & 0.126 & 0.126 & -137.9 & 1.136 & $0.9\cdot 10^{11}$ \\
 & S2 & -1.133 & 0.845 & 1.225 & 1.225 & -148.7 & 0.12 & $1.0\cdot 10^{8}$ \\
\hline
2009-09-16 & C & 0.018 & -0.009 & 0.064 & 0.04 & -43.5 & 3.524 & $1.73\cdot 10^{12}$ \\
 & S1 & -0.104 & 0.064 & 0.035 & 0.035 & -139.2 & 1.045 & $1.06\cdot 10^{12}$ \\
 & J1 / J1 & -0.184 & 0.104 & 0.086 & 0.086 & -140.2 & 0.348 & $0.59\cdot 10^{11}$ \\
 & S2 & -1.4 & 0.48 & 0.895 & 0.895 & -149.4 & 0.054 & $0.84\cdot 10^{8}$ \\
\hline
2009-11-28 & C & 0.033 & -0.01 & 0.056 & 0.024 & -50.1 & 2.964 & $2.78\cdot 10^{12}$ \\
 & S1 & -0.107 & 0.059 & 0.08 & 0.08 & -152.6 & 1.804 & $3.53\cdot 10^{11}$ \\
 & S2 & -1.4 & 0.693 & 0.89 & 0.89 & -143.6 & 0.078 & $1.23\cdot 10^{8}$ \\
\hline
2010-02-10 & C & -0.004 & -0.009 & 0.057 & 0.034 & -40.9 & 2.422 & $1.58\cdot 10^{12}$ \\
 & S1 & -0.142 & 0.057 & 0.09 & 0.09 & -136.7 & 1.338 & $2.08\cdot 10^{11}$ \\
 & S2 & -1.419 & 0.535 & 1.234 & 1.234 & -152.0 & 0.106 & $0.87\cdot 10^{8}$ \\
\hline
2010-04-07 & C & 0.021 & -0.008 & 0.112 & < 0.032 & 23.6 & 2.247 & $>0.8\cdot 10^{12}$ \\
 & S1 & -0.101 & 0.061 & 0.153 & 0.153 & -142.8 & 1.972 & $1.05\cdot 10^{11}$ \\
 & S2 & -1.494 & 0.692 & 0.682 & 0.682 & -157.7 & 0.115 & $3.1\cdot 10^{8}$ \\
\hline
2010-04-10 & C & -0.001 & 0.011 & 0.105 & 0.061 & -13.4 & 2.847 & $0.56\cdot 10^{12}$ \\
 & S1 & -0.137 & 0.085 & 0.098 & 0.098 & -149.4 & 1.408 & $1.81\cdot 10^{11}$ \\
 & -- / J1 & -0.306 & 0.407 & 0.302 & 0.302 & -155.4 & 0.041 & $0.56\cdot 10^{9}$ \\
 & S2 & -1.492 & 0.726 & 0.594 & 0.594 & -156.9 & 0.111 & $3.93\cdot 10^{8}$ \\
\hline
2010-04-15 & C & 0.023 & -0.029 & 0.091 & 0.057 & -7.8 & 2.45 & $0.58\cdot 10^{12}$ \\
 & S1 & -0.107 & 0.052 & 0.119 & 0.119 & -153.8 & 1.537 & $1.35\cdot 10^{11}$ \\
 & S2 & -1.4 & 0.735 & 0.781 & 0.781 & -168.3 & 0.128 & $2.62\cdot 10^{8}$ \\
\hline
2010-05-19 & C & 0.003 & -0.002 & 0.073 & 0.05 & -20.8 & 2.009 & $0.69\cdot 10^{12}$ \\
 & S1 & -0.125 & 0.078 & 0.084 & 0.084 & -143.5 & 0.889 & $1.58\cdot 10^{11}$ \\
 & J1 / -- & -0.224 & 0.077 & 0.106 & 0.106 & -158.0 & 0.066 & $0.73\cdot 10^{10}$ \\
 & -- / J1 & -0.578 & 0.147 & 0.491 & 0.491 & -155.6 & 0.021 & $1.1\cdot 10^{8}$ \\
 & S2 & -1.496 & 0.685 & 0.466 & 0.466 & -150.1 & 0.075 & $4.35\cdot 10^{8}$ \\
\hline
2010-08-01 & C & -0.003 & 0.001 & 0.067 & 0.044 & -17.4 & 2.321 & $1.0\cdot 10^{12}$ \\
 & S1 & -0.126 & 0.076 & 0.055 & 0.055 & -154.2 & 0.992 & $4.1\cdot 10^{11}$ \\
 & J1 / -- & -0.213 & 0.083 & 0.073 & 0.073 & -155.9 & 0.12 & $2.83\cdot 10^{10}$ \\
 & S2 & -1.598 & 0.67 & 0.474 & 0.474 & -145.2 & 0.085 & $4.71\cdot 10^{8}$ \\
\hline
2010-09-18 & C & -0.077 & 0.053 & 0.09 & 0.055 & -11.3 & 2.609 & $0.66\cdot 10^{12}$ \\
 & S1 & -0.199 & 0.139 & 0.117 & 0.117 & -135.0 & 1.336 & $1.23\cdot 10^{11}$ \\
 & S2 & -1.701 & 0.777 & 0.971 & 0.971 & -161.9 & 0.101 & $1.34\cdot 10^{8}$ \\
\hline
2010-12-04 & C & -0.001 & -0.005 & 0.07 & 0.036 & -24.2 & 2.593 & $1.29\cdot 10^{12}$ \\
 & S1 & -0.117 & 0.071 & 0.041 & 0.041 & -154.2 & 0.687 & $0.5\cdot 10^{12}$ \\
 & J1 / J2 & -0.197 & 0.089 & 0.078 & 0.078 & -144.0 & 0.13 & $2.66\cdot 10^{10}$ \\
 & S2 & -1.648 & 0.702 & 0.88 & 0.88 & -162.8 & 0.083 & $1.34\cdot 10^{8}$ \\
\hline
2011-02-04 & C & 0.011 & -0.005 & 0.065 & 0.042 & -38.0 & 2.679 & $1.21\cdot 10^{12}$ \\
 & S1 & -0.1 & 0.068 & 0.049 & 0.049 & -142.9 & 0.739 & $3.82\cdot 10^{11}$ \\
 & J1 / J2 & -0.198 & 0.077 & 0.056 & 0.056 & -146.3 & 0.08 & $3.17\cdot 10^{10}$ \\
 & S2 & -1.663 & 0.739 & 0.521 & 0.521 & -141.3 & 0.048 & $2.2\cdot 10^{8}$ \\
\hline
2011-04-21 & C & 0.005 & -0.004 & 0.069 & 0.045 & -24.0 & 3.571 & $1.44\cdot 10^{12}$ \\
 & S1 & -0.105 & 0.081 & 0.079 & 0.079 & -141.5 & 1.486 & $3.0\cdot 10^{11}$ \\
 & S2 & -1.681 & 0.729 & 0.819 & 0.819 & -147.2 & 0.102 & $1.9\cdot 10^{8}$ \\
\hline
2011-06-12 & C & 0.011 & -0.01 & 0.052 & 0.05 & -85.0 & 3.52 & $1.71\cdot 10^{12}$ \\
 & S1 & -0.103 & 0.062 & 0.081 & 0.081 & -144.3 & 1.363 & $2.58\cdot 10^{11}$ \\
 & S2 & -1.675 & 0.718 & 0.522 & 0.522 & -146.0 & 0.106 & $4.88\cdot 10^{8}$ \\
\hline
2011-08-23 & C & 0.013 & -0.005 & 0.103 & 0.078 & -17.9 & 3.956 & $0.62\cdot 10^{12}$ \\
 & S1 & -0.112 & 0.086 & 0.046 & 0.046 & -135.6 & 0.846 & $4.99\cdot 10^{11}$ \\
 & J1 / J2 & -0.279 & -0.095 & 0.204 & 0.204 & -148.8 & 0.089 & $2.65\cdot 10^{9}$ \\
 & S2 & -1.622 & 0.775 & 0.675 & 0.675 & -142.6 & 0.104 & $2.85\cdot 10^{8}$ \\
\hline
2011-10-16 & C & 0.003 & -0.001 & 0.045 & 0.037 & -33.2 & 3.57 & $2.7\cdot 10^{12}$ \\
 & S1 & -0.111 & 0.072 & 0.067 & 0.067 & -142.5 & 1.535 & $4.28\cdot 10^{11}$ \\
 & S2 & -1.778 & 0.836 & 0.582 & 0.582 & -156.5 & 0.112 & $4.15\cdot 10^{8}$ \\
\hline
2012-01-27 & C & 0.02 & -0.013 & 0.076 & 0.05 & -5.2 & 2.753 & $0.91\cdot 10^{12}$ \\
 & S1 & -0.096 & 0.057 & 0.093 & 0.093 & -141.2 & 1.311 & $1.91\cdot 10^{11}$ \\
 & S2 & -1.752 & 0.793 & 0.493 & 0.493 & -150.4 & 0.096 & $4.92\cdot 10^{8}$ \\
\hline
2012-04-02 & C & 0.012 & -0.012 & 0.08 & 0.039 & -13.6 & 2.234 & $0.89\cdot 10^{12}$ \\
 & S1 & -0.098 & 0.07 & 0.097 & 0.097 & -146.5 & 1.021 & $1.35\cdot 10^{11}$ \\
 & J1 / J2 & -0.284 & 0.123 & 0.3 & 0.3 & -142.6 & 0.05 & $0.7\cdot 10^{9}$ \\
 & S2 & -1.669 & 0.815 & 0.517 & 0.517 & -141.5 & 0.087 & $4.05\cdot 10^{8}$ \\
\hline
2012-07-04 & C & -0.002 & 0.005 & 0.083 & 0.029 & -32.0 & 2.411 & $1.24\cdot 10^{12}$ \\
 & S1 & -0.122 & 0.085 & 0.078 & 0.078 & -145.0 & 0.473 & $0.97\cdot 10^{11}$ \\
 & J1 / J2 & -0.375 & 0.176 & 0.173 & 0.173 & -160.4 & 0.028 & $1.17\cdot 10^{9}$ \\
 & S2 & -1.744 & 0.856 & 0.531 & 0.531 & -152.5 & 0.087 & $3.87\cdot 10^{8}$ \\
\hline
2012-10-28 & C & 0.018 & -0.032 & 0.112 & 0.025 & -37.7 & 2.04 & $0.9\cdot 10^{12}$ \\
 & S1 & -0.045 & 0.057 & 0.156 & 0.156 & -144.5 & 0.994 & $0.51\cdot 10^{11}$ \\
 & S2 & -1.725 & 0.759 & 0.508 & 0.508 & -163.8 & 0.095 & $4.59\cdot 10^{8}$ \\
\hline
2013-01-15 & C & 0.034 & -0.009 & 0.052 & 0.036 & 22.5 & 1.344 & $0.9\cdot 10^{12}$ \\
 & S1 & -0.067 & 0.081 & 0.065 & 0.065 & -170.2 & 0.782 & $2.31\cdot 10^{11}$ \\
 & J1 / J3 & -0.21 & 0.088 & 0.098 & 0.098 & -159.9 & 0.023 & $3.04\cdot 10^{9}$ \\
 & -- / J2 & -0.629 & 0.323 & 0.252 & 0.252 & -177.9 & 0.009 & $1.75\cdot 10^{8}$ \\
 & S2 & -1.728 & 0.856 & < 0.393 & < 0.393 & -154.1 & 0.063 & $>0.51\cdot 10^{9}$ \\
\hline
2013-04-16 & C & 0.024 & -0.019 & 0.047 & 0.038 & 21.3 & 1.18 & $0.83\cdot 10^{12}$ \\
 & S1 & -0.088 & 0.09 & 0.082 & 0.082 & -140.7 & 0.769 & $1.42\cdot 10^{11}$ \\
 & S2 & -1.768 & 0.702 & 1.028 & 1.028 & -157.7 & 0.089 & $1.05\cdot 10^{8}$ \\
\hline
2013-06-30 & C & -0.003 & -0.008 & 0.07 & 0.022 & -22.6 & 1.088 & $0.87\cdot 10^{12}$ \\
 & S1 & -0.133 & 0.103 & 0.096 & 0.096 & -145.4 & 0.417 & $0.57\cdot 10^{11}$ \\
 & J1 / J3 & -0.251 & 0.355 & 0.105 & 0.105 & -150.9 & 0.018 & $2.03\cdot 10^{9}$ \\
 & S2 & -1.774 & 0.772 & 0.5 & 0.5 & -153.4 & 0.055 & $2.74\cdot 10^{8}$ \\
\hline
2013-08-26 & C & 0.004 & -0.006 & 0.115 & 0.034 & -23.8 & 1.921 & $0.61\cdot 10^{12}$ \\
 & S1 & -0.113 & 0.141 & 0.179 & 0.179 & -143.6 & 0.478 & $1.87\cdot 10^{10}$ \\
 & S2 & -1.766 & 0.699 & 0.623 & 0.623 & -155.9 & 0.091 & $2.92\cdot 10^{8}$ \\
\hline
2013-12-16 & C & 0.005 & -0.001 & 0.065 & 0.03 & 70.1 & 2.551 & $1.67\cdot 10^{12}$ \\
 & S1 & -0.096 & 0.031 & 0.101 & 0.101 & -149.3 & 0.29 & $3.53\cdot 10^{10}$ \\
 & J1 / J3 & -0.307 & 0.19 & 0.199 & 0.199 & -148.8 & 0.029 & $0.91\cdot 10^{9}$ \\
 & S2 & -1.75 & 0.785 & 0.576 & 0.576 & -152.2 & 0.076 & $2.84\cdot 10^{8}$ \\
\hline
2014-02-24 & C & 0.004 & -0.004 & 0.034 & 0.022 & -44.2 & 2.978 & $0.5\cdot 10^{13}$ \\
 & S1 & -0.065 & 0.056 & 0.091 & 0.091 & -145.9 & 1.021 & $1.53\cdot 10^{11}$ \\
 & S2 & -1.841 & 0.987 & 0.446 & 0.446 & -160.3 & 0.059 & $3.68\cdot 10^{8}$ \\
\hline
2014-07-28 & C & 0.006 & -0.006 & 0.08 & 0.037 & -38.8 & 3.136 & $1.32\cdot 10^{12}$ \\
 & S1 & -0.053 & 0.038 & 0.131 & 0.131 & -139.2 & 1.256 & $0.91\cdot 10^{11}$ \\
 & S2 & -1.693 & 0.798 & 0.477 & 0.477 & -149.7 & 0.077 & $4.21\cdot 10^{8}$ \\
\hline
2014-11-15 & C & 0.029 & -0.009 & 0.078 & 0.026 & -47.5 & 2.593 & $1.58\cdot 10^{12}$ \\
 & S1 & -0.046 & 0.004 & 0.073 & 0.073 & -147.0 & 1.876 & $4.45\cdot 10^{11}$ \\
 & -- & -0.181 & 0.071 & 0.128 & 0.128 & -153.4 & 0.053 & $4.03\cdot 10^{9}$ \\
 & S2 & -1.74 & 0.825 & 0.285 & 0.285 & -156.1 & 0.037 & $0.57\cdot 10^{9}$ \\
\hline
2014-12-29 & C & 0.029 & -0.006 & 0.085 & 0.03 & -58.0 & 3.34 & $1.66\cdot 10^{12}$ \\
 & S1 & -0.063 & 0.005 & 0.06 & 0.06 & -147.0 & 1.887 & $0.66\cdot 10^{12}$ \\
 & -- & -0.245 & 0.083 & < 0.097 & < 0.097 & -153.4 & 0.021 & $>2.77\cdot 10^{9}$ \\
 & S2 & -1.741 & 0.819 & 0.444 & 0.444 & -156.1 & 0.048 & $3.02\cdot 10^{8}$ \\
\hline
2015-04-11 & C & 0.01 & -0.015 & 0.071 & 0.038 & -43.6 & 3.317 & $1.54\cdot 10^{12}$ \\
 & S1 & -0.101 & 0.023 & 0.055 & 0.055 & -154.1 & 2.426 & $1.0\cdot 10^{12}$ \\
 & J4 & -0.168 & -0.016 & 0.088 & 0.088 & -137.6 & 0.562 & $0.9\cdot 10^{11}$ \\
 & S2 & -1.759 & 0.861 & 0.738 & 0.738 & -176.1 & 0.053 & $1.22\cdot 10^{8}$ \\
\hline
2015-06-09 & C & 0.04 & -0.018 & 0.066 & 0.027 & -57.0 & 3.389 & $2.35\cdot 10^{12}$ \\
 & S1 & -0.083 & 0.03 & 0.053 & 0.053 & -142.5 & 2.4 & $1.05\cdot 10^{12}$ \\
 & J4 & -0.206 & -0.024 & 0.108 & 0.108 & -144.0 & 0.316 & $3.4\cdot 10^{10}$ \\
\hline
2015-08-01 & C & 0.036 & -0.006 & 0.053 & 0.045 & -69.3 & 3.398 & $1.79\cdot 10^{12}$ \\
 & S1 & -0.089 & 0.032 & 0.071 & 0.071 & -149.3 & 2.399 & $0.6\cdot 10^{12}$ \\
 & J4 & -0.237 & -0.026 & 0.201 & 0.201 & -142.3 & 0.276 & $0.85\cdot 10^{10}$ \\
\hline
2015-12-05 & C & 0.051 & -0.013 & 0.102 & 0.044 & -13.4 & 2.284 & $0.64\cdot 10^{12}$ \\
 & S1 & -0.061 & 0.015 & 0.079 & 0.079 & -144.9 & 2.319 & $4.66\cdot 10^{11}$ \\
 & J4 & -0.307 & -0.029 & 0.313 & 0.313 & -144.2 & 0.219 & $2.79\cdot 10^{9}$ \\
 & S2 & -1.844 & 0.48 & 2.214 & 2.214 & -163.2 & 0.078 & $1.98\cdot 10^{7}$ \\
\hline
2016-01-31 & C & 0.067 & -0.066 & 0.109 & 0.026 & -13.2 & 1.92 & $0.84\cdot 10^{12}$ \\
 & S1 & -0.044 & -0.01 & 0.08 & 0.08 & -144.9 & 2.623 & $0.51\cdot 10^{12}$ \\
 & J4 & -0.348 & -0.053 & 0.216 & 0.216 & -144.2 & 0.172 & $4.6\cdot 10^{9}$ \\
\hline
2016-04-22 & C & 0.055 & -0.036 & 0.114 & 0.034 & -7.1 & 2.02 & $0.65\cdot 10^{12}$ \\
 & S1 & -0.06 & 0.035 & 0.086 & 0.086 & -144.9 & 2.336 & $3.93\cdot 10^{11}$ \\
 & J4 & -0.34 & -0.09 & 0.418 & 0.418 & -133.7 & 0.159 & $1.14\cdot 10^{9}$ \\
 & S2 & -1.219 & 0.835 & 0.45 & 0.45 & -169.7 & 0.029 & $1.76\cdot 10^{8}$ \\
\hline
2016-07-04 & C & 0.046 & -0.044 & 0.116 & 0.036 & -11.3 & 2.262 & $0.68\cdot 10^{12}$ \\
 & S1 & -0.07 & 0.035 & 0.101 & 0.101 & -143.8 & 2.28 & $2.77\cdot 10^{11}$ \\
 & J4 & -0.429 & -0.125 & 0.301 & 0.301 & -147.7 & 0.15 & $2.06\cdot 10^{9}$ \\
\hline
2016-09-05 & C & 0.032 & -0.028 & 0.094 & 0.039 & -15.2 & 3.382 & $1.16\cdot 10^{12}$ \\
 & S1 & -0.082 & 0.065 & 0.105 & 0.105 & -150.9 & 2.292 & $2.59\cdot 10^{11}$ \\
 & J4 & -0.497 & -0.09 & 0.222 & 0.222 & -141.8 & 0.319 & $0.81\cdot 10^{10}$ \\
\hline
2016-10-23 & C & 0.021 & -0.02 & 0.093 & 0.036 & -11.2 & 3.216 & $1.19\cdot 10^{12}$ \\
 & S1 & -0.089 & 0.063 & 0.1 & 0.1 & -139.0 & 1.696 & $2.12\cdot 10^{11}$ \\
 & J4 & -0.556 & -0.086 & 0.185 & 0.185 & -150.3 & 0.332 & $1.21\cdot 10^{10}$ \\
\hline
2016-12-23 & C & 0.015 & -0.041 & 0.11 & 0.047 & -18.6 & 3.824 & $0.93\cdot 10^{12}$ \\
 & S1 & -0.081 & 0.046 & 0.102 & 0.102 & -155.1 & 2.13 & $2.58\cdot 10^{11}$ \\
 & J4 & -0.629 & -0.089 & 0.176 & 0.176 & -150.0 & 0.265 & $1.07\cdot 10^{10}$ \\
\hline
2017-02-04 & C & 0.031 & -0.042 & 0.11 & 0.045 & -19.5 & 3.745 & $0.95\cdot 10^{12}$ \\
 & S1 & -0.062 & 0.038 & 0.085 & 0.085 & -153.7 & 2.313 & $4.02\cdot 10^{11}$ \\
 & J5 & -0.172 & 0.063 & 0.161 & 0.161 & -168.0 & 0.175 & $0.84\cdot 10^{10}$ \\
 & J4 & -0.646 & -0.093 & 0.181 & 0.181 & -158.3 & 0.28 & $1.07\cdot 10^{10}$ \\
 & S2 & -1.166 & 0.4 & 1.3 & 1.3 & -163.9 & 0.101 & $0.75\cdot 10^{8}$ \\
\hline
2017-04-16 & C & 0.0 & -0.007 & 0.13 & 0.048 & -27.2 & 2.714 & $0.55\cdot 10^{12}$ \\
 & S1 & -0.094 & 0.062 & 0.084 & 0.084 & -153.2 & 1.089 & $1.91\cdot 10^{11}$ \\
 & J5 & -0.289 & 0.077 & 0.249 & 0.249 & -154.7 & 0.062 & $1.25\cdot 10^{9}$ \\
 & J4 & -0.753 & -0.037 & 0.185 & 0.185 & -157.8 & 0.251 & $0.91\cdot 10^{10}$ \\
 & S2 & -1.515 & 0.624 & 0.848 & 0.848 & -160.7 & 0.032 & $0.56\cdot 10^{8}$ \\
\hline
2017-06-08 & C & 0.015 & -0.048 & 0.137 & 0.054 & -28.3 & 4.085 & $0.69\cdot 10^{12}$ \\
 & S1 & -0.074 & -0.011 & 0.062 & 0.062 & -153.9 & 1.158 & $3.75\cdot 10^{11}$ \\
 & J5 & -0.228 & 0.003 & 0.236 & 0.236 & -152.8 & 0.168 & $3.76\cdot 10^{9}$ \\
 & J4 & -0.806 & -0.076 & 0.167 & 0.167 & -154.4 & 0.395 & $1.76\cdot 10^{10}$ \\
 & S2 & -1.378 & 0.598 & 0.744 & 0.744 & -144.5 & 0.064 & $1.45\cdot 10^{8}$ \\
\hline
2017-08-06 & C & 0.041 & -0.053 & 0.141 & 0.013 & -31.5 & 1.953 & $1.32\cdot 10^{12}$ \\
 & S1 & -0.03 & 0.019 & 0.086 & 0.086 & -156.8 & 2.151 & $3.63\cdot 10^{11}$ \\
 & J5 & -0.273 & -0.011 & 0.309 & 0.309 & -155.3 & 0.084 & $1.09\cdot 10^{9}$ \\
 & J4 & -0.855 & -0.022 & 0.191 & 0.191 & -158.3 & 0.28 & $0.95\cdot 10^{10}$ \\
\hline
2017-11-06 & C & 0.034 & -0.042 & 0.129 & 0.046 & -43.4 & 2.892 & $0.61\cdot 10^{12}$ \\
 & S1 & -0.05 & -0.004 & < 0.074 & < 0.074 & -148.0 & 1.294 & $>2.94\cdot 10^{11}$ \\
 & J5 & -0.366 & 0.036 & 0.372 & 0.372 & -157.4 & 0.064 & $0.57\cdot 10^{9}$ \\
 & J4 & -0.91 & 0.013 & 0.288 & 0.288 & -166.9 & 0.193 & $2.91\cdot 10^{9}$ \\
 & S2 & -1.374 & 0.634 & 0.538 & 0.538 & -149.2 & 0.048 & $2.07\cdot 10^{8}$ \\
\hline
2018-02-17 & C & 0.081 & -0.068 & 0.083 & 0.029 & -34.3 & 1.431 & $0.75\cdot 10^{12}$ \\
 & S1 & -0.044 & 0.034 & 0.078 & 0.078 & -154.7 & 2.38 & $4.9\cdot 10^{11}$ \\
 & J5 & -0.239 & 0.116 & 0.164 & 0.164 & -155.4 & 0.189 & $0.87\cdot 10^{10}$ \\
 & J4 & -1.0 & 0.11 & 0.185 & 0.185 & -148.3 & 0.156 & $0.57\cdot 10^{10}$ \\
 & S2 & -1.365 & 0.649 & 0.961 & 0.961 & -166.6 & 0.081 & $1.1\cdot 10^{8}$ \\
\hline
2018-04-19 & C & 0.076 & -0.064 & 0.126 & 0.037 & -16.9 & 2.208 & $0.59\cdot 10^{12}$ \\
 & S1 & -0.052 & 0.038 & 0.081 & 0.081 & -154.1 & 2.915 & $0.56\cdot 10^{12}$ \\
 & J5 & -0.317 & 0.058 & 0.135 & 0.135 & -146.6 & 0.317 & $2.16\cdot 10^{10}$ \\
 & J4 & -1.094 & 0.197 & 0.253 & 0.253 & -132.8 & 0.217 & $4.22\cdot 10^{9}$ \\
 & S2 & -1.401 & 0.779 & 0.524 & 0.524 & -145.7 & 0.05 & $2.29\cdot 10^{8}$ \\
\hline
2018-06-16 & C & 0.043 & -0.029 & 0.142 & 0.053 & -29.2 & 2.039 & $3.38\cdot 10^{11}$ \\
 & S1 & -0.081 & 0.029 & 0.079 & 0.079 & -154.1 & 1.301 & $2.6\cdot 10^{11}$ \\
 & J5 & -0.394 & 0.123 & < 0.25 & < 0.25 & -146.6 & 0.196 & $>3.92\cdot 10^{9}$ \\
 & J4 & -1.168 & 0.256 & 0.308 & 0.308 & -132.8 & 0.14 & $1.85\cdot 10^{9}$ \\
 & S2 & -1.533 & 0.842 & 0.413 & 0.413 & -145.7 & 0.017 & $1.25\cdot 10^{8}$ \\
\hline
2018-08-26 & C & 0.051 & -0.069 & 0.117 & 0.02 & -12.4 & 1.02 & $0.53\cdot 10^{12}$ \\
 & S1 & -0.056 & 0.029 & 0.113 & 0.113 & -154.1 & 1.327 & $1.31\cdot 10^{11}$ \\
 & J5 & -0.462 & 0.156 & 0.338 & 0.338 & -146.6 & 0.13 & $1.42\cdot 10^{9}$ \\
 & J4 & -1.249 & 0.331 & 0.386 & 0.386 & -132.8 & 0.083 & $0.69\cdot 10^{9}$ \\
 & S2 & -1.821 & 0.838 & 0.545 & 0.545 & -145.7 & 0.015 & $0.64\cdot 10^{8}$ \\
\hline
2018-12-08 & C & 0.07 & -0.106 & 0.095 & < 0.012 & -33.2 & 1.272 & $>1.38\cdot 10^{12}$ \\
 & S1 & -0.015 & 0.036 & 0.077 & 0.077 & -152.9 & 3.517 & $0.74\cdot 10^{12}$ \\
 & -- & -0.153 & 0.024 & 0.232 & 0.232 & -152.9 & 0.326 & $0.76\cdot 10^{10}$ \\
 & J5 & -0.561 & 0.171 & 0.36 & 0.36 & -168.2 & 0.151 & $1.46\cdot 10^{9}$ \\
 & J4 & -1.303 & 0.404 & 0.456 & 0.456 & -153.4 & 0.125 & $0.75\cdot 10^{9}$ \\
 & S2 & -1.591 & 0.859 & < 0.26 & < 0.26 & -152.2 & 0.014 & $>2.58\cdot 10^{8}$ \\
\end{longtable}
}




\end{appendix}

\end{document}